\documentclass[reprint,superscriptaddress, twocolumn, notitlepage, floatfix]{revtex4-1}
\usepackage[utf8]{inputenc}
\pdfoutput=1
\usepackage{times}
\usepackage[dvipsnames]{xcolor}
\usepackage{array}
\usepackage{verbatim}
\usepackage{multirow}
\usepackage{amsmath}
\usepackage{amssymb}
\usepackage{dsfont}
\usepackage{graphicx}
\usepackage{esint}
\usepackage[unicode=true,
 bookmarks=true,bookmarksnumbered=true,bookmarksopen=false,
 breaklinks=true,pdfborder={0 0 0},backref=false,colorlinks=true]
 {hyperref}
\usepackage{enumitem}
\usepackage{titlesec}
\usepackage[normalem]{ulem}
\usepackage{pifont}
\usepackage{circledsteps}
\hypersetup{
    linkcolor=purple,          % color of internal links (change box color with linkbordercolor)
    citecolor=blue,        % color of links to bibliography
    filecolor=magenta,      % color of file links
    urlcolor=magenta           % color of external links
}

\usepackage{zref-clever}
\newcommand{\cref}[1]{\zcref{#1}}

\zcRefTypeSetup{equation}{
    Name-sg=\textcolor{black}{Eq.},
    name-sg=\textcolor{black}{Eq.},
    Name-pl=\textcolor{black}{Eqs.},
    name-pl=\textcolor{black}{Eqs.}
}
\zcRefTypeSetup{section}{
    Name-sg=\textcolor{black}{Sec.},
    name-sg=\textcolor{black}{Sec.},
    Name-pl=\textcolor{black}{Secs.},
    name-pl=\textcolor{black}{Secs.}
}
\zcRefTypeSetup{appendix}{
    Name-sg=\textcolor{black}{App.},
    name-sg=\textcolor{black}{App.},
    Name-pl=\textcolor{black}{Apps.},
    name-pl=\textcolor{black}{Apps.}
}
\zcRefTypeSetup{figure}{
    Name-sg=\textcolor{black}{Fig.},
    name-sg=\textcolor{black}{Fig.},
    Name-pl=\textcolor{black}{Figs.},
    name-pl=\textcolor{black}{Figs.}
}
\zcRefTypeSetup{table}{
    Name-sg=\textcolor{black}{Table},
    name-sg=\textcolor{black}{Table},
    Name-pl=\textcolor{black}{Tables},
    name-pl=\textcolor{black}{Tables}
}

\renewcommand{\paragraph}[1]{\vspace{0.2cm}{\textit{#1}---\!}} % state-of-the-art PRL style
\def\ie{i.e.,\ }

\usepackage{simpler-wick}
\allowdisplaybreaks[1] % page breaks are allowed, but avoided if possible
\newcommand{\mbf}{\mathbf}

\newcommand{\mcl}{\mathcal}

\newcommand{\mrm}{\mathrm}

\newcommand{\td}{\widetilde}
\newcommand{\ovl}{\overline}

\def\pare#1{\left( #1 \right)}
\def\brak#1{\left[#1\right]}

\def\bra#1{\langle #1 |}
\def\ket#1{| #1 \rangle}

\def\Im{\mathrm{Im}}
\def\Re{\mathrm{Re}}
\def\sgn{\mathrm{sgn}}
\def\ii{\mathrm{i}}
\def\Tr{\mathrm{Tr}}

\def\SUt{{\mathrm{SU}(2)}}
\def\SUf{{\mathrm{SU}(4)}}
\def\Uo{{\mathrm{U}(1)}}
\def\Ut{{\mathrm{U}(2)}}
\def\Uf{{\mathrm{U}(4)}}
\def\Zt{{\mathrm{Z}_2}}

\def\dd{\mathrm{d}}

\def\kk{\mathbf{k}}

\def\qq{\mathbf{q}}
\def\RR{\mathbf{R}}
\def\spin{\varsigma}

\def\PP{\mathbb{P}}

\def\UU{\mathcal{U}}

\def\TK{T_{\rm K}}

\begin{document}
\title{Spin-Valley Anderson Impurity for Moir\'e Systems: Fermi Liquid, Pairing, and Pseudogap}

\author{Yi-Jie Wang}
\thanks{These authors contributed equally to this work.}
\affiliation{International Center for Quantum Materials, School of Physics, Peking University, Beijing 100871, China}

\author{Geng-Dong Zhou}
\thanks{These authors contributed equally to this work.}
\affiliation{International Center for Quantum Materials, School of Physics, Peking University, Beijing 100871, China}

\author{Hyunsung Jung}
\affiliation{Department of Physics and Astronomy, Seoul National University, Seoul 08826, Korea}

\author{Seongyeon Youn}
\affiliation{Department of Physics and Astronomy, Seoul National University, Seoul 08826, Korea}
\affiliation{Center for Theoretical Physics, Seoul National University, Seoul 08826, Korea}

\author{Seung-Sup B.~Lee}
\email{sslee@snu.ac.kr}
\affiliation{Department of Physics and Astronomy, Seoul National University, Seoul 08826, Korea}
\affiliation{Center for Theoretical Physics, Seoul National University, Seoul 08826, Korea}
\affiliation{Institute for Data Innovation in Science, Seoul National University, Seoul 08826, Korea}

\author{Zhi-Da Song}
\email{songzd@pku.edu.cn}
\affiliation{International Center for Quantum Materials, School of Physics, Peking University, Beijing 100871, China}
\affiliation{Hefei National Laboratory, Hefei 230088, China}
\affiliation{Collaborative Innovation Center of Quantum Matter, Beijing 100871, China}

\date{\today}

\begin{abstract}
Recent experiments support that the magic-angle graphene can be modeled by a periodic array of correlated quantum impurities, immersed in a Dirac sea. 
This work \textit{analytically} tackles a spin-valley Anderson impurity, featuring a general (anti-)Hund's interaction ($J_D$, $J_S$) that can originate from electron-phonon couplings. 
We derive its full phase diagram, which encompasses rich continuous local phase transitions, and presents a unified origin for pairing potential and pseudogap. 
In particular, $J_D$ favors a valley doublet, and we show it drives a BKT transition out of heavy Fermi liquid, to an anisotropic doublet phase exhibiting a non-analytic zero-energy kink in the impurity spectral function. 
$J_S$ drives a second-order transition out of heavy Fermi liquid, to a local singlet phase, with a non-Fermi liquid critical point. 
We analyze the pairing potential across the phase diagram, and unveil their ubiquitous existence triggered by the (anti-)Hund's multiplet splitting. 
Crucially, we show the pseudogap shoulders in the spectral function represent multiplet excitations induced by an injected electron or hole. 
All results are obtained analytically, using techniques including bosonization--refermionization, with further verification by numerical renormalization group calculations. 
Then we derive the correlation self-energy ans\"atze that account for pseudogap, and apply to the magic-angle graphene lattice. 
% \old{Based on exact solutions, we then obtain correlation self-energy ans\"atze, which can capture the pseudogap in the lattice problem. }
\end{abstract}

\maketitle

\paragraph{Introduction}
Moir\'e hetero-structures have opened up a new stage to engineer electronic flat bands, providing thrilling new possibilities to study exotic correlations besides conventional materials \cite{andrei_graphene_2020, mak_semiconductor_2022, nuckolls_microscopic_2024}. 
In a variety of systems, flat bands originate from the formation of local orbitals at the moir\'e length scale \cite{Wu_Hubbard_2018, Koshino_maximally_2018, haule_2019_mott, Liu_2019_pseudoLL, calderon_2020_interaction, Reddy_2023_artificial, liu_ideal_2025}, akin to the atomic $d$ or $f$ shells, but with richer inner degrees of freedom such as layer and valley. 
More importantly, with an underlying lattice, electrons interact not only through the Coulomb repulsion, but also through microscopic processes such as phonons \cite{chen_strong_2024, birkbeck_quantum_2025}, which act non-trivially on the new degrees of freedom. 
These aspects imply that, even for a model as simple as a local orbital, new physics is yet to be explored. 

One paradigmatic moir\'e material is the magic-angle twisted bilayer/trilayer graphene (MATBG/TTG) \cite{BM_2011}, where correlated (Chern) insulators \cite{Cao_2018_CI, Lu_2019_superconductors, Choi_2019_electronic, Kerelsky_2019_maximized, Jiang_2019_charge, Xie_2019_spectroscopic, Nuckolls_2020_strongly, Nuckolls_2023_quantum, kim_imaging_2023}, unconventional superconductivity \cite{Cao_2018_CI, Lu_2019_superconductors, Yankowitz_2019_tuning, Arora_2020_superconductivity, Saito_independent_2020, Stepanov_untying_2020, Liu_tuning_2021, Cao_2021_nematicity, cao_pauli-limit_2021, park_tunable_2021, hao_electric_2021, kim_evidence_2022, liu_isospin_2022, gao_2024_doubleedgedrole, tanaka_superfluid_2025, banerjee_superfluid_2025} with pseudogaps \cite{Oh_2021_evidence, park_2025_simultaneoustransporttunnelingspectroscopy, kim_2025_resolvingintervalleygapsmanybody}, and strange metal transport \cite{Polshyn_2019_Large, Cao_2020_strange, Jaoui_2022_quantum} are discovered. 
It is then realized that the topological flat bands \cite{Zou_2018_band, song_all_2019, Po_2019_faithful, Ahn_2019_NNfail, wang_chiral_2021} can be disentangled into moir\'e local orbitals ($f$) \cite{haule_2019_mott, Liu_2019_pseudoLL, calderon_2020_interaction} hybridizing with itinerant Dirac bands ($c$) \cite{song_magic-angle_2022, shi_heavy-fermion_2022, Yu_2023_THF_TSTG, singh_topological_2024, herzog_2025_efficient}. 
Coulomb repulsion generates a large Hubbard $U\!\sim\!60$meV on each $f$ orbital, promoting the formation of local moments. 
Fermi liquid (FL) phases of heavy fermion or mixed-valence types can form via the Kondo screening by the $c$ electrons. 
Various phenomena get explained within this framework \cite{Chou_2023_Kondo, zhou_kondo_2024, Rai_2023_DMFT, Hu_2023_Kondo, Hu_2023_Symmetric, Chou_2023_scaling, Datta_2023_heavy, Lau_2023_topological, calugaru_thermoelectric_2024, herzog_2025_kekule, crippa_2025_dynamicalcorrelation, calugaru_2025_obtainingspectral}, including the Pomeranchuk effect \cite{Rozen_2021_entropic, Saito_2021_isospin, zhang_2025_heavyfermions}, cascade transitions in scanning tunneling microscope (STM) spectrum \cite{Zondiner_2020_cascade, Wong_2020_cascade} and compressibility \cite{liu_isospin_2022,hu_2024_linkcascade,zhang_2025_heavyfermions}. 
The coexistence of correlated $f$ and light $c$ is supported by thermoelectric transport \cite{merino_interplay_2025} and the quantum twisting microscopy \cite{xiao_2025_interacting}. 
The Kondo resonance is also recently observed in STM \cite{kim_2025_resolvingintervalleygapsmanybody}. 

The superconducting gap coexists with a larger pseudogap \cite{park_2025_simultaneoustransporttunnelingspectroscopy, kim_2025_resolvingintervalleygapsmanybody}, which appears at an energy scale of 1meV, comparable to the phonon-mediated electron-electron interaction $J$ \cite{Wu_2018_SCop, Lian_2019_SCac, Koshino_epc_2020, Cea_2021_coulomb, liu_moire_2022, Liu_2023_electronkphonon,zhu_2025_microscopictheory,lau_oscillate_2025}. 
In the Hilbert space of a local $f$ orbital, $J$ induces anti-Hund's splitting favoring spin-singlet configurations \cite{Dodaro_2018_phases, Angeli_2019_valleyJT, Andrea_2022_local, wang_2025_epc, shi_2025_optical}, in stark contrast to the atomic $f$ shells, where the conventional Hund's rule governs. 
Based on the assumption that a local FL emerges at an energy below $\mcl{O}(J)$, a previous work demonstrates that quasiparticles experience an attractive renormalized interaction \cite{wang_molecular_2024}. 
In addition, recent dynamical mean-field theory (DMFT) works also show that $J$ leads to pseudogaps of size $\mcl{O}(J)$ and different quantum phases \cite{youn_hundness_2024}. 
These studies strongly indicate that the anti-Hund's coupling can account for both pairing and pseudogap.

In this context, we consider a spin-valley Anderson impurity model (SVAIM) with a general valley-anisotropic (anti-)Hund's interaction, which can describe a correlated orbital in the hetero-strained MATBG/TTG \cite{Parker_2021_strain, Wagner_2022_global}. 
We study its full phase diagram analytically, with support by numerical renormalization group (NRG) calculations, and analyze the occurrence of pairing and pseudogap. 
Analytical solutions to the finite-size many-body spectrum and RG flows are provided in a companion work \cite{wang_bosonization_2025}.

\begin{figure}[t]
    \centering
    \includegraphics[width=1.0\linewidth]{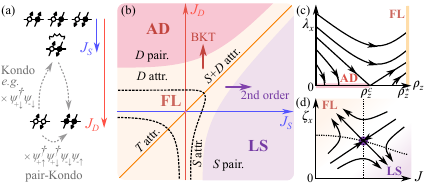}
    \caption{
    (a) Two-electron impurity states. 
    White and black circles indicate valleys $l\!=\!\pm$, respectively, and arrows indicate spin $s \!=\! {\uparrow} {\downarrow}$.
    \cref{eq:H_AH} lowers the energy of the singlet $S$ (the valley doublet $D$) by $J_S$ ($J_D$), compared to the spin triplet $T$. 
    (b) Schematic phase diagram of SVAIM.
    The FL phase is separated from the anisotropic doublet (AD) phase by a BKT transition, and separated from the local singlet (LS) phase by a second-order transition. 
    Dashed lines mark the crossover boundaries within the FL phase, beyond which renormalized interactions turn attractive (``attr.'') in certain channels ($S$, $D$, or $T$). 
    AD and LS also exhibit enhanced pairing (``pair.'') susceptibilities in the corresponding channels, despite no quasiparticle exists. 
    (c,d) Schematic RG flows for the BKT and second-order transitions, respectively. 
    For NRG results corresponding to panels (b)--(d), see End Matter and \cref{app:NRG} in Supplementary Material (SM) \cite{supplement}. 
    }
    \label{fig:intro}
\end{figure}

\paragraph{Model}
Hetero-strain in MATBG/TTG lifts the otherwise degenerate orbital angular momenta of $f$ \cite{song_magic-angle_2022} into bonding and anti-bonding levels \cite{herzog_2025_efficient}, while leaving the valley ($l \!=\! \pm$) and spin ($s \!= \uparrow\downarrow$) symmetries intact. 
Upon electron (hole) doping, the bonding (anti-bonding) level remains frozen \cite{herzog_2025_kekule, crippa_2025_dynamicalcorrelation}, thus it suffices to model the active level with electron operator $f_{ls}$. 
We introduce Pauli matrices $\sigma^{\mu}$ and $\spin^{\nu}$ ($\mu,\nu \!=\! 0,x,y,z$) for valley and spin, respectively. 
Besides charge-$\Uo$ symmetry generated by $\sigma^0 \spin^0$, there are spin-$\SUt$ symmetry generated by $\spin^{x,y,z}$, valley-$\Uo$ symmetry generated by $\sigma^z$, and a $C_{2z}$ symmetry that swaps the two valleys represented by $\sigma^x$. 
% The model preserves charge-$\Uo$ symmetry generated by $\sigma^0 \spin^0$, spin-$\SUt$ symmetry generated by $\spin^{x,y,z}$, valley-$\Uo$ symmetry generated by $\sigma^z$, and a $C_{2z}$ symmetry that interchanges the two valleys represented by $\sigma^x$.

The SVAIM is described by $H \!=\! H_0 \!+\! H_{\rm imp} \!+\! H_{\rm c}$. 
The bath Hamiltonian $H_0 \!=\! \int \! \dd x \sum_{ls} \! \psi^\dagger_{ls}(x) (\ii \partial_x) \psi_{ls}(x)$ is chosen as a chiral fermion for the convenience of analytical treatment, $H_{\rm c} \!=\! \sqrt{2\Delta_0} \sum_{ls} f^\dagger_{ls} \psi_{ls}(0)$ is the hybridization between the impurity and bath states, and $H_{\rm imp} \!=\! \frac{U}{2}(N-2)^2 \!+\! H_{\rm AH}$ is the impurity Hamiltonian. 
$U$ is the Hubbard repulsion, $N$ counts the impurity electron number, and $H_{\rm AH}$ is a general symmetry-allowed (anti-)Hund's interaction. 
By symmetry, the two-electron subspace can split into a spin triplet ($T$), a valley doublet ($D$) carrying total valley charge $L^z \!=\! \pm 2$, and a singlet ($S$) [\cref{fig:intro}(a)]. 
Therefore, we parametrize
\begin{align}  \label{eq:H_AH}
H_{\rm AH} = -\frac{J_S}{2} \sum_{ll'} f^\dagger_{l\uparrow} f^\dagger_{\ovl{l}\downarrow} f_{\ovl{l'} \downarrow} f_{l' \uparrow} - J_D \sum_{l} f^\dagger_{l\uparrow} f^\dagger_{l\downarrow} f_{l\downarrow} f_{l\uparrow} \ ,
\end{align}
which lowers the energy of $S,D$ relative to $T$ by $J_{S,D}$, respectively. 
$J_{S,D} \!>\! 0$ thus corresponds to an anti-Hund's rule. 
Since $S \oplus D$ forms the ``valley triplet'' of a valley-$\SUt$ group generated by $\sigma^{x,y,z}$, $J_S \!\neq\! J_D$ describes valley-anisotropy. 
As $J_{S,D}$ originate from phonon-mediated interactions, they are much weaker than the Coulomb repulsion $U$.

At energy $\omega \!\ll\! \mcl{O}(U)$, charge fluctuations on $f$ get frozen, turning into virtual processes that induce a Kondo coupling, $|\Xi\rangle \langle\Xi'| \ \psi^\dagger \sigma^{\mu} \spin^{\nu} \psi $ (\cref{sec:Kondo-coupling-general} in SM \cite{supplement}), where $|\Xi\rangle$ runs over two-electron states, and $:\cdots:$ normal-orders bilinear operators of bath electrons. 
The Kondo couplings are anti-ferromagnetic ($>\!0$) and of order $\mcl{O}(\frac{\Delta_0}{U})$. 

Several limits are well studied. 
$J_S\!=\!J_D\!=\!0$ enjoys a full $\SUf$ symmetry, and one channel of $\SUf$ bath is known to exactly screen the $\SUf$ impurity moment \cite{Andrei_1983_Solution, Parcollet_1998_overscreened, zhou_kondo_2024}. 
Increasing $J_S\!=\!J_D$ in either sign breaks $\SUf$ into commuting spin-$\SUt$ and valley-$\SUt$ groups. % As $J_S \!=\! J_D \!>\! 0$ grows, $T$ gradually disappears from the low-energy space. 
% Consequently, the bath spins $s\!=\!\uparrow\downarrow$ degrade to two degenerate channels that carry valley-$\SUt$ moments to screen the valley triplet. 
% The solution is also a FL \cite{nozieres_kondo_1980, Andrei_1983_Solution}.
As $J_S \!=\!J_D \!<\! 0$ becomes more negative, $S,D$ gradually disappears from the low-energy space, leaving an effective $\SUt$ spin-1 two-channel Kondo model, which is FL at low energy \cite{nozieres_kondo_1980, Andrei_1983_Solution}.
Physics at $J_S \!=\!J_D \!>\!0$ is equivalent to $J_S\!=\!J_D\!<\!0$ with the roles of ``spin'' and ``valley'' exchanged. 
Since removing or recovering either triplet does not interrupt the exact screening, we conclude that the full diagonal line $J_S\!=\!J_D$ is FL. 

As valley-$\SUt$ is not guaranteed in real materials, $J_{S}\neq J_D$. 
Depending on which multiplet is the lowest, we divide the parameter space into three regimes. 
In the triplet regime ($J_{S,D} \!<\! 0$), splitting occurs in the high-energy subspace, not affecting the low-energy FL phase. 
In the doublet ($0 \!<\! J_D,J_S \!<\! J_D$) or singlet ($0 \!<\! J_S$, $J_D \!<\! J_S$) regimes, however, splitting can eventually remove the Kondo resonance. 

\paragraph{FL phase}
The FL phase manifests a coherent Kondo peak in the impurity spectral function $A_f(\omega)$ at $\omega \!=\! 0$, which adiabatically evolves from the non-interacting resonant level ($U \!=\! J_{S,D} \!=\!0$) and indicates the formation of heavy quasiparticles at energies lower than the Kondo temperature $\TK$. 
$\TK$ is the inverse of quasiparticle lifetime due to hybridization with the bath. 
It decreases exponentially when increasing $\frac{U}{\Delta_0}$, and defines a universal energy scale at low energy. 
By symmetry, the renormalized interactions between quasiparticles obey the same form as the bare ones, and we denote them as $\td{U}$ and $\td{J}_{S,D}$. 
Due to the continuous symmetry generated by $\sigma^0\spin^0$ (charge), $\sigma^0\spin^z$ (spin), and $\sigma^z\varsigma^0$ (valley), the corresponding quasi-particle susceptibilities $\chi_{c,s,v}$\cite{hewson_fermi_1993, hewson_renormalized_1993, hewson_renormalized_2004, nishikawa_renormalized_2010} can be exactly obtained by Ward identities \cite{yamada_perturbation_1975, yamada_perturbation_1975-1, yoshimori_perturbation_1976} as 
\begin{align}    \label{eq:Ward}
    \frac{\pi \TK}{4}\begin{pmatrix}
        \chi_c \\
        \chi_s \\
        \chi_v \\
    \end{pmatrix} = 1  -  \frac{1}{\pi \TK} 
    \pare{
    \begin{array}{rrr}
        3 & -1 & -\frac{1}{2} \\
        -1 & 1 & \frac{1}{2} \\
        -1 & -1 & \frac{1}{2} \\
    \end{array}
    }
 \! \begin{pmatrix}
        \td{U} \\
        \td{J}_D \\ 
        \td{J}_S \\
    \end{pmatrix} \ ,
\end{align}
as detailed in \cref{sec:effective-interaction-1} in SM \cite{supplement}.
Knowledge of $\chi_{c,s,v}$ allows us to constrain $\td{U}$ and $\td{J}_{S,D}$. 

The strong $U$ freezes charge fluctuation at the Kondo energy scale, implying $\chi_c \!\ll\! T^{-1}_{\rm K}$. 
In the $\SUf$ symmetric limit, since $\td{J}_D \!=\! \td{J}_S \!=\! 0$, this constraint fixes $\td{U} \!=\! \frac{\pi\TK}{3}$, meaning all channels are repulsive.  
Increasing $J_D \!=\! J_S$ until $J_S \! \gg\! \TK$, as $S \oplus D$ does not carry spin, $\chi_s \!\ll\! T^{-1}_{\rm K}$ also freezes, leading to an attraction $\td{U} \!-\! \td{J}_S \!=\! -\frac{\pi\TK}{3}$ in the $S\!\oplus\! D$ channel. 
Between the two limits, $\chi_{s}$ interpolates smoothly, implying a $J_\star$ where the $S \oplus D$ channel turns attractive, as marked by a dashed line in \cref{fig:intro}(b). 
Reversely, the $T$ channel turns attractive beyond $-J_\star$ \cite{nishikawa_renormalized_2010}.

\paragraph{Doublet regime}
We consider the low-energy theory involving bath and the valley doublet, spanned by $|L^z\rangle$ ($L^z = 2,\ovl{2}$), with Pauli matrices $\Lambda_z \!=\! |2\rangle\langle 2| \!-\! |\ovl{2}\rangle\langle\ovl{2}|$, and $\Lambda_+ \!=\! \Lambda_-^\dagger \!=\! |2\rangle \langle\ovl{2}|$. 
The only symmetry-allowed Kondo coupling reads $H_z \!=\! (2\pi\lambda_z) \Lambda_z :\! \psi^\dagger \sigma^z \spin^0 \psi \!:\!|_{x=0}$, where $\lambda_z \!\sim\! \frac{\Delta_0}{U} \!>\! 0$ is anti-ferromagnetic. 
Crucially, $\Lambda_\pm$ cannot appear in the Kondo coupling, as they alter the impurity valley-charge by $\pm 4$, which cannot be compensated by a bilinear bath electron operator. %
Nonetheless, two successive Kondo scatterings can first excite $|L^z\rangle$ to the $S$ or $T$ manifold, and then lower it to $|\ovl{L^z}\rangle$ [\cref{fig:intro}(a)]. % Nonetheless, two successive Kondo scatterings can scatter $|L^z\rangle$ to $|\ovl{L^z}\rangle$ via intthe $S$ or $T$ manifold, and then lower it to  [\cref{fig:intro}(a)]. 
Such virtual multiplet fluctuations couple $\Lambda_\pm$ to a quartic bath operator, which scatters an electron pair at once. 
We thus dub it as the pair-Kondo (PK) coupling. 
By symmetries, it must take the form of 
\begin{align} \label{eq:Hx-maintext}
    H_x \!=\! (2\pi)^2{\lambda_x}{x_c} \!\cdot\! \Lambda_+ \!\cdot\! \psi^\dagger_{-\uparrow} \psi^\dagger_{-\downarrow} \psi_{+\downarrow} \psi_{+\uparrow}\Big|_{x=0} \!+\! h.c.\ ,
\end{align}
where $\lambda_x$ is real, and $x_c$ is a microscopic length scale. 
A second-order perturbation theory estimates $\lambda_x \!\sim\! \mcl{O}(\frac{\Delta_0^2}{U^2} \frac{1}{x_c J})$.
Hereafter we always reserve $J$ for the minimal multiplet excitation energy, which is $J \!=\! \min\{ J_D, J_D\!-\!J_S \}$ in the doublet regime. 
The sign of $\lambda_x$ does not affect the physics, as it can be changed by the gauge transformation $\ii\Lambda_z$.  
In sum, the effective Hamiltonian in the doublet regime is $H_{\rm PK} \!=\! H_0 \!+\! H_z \!+\! H_x$. 

We bosonize the chiral bath as $\psi_{ls}(x) \!\sim\! \frac{e^{-\ii \phi_{ls}(x)}}{\sqrt{2\pi x_c}}$ \cite{toulouse_1969, Mapping_Emery_1992, Kotliar_toulouse_1996, von_delft_bosonization_1998, vonDelft_1998_finitesize, zarand_analytical_2000, giamarchi2003quantum, Schiller_phasediagram_2008, Krishnan_2024_kondo}. (See Ref.~\cite{wang_bosonization_2025} for details.) %where $\frac{1}{2\pi} \partial_x \phi_{ls}(x) = :\! \psi^\dagger_{ls}(x) \psi_{ls}(x) \!:$ represents the electron density and $e^{-\ii \phi_{ls}(x)}$ serves as a Jordan-Wigner string that implements fermion anti-commutation within the same flavor. 
In $H_{\rm PK}$, only one combination of boson fields, $\phi_v \!=\!\frac{1}{2} \!\sum_{ls} l \!\cdot\! \phi_{ls}$, couples to the impurity, which corresponds to the fluctuation of valley charges. 
The remaining three orthogonal channels decouple, including densities of electric charge ($\phi_c$), spin ($\phi_s$), and valley-contrasting spin ($\phi_{vs}$). 
In subspaces that diagonalize $\Lambda_z \!=\! \pm$, $H_z$ generates a phase shift of $l\Lambda_z\rho_z$ to each electron flavor $ls$, where $\rho_z \!=\! \frac{{\arctan}(\pi\lambda_z)}{\pi} \!\in\!(0,\frac{1}{2})$
\cite{Andrei_1983_Solution, vonDelft_1998_finitesize, Krishnan_2024_kondo}. 
A unitary transformation $\UU \!=\! e^{\ii 2\rho_z \Lambda_z\phi_v(0)}$ is then implemented to absorb this phase shift, such that $\ovl{H}_{\rm PK} \!=\! \UU H_{\rm PK} \UU^\dagger \!=\! H_0 \!+\! \ovl{H}_x$. 
$H_0$ takes the same form, while the PK term becomes
\begin{align}    \label{eq:ovlHx}
    \ovl{H}_x \!=\! \UU H_x \UU^\dagger \!=\! \frac{\lambda_x}{x_c} \!\cdot\! \Lambda_+ \!\cdot\! e^{-\ii\gamma\phi_v(0)} \!+\! h.c.\ ,
\end{align} 
where $\gamma \!=\! 2 \!-\! 4\rho_z$. The vertex operator $e^{\ii \gamma \phi_v}$ has auto-correlation $ |\tau \!-\! \tau'|^{-\gamma^2}$ in the temporal domain, and therefore has scaling dimension $[e^{\ii \gamma \phi_v(x)}] \!=\! \frac{\gamma^2}{2}$. 
Given $[\ovl H_x]=1$, there must be $[\lambda_x]=1-\frac{\gamma^2}2$, suggesting the RG flow
\begin{align}   \label{eq:dlx}
    \frac{\dd\lambda_x}{\dd \ell} \!=\! \left(\! 1 \!-\!\frac{\gamma^2}{2} \!\right)  \lambda_x \!=\! (-1+8\rho_z-8\rho_z^2) \lambda_x \ . 
\end{align}
$\rho_z^c \!=\!\frac{1}{2} \!-\! \frac{1}{2\sqrt{2}} \!\approx\! 0.1464$ is thus a critical value, above which $\lambda_x$ turns relevant. 

$\rho_z$ scales, too. 
$\ovl H_x$ contributes a factor of $\langle T_\tau e^{-\int\dd\tau \ovl{H}_x(\tau)} \rangle_0$ to the partition function, which can be expanded perturbatively in $\lambda_x$.
The result can be mapped to a classical Coulomb gas \cite{wang_bosonization_2025}, where each flipping event $\Lambda_\pm$ is mapped to a particle on the $\tau$ axis with ``electric charge'' $\pm$, respectively, created with probability (fugacity) $\lambda_x$. 
$\gamma^2$ determines the inter-event correlations and is the effective Coulomb strength. 
RG proceeds as two particles move close to form a dipole, which screens the Coulomb interaction among remaining particles, implying $\frac{\dd(\gamma^2)}{\dd \ell} \!\propto\! - \lambda_x^2 \gamma^2$. 
Further examination finds the prefactor as $4$, namely,
\begin{align}   \label{eq:drz}
    \frac{\dd \rho_z}{\dd\ell} \!=\! (1 \!-\! 2\rho_z) \lambda_x^2 \ . 
\end{align}
$\rho_z$ always grows since $\rho_z \!\in\! (0,\frac{1}{2})$. 
\cref{eq:dlx,eq:drz} are exact in $\rho_z$ but approximate to $\mcl{O}(\lambda_x^3)$ order. 
The RG flow is drawn in \cref{fig:intro}(c), belonging to the Berezinskii--Kosterlitz--Thouless (BKT) type \cite{Galpin_2005_quantumphasetransition, Galpin_2006_renormalization}. A continuous fixed line extends along $\lambda_x \!=\! 0, \rho_z \!<\! \rho_z^c$, which we term as the AD phase. 
Beyond $\rho_z^c$, $\lambda_x$ grows into a strong-coupling regime, which is a FL phase confirmed by the exact solution below. 

\paragraph{FL in doublet regime}
At $\rho_z^\star \!=\! \frac{1}{4}$ and arbitrary $\lambda_x$, the vertex operator appearing in \cref{eq:ovlHx} reads $e^{-\ii \phi_v(x)}$, hence can be refermionized as $\psi_v(x) \!\sim\! \frac{e^{-\ii \phi_v(x)}}{\sqrt{2\pi x_c}}$ \cite{toulouse_1969, Mapping_Emery_1992, Kotliar_toulouse_1996, von_delft_bosonization_1998, vonDelft_1998_finitesize, zarand_analytical_2000}. 
$\Lambda_-$ can also be mapped into a fermion $f_v$ after attaching a Jordan-Wigner string. 
Therefore, $\ovl{H}_x \!=\! \sqrt{\frac{2\pi}{x_c}}\lambda_x f^\dagger_v \psi_v(0) \!+\! h.c.$, describing a resonant level $f_v$ that hybridizes with $\psi_v$. The resonance linewidth $\frac{\pi \lambda_x^2}{x_c}$ is identified as $\TK$. 
As detailed in Ref.~\cite{wang_bosonization_2025}, $\ovl{H}_{\rm PK}$ is now exactly solvable. We find that at $T \!\ll\! \TK$, the impurity entropy freezes to $0$, and the static longitudinal susceptibility $\chi_z$ saturates to $\mcl{O}(\TK^{-1})$, implying exact screening. 
The impurity dynamic susceptibilities of $\Lambda_z$ and $\Lambda_\pm$ (denoted as $\Im \chi_z(\omega \!+\! \ii0^+)$ and $\Im \chi_x(\omega \!+\! \ii0^+)$, respectively) scale as $\sim \! \omega$ at $\omega \ll \TK$, 
also confirming the FL behaviors. 

Since bringing down $S,T$ states into the low-energy space does not interrupt the exact screening, FL in the doublet regime can cross over to FL in other regimes. 
Nonetheless, the renormalized interaction behaves differently. 
A special limit is $J_S \!=\! 0$, where the global spin-$\SUt$ is enhanced into two independent $\SUt$ rotations in $l \!=\! \pm$ valleys, generated by $\frac{\sigma^0 \pm \sigma^z}{2} \spin^{x,y,z}$. 
Such symmetry locks $S$ and $T$ as degenerate, hence $\td{J}_S \!=\!0$. 
When $\TK \!\ll\! J_D$, $\chi_c,\chi_s \!\ll\! \TK^{-1}$, and solving \cref{eq:Ward} shows that $D$ is the only attractive channel, with $\td{U}\!-\!\td{J}_D \!=\! -\pi \TK$ \cite{nishikawa_convergence_2012}. 
As splitting $J_S \!\not=\! 0$ in the high-energy end should not affect low-energy physics, the $D$ channel will remain attractive as long as $\TK \!\ll\! J$.

\paragraph{Anisotropic doublet}
At the fixed line ($\lambda_x=0$, $\rho_z<\rho_z^c$), $\Lambda_z \!=\! \pm$ is conserved, implying its static susceptibility to exhibit the Curie's law, $\chi_v \!\sim\! T^{-1}$. 
On the other hand, $\Lambda_\pm$ is dressed by $\UU \Lambda_\pm \UU^\dagger \!=\! \Lambda_\pm e^{\pm \ii 4 \rho_z \phi_v(0)}$, with $\UU$ defined above \cref{eq:ovlHx}, implying its correlation function to scale as $\chi_x(\tau) \!\sim\! |\tau|^{-(4\rho_z)^2}$. 
Therefore, the dynamic susceptibility scales in a non-universal power law, $\Im \chi_x(\omega \!+\!\ii0^+) \!\sim\! \sgn(\omega) |\omega|^{16\rho_z^2-1}$, and the static susceptibility diverges as $\chi_{x} \!\sim\! T^{16\rho_z^2-1}$.

\begin{figure}[tb]
    \centering
    \includegraphics[width=1\linewidth]{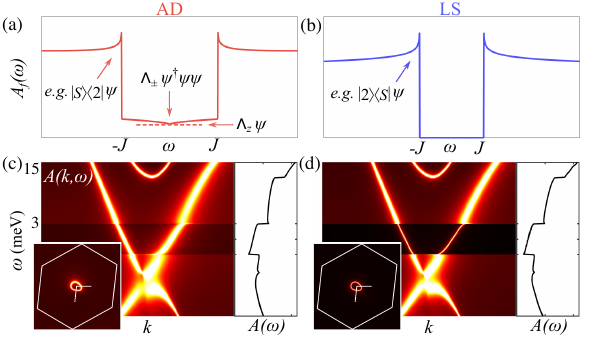}
    \caption{
    (a, b) $A_f(\omega)$ obtained from bosonization. 
    Pseudogap shoulders ($A_f^{(3)}$) are multiplet excitations, induced by scattering a bath electron ($\omega\!\le\!-J$) or hole ($\omega\!\ge\!J$). They are symmetrically pinned around the Fermi energy. 
    For AD, residual longitudinal coupling contributes a constant background ($A^{(1)}_f$, dashed line), while the irrelevant PK coupling contributes a non-analytic kink ($A^{(2)}_f$) above it. 
    (c, d) Lattice spectral function $A(k,\omega)$ in one valley, obtained using the \textit{ans\"atze} of $\Sigma_f(\omega)$ derived from single impurity. 
    Insets are contours at $\omega \!=\! 0$, with hexagons denoting the strained moir\'e Brillouin zone, and white lines indicating the $k$-path of main figures. 
    AD does not have a well-defined Fermi surface, while LS is a Fermi liquid of $c$ electrons. 
    $A(\omega) \!=\! \int\frac{\mrm{d}^2k}{(2\pi)^2} A(k,\omega)$ is the total density of states.}
    \label{fig:Af}
\end{figure}

The impurity spectral function $A_f(\omega)$ is proportional to the scattering $\mcl{T}$-matrix of bath electrons \cite{Costi_2000_Kondo,Bulla_2008_nrg,Moca_2019_quantumcriticality}, and the latter remains well-defined in the downfolded model $H_{\rm PK}$. 
According to the equation of motion $[H_{z}\!+\!H_x, \psi_{+\uparrow}(0)]$, $\psi_{+\uparrow}(0)$ scatters into two pieces: $\td{f}^{(1)}_{+\uparrow} \!=\! \lambda_z \Lambda_z \psi_{+\uparrow} \big|_{x=0}$ and $\td{f}^{(2)}_{+\uparrow} \!=\! (2\pi\lambda_x x_c) \Lambda_- \psi^\dagger_{+\downarrow} \psi_{-\downarrow} \psi_{-\uparrow}\big|_{x=0}$, where $\lambda_{z,x}$ are the un-renormalized parameters in $H_{\rm PK}$. 
The $\mcl{T}$-matrix is then given by the Green's function of $\td{f}^{(1)} \!+\! \td{f}^{(2)}$, whose long-time behavior is governed by the AD fixed point Hamiltonian. 
As $\Lambda_z$ is conserved there, $\td{f}^{(1,2)}_{+\uparrow}$ do not mix. 
The time-evolution of $\td{f}^{(1)}_{+\uparrow}$ is solely governed by $\psi_{+\uparrow}$, which produces a spectrum proportional to the bath density of states, $A_f^{(1)}(\omega) \!\sim\! \mrm{const}$. 
Conversely, as $\UU \td{f}^{(2)}_{+\uparrow} \UU^\dagger \!\sim\! \Lambda_- e^{-\frac{\ii}{2}(\phi_c+\phi_s+\phi_{vs})} e^{\ii(\frac{3}{2}-4\rho_z) \phi_v}|_{x=0}$, $\td{f}^{(2)}_{+\uparrow}$ is dressed by a non-universal phase factor. 
Its Green's function hence scales as $|\tau|^{-\alpha_2}$ with $\alpha_2 \!=\! \frac{3}{4} \!+\! (\frac{3}{2} \!-\! 4\rho_z)^2$, implying a spectral function $A_f^{(2)}(\omega) \!\sim\! |\omega|^{\alpha_2-1}$. 
$0 \!<\! \rho_z \!<\! \rho_z^c$ maps to $2 \!>\! \alpha_2\!-\!1 \!>\! 2\!-\!\sqrt{2} \! \approx \! 0.5858$ monotonically, hence near the BKT transition $\rho_z^c$, $A_f^{(2)}(\omega)$ behaves as a non-analytic kink depicted in \cref{fig:Af}(a), which contrasts significantly to the Kondo peak in FL.

Local pairing susceptibility in the $D$ channel is found enhanced in the AD phase \cite{youn_hundness_2024}. 
We find this is due to the residual PK coupling $\lambda_x$ [\cref{eq:ovlHx}] at intermediate energy scales, which couples the impurity to a bath electron pair in the $D$ channel, and allows such pair excitations to lower energy by forming a singlet with the impurity (\cref{sec:effective-interaction-2} in SM \cite{supplement}). 
Meanwhile, an individual bath electron cannot benefit from such effect. 
Therefore, as the PK model inherits the symmetry charges of the Anderson model, while the residual charge fluctuation on $f$ has been absorbed into the bath, such a pairing enhancement in the bath also reflects a pairing enhancement on $f$. 
It will be interesting for future work to investigate whether such ``attraction'' can lead to superconductivity on the lattice. 

\paragraph{Pseudogap}
Following the same reasoning, we investigate $A_f(\omega)$ at $\omega \!\sim\! \mcl{O}(J)$, where the effective theory is the Kondo Hamiltonian. 
For simplicity, let us assume $0 \!<\! J_S \!<\! J_D$, and first include $S$ into the low-energy space, so that the Kondo Hamiltonian reads $H_{\rm K} \!=\! H_0 \!+\! H_{J} \!+\! H_z \!+\! H_{x0}$. 
$H_J \!=\! J |S\rangle\langle S|$ denotes the multiplet excitation with $J \!=\! J_D\!-\!J_S$, and the Kondo coupling between $S$ and $D$ reads $H_{x0} \!=\! (2\pi\zeta_x) \Theta_+ :\!\psi^\dagger \sigma^- \spin^0 \psi\!: \!+ h.c.$, where 
$\sigma^{\pm} \!=\! \frac{\sigma^x \pm \ii \sigma^y}{2}$, and $\Theta_+ \!=\! \Theta_-^\dagger \!=\! |S\rangle \! \langle \ovl{2}| \!+\! |{2}\rangle \! \langle S|$. 
Therefore, $\psi_{+\uparrow}(0)$ also scatters into $\td{f}^{(3)}_{+\uparrow} \!=\! (2\pi\zeta_x) \Theta_- \psi_{-\uparrow}$, whose motion will contribute an $A_f^{(3)}(\omega)$. 
The phase shift dresses $\td{f}^{(3)}_{+\uparrow}$ into $\UU \td{f}^{(3)}_{+\uparrow} \UU^\dagger \!\sim\! \Theta_- \!\cdot\! e^{-\frac{\ii}{2}(\phi_{c}+\phi_s-\phi_{vs})} e^{\ii(\frac{1}{2} - 2\rho_z)\phi_v} \big|_{x=0}$.
Since $\Theta_-$ excites $D$ to $S$, the minimal energy cost is $J$, leading to a factor $\theta(|\omega|-J)$ in $A^{(3)}_f(\omega)$. 
Meanwhile, the correlation function of the bath part scales as $|\tau|^{-\alpha_3}$, with $\alpha_3 \!=\! \frac{3}{4} \!+\! (\frac{1}{2} \!-\! 2\rho_z)^2 \!<\! 1$. 
Consequently, we find $A_f^{(3)}(\omega) \!\sim\! \theta(|\omega| \!-\! J) \big||\omega| \!-\! J \big|^{\alpha_3-1}$, forming the pseudogap shoulder [\cref{fig:Af}(a))]. 
Since terms irrelevant at the AD fixed point can be important at such a high energy scale, quantitative behaviors around the shoulders can be altered. 
For example, the sharp step function $\theta(|\omega|-J)$ could be broadened. 
Further including $T$ will bring in another set of shoulders at $\omega \!=\! \pm J_D$.

\paragraph{Singlet regime}
Unlike the doublet regime, if the low-energy space is restricted to $S$, the impurity will have no internal degrees of freedom, hence decouple from the bath. 
We term this phase as LS. 
When $J_D>0$, we perform a similar RG analysis as in the doublet regime, by including both $S$ and $D$ states. 
We find that FL and LS are separated by an unstable fixed point as shown by \cref{fig:intro}(d) \cite{wang_bosonization_2025}.  
This phase transition is consistent with previous NRG studies in similar models \cite{fabrizio_nontrivial_2003, leo_spectral_2004, nishikawa_convergence_2012}, where the critical point is found to be described by a non-Fermi liquid with impurity entropy $\ln \! \sqrt{2}$ \cite{fabrizio_nontrivial_2003}. 
When $J_D \!<\! 0$, the low-energy space consists of $S \oplus T$, and the phase transition should be equivalent to that in the two-impurity Kondo problem \cite{Jayaprakash_1981_2IK, Jones_1987_study, Jones_1988_lowT, Jones_1989_critical, Affleck_1992_exactcriticaltheory, Gan_1995_mapping, mitchell_universal_2012, mitchell_2channel_2012}, which was also found to be second-order.

As the $S$ state carries no symmetry charge, $\chi_{c,s,v} \!\ll\! T_{\rm K}^{-1}$ all freeze in FL if $\TK \!\ll\! J$. 
Solving \cref{eq:Ward} finds the $S$ channel is the only attractive one, with $\td{U} \!-\! \td{J}_S \!=\! -3\pi \TK$ \cite{leo_spectral_2004}. 
In the LS phase, Ref. \cite{youn_hundness_2024} also finds the local pairing susceptibility in $S$ channel gets enhanced. 
Deep in LS phase, this can be shown by a perturbative calculation that integrates out the multiplet fluctuations, where the attractive strength is of $\mcl{O}(\frac{\Delta_0^2}{U^2} \frac{1}{x_c J})$ (\cref{sec:effective-interaction-2} in SM \cite{supplement}). 

The fixed-point Hamiltonian of LS only contains $H_0$, hence $A_f(\omega) \!\to\! 0$ as $\omega \!\to\! 0$, in stark contrast to the in-gap excitations of AD. 
(Particle-hole asymmetry will lead to a small finite $A_f(0)$, see \cref{app:AfSig-Af} in SM \cite{supplement}, but does not affect the pole in $\Sigma_f$.)
The pseudogap shoulders, however, form by the same mechanism as in AD, as multiplet excitations to the $D$ or $T$ states induced by scattering a bath electron or hole [\cref{fig:Af}(b)]. 

\paragraph{Discussion}
The single-impurity solutions [\cref{fig:intro}(b)] shed light into the lattice problem of periodic spin-valley Anderson impurities. 
% The single-impurity phase diagram [\cref{fig:intro}(b)] provides useful insights into the correlated phases in moir\'e lattices described by models of the heavy fermion type \cite{song_magic-angle_2022, shi_heavy-fermion_2022, Yu_2023_THF_TSTG, singh_topological_2024, herzog_2025_efficient, liu_ideal_2025}. 
In separate papers~\cite{youn_hundness_2024, youn_hundness_2025}, we show that quantum phase transitions into the AD and LS phases can indeed appear in the DMFT solution of MATBG \cite{song_magic-angle_2022} at filling fractions $\nu \!\approx\! \pm 2$, if the corresponding anti-Hund's rule is present. 
For a sketchy understanding to the lattice spectral function $A(k,\omega)$, we construct analytic \textit{an\"satz} for the correlation self-energy $\Sigma_f(\omega)$ that reproduces the single-impurity $A_f(\omega)$ (See End Matter), and insert it into the lattice Green's function. 
Local Kondo FL leads to a heavy Fermi liquid on the lattice, with a Fermi volume $\frac{\nu}{4}$, where both $f$ and $c$ electrons contribute \cite{Chou_2023_Kondo, zhou_kondo_2024, Rai_2023_DMFT, Hu_2023_Kondo, Hu_2023_Symmetric, Chou_2023_scaling, Datta_2023_heavy, Lau_2023_topological}. 
If $\TK \!\ll\! J$, pseudogap shoulders at $\omega\!\sim\!\mcl{O}(J)$ [similar to \cref{fig:Af}(a,b)] due to multiplet excitations can also be found, besides the Kondo resonance peak at $\omega \!=\! 0$. 
Increasing $J_S$ locks each impurity into a LS, and we find $\Sigma_f \!\sim\! \frac{1}{\omega}$ at $\omega\!\ll\!J$, which gaps out the heavy $f$ quasiparticles. 
Correspondingly, the Fermi volume jumps to $\frac{\nu - 2\sgn(\nu)}{4}$ [\cref{fig:Af}(d)] \cite{zhang_2020_spinliquids}, where $\nu_f \!\approx\! \pm 2$ disappears. 
If $\nu \!=\! 2$, the lattice forms a symmetric Mott phase \cite{youn_hundness_2024, youn_hundness_2025, wang_2025_epc, zhao_mixed_2025}. 
Contrarily, on increasing $J_D$ into the AD phase, an unscreened doublet per unit cell remains. 
At finite temperature where no spontaneous symmetry breaking occurs, the in-gap excitations of $A_f$ pervade the Brillouin zone, and also incur a finite lifetime to the $c$ bands [\cref{fig:Af}(c)]. 

On lowering temperature, superconductivity may develop from different normal states, due to the local pairing potential in channels summarized in \cref{fig:intro}(b), forming a lower superconductivity tunneling gap. 
Together with the higher pseudogap [\cref{fig:Af}], they provide potential explanation to the two-gap structure found by Refs. \cite{park_2025_simultaneoustransporttunnelingspectroscopy, kim_2025_resolvingintervalleygapsmanybody}. 
Simultaneously, in the anti-Hund's regime ($J_{S,D} \!>\! 0$), the valley moments spanned by $S \!\oplus\! D$ also couple to one another via Ruderman--Kittel--Kasuya--Yosida (RKKY) interactions. 
When the RKKY interaction is strong, the valley moments can align, leading to spontaneous symmetry breaking into either valley-polarized or inter-valley-coherent (IVC) \cite{Nuckolls_2023_quantum, kim_imaging_2023} states, which can coexist with superconductivity. 
To clarify the interplay between superconductivity and IVC orders will be a crucial next step toward a complete theory of MATBG/TTG.

\paragraph{Note added}
After posting this work, we became aware of a related study \cite{zhao_2025_rvb} that shares a similar perspective on the origin of pairing and pseudogap in MATBG. 

\begin{acknowledgments}
We thank Hyunjin Kim and Jeong Min Park for fruitful discussions. 
Z.-D.~S., Y.-J.~W., and G.-D.~Z.~were supported by National Natural Science Foundation of China (General Program No.~12274005), National Key Research and Development Program of China (No.~2021YFA1401900), and Quantum Science and Technology-National Science and Technology Major Project (No.~2021ZD0302403). 
H.~J., S.~Y., and S.-S.~B.~L.~were supported by the National Research Foundation of Korea (NRF) grants funded by the Korean government (MSIT: No.~RS-2023-00214464, No.~RS-2023-00258359, No.~RS-2023-NR119931, No.~RS-2024-00442710; MEST: No.~2019R1A6A1A10073437), the Global-LAMP Program funded by the Ministry of Education (No.~RS-2023-00301976), and Samsung Electronics Co., Ltd.~(No.~IO220817-02066-01).
\end{acknowledgments}


% \clearpage

% "End Matter" title setting closest to the actual PRL prints
\onecolumngrid% Temporarily switch to single-column format
\vspace{10pt}% Add some vertical space before the title
\begin{center}% Center the text horizontally across the full page width
    \fontsize{12}{16}\selectfont% Set the font size (e.g., 12pt with 16pt baseline skip)
    \textbf{End Matter}% Print the title in bold
\end{center}%
\twocolumngrid% Switch back to the standard two-column format

\paragraph{\textit{Ans\"atze} for correlation self-energy in AD and LS phases}
Hubbard bands due to $f$ charge fluctuation can be approximated as poles as $A_f^{(\rm at)}(\omega) \!=\! \frac{1}{2} \delta(\omega \!+\! \Delta E) + \frac{1}{2} \delta(\omega \!-\! \Delta E)$ \cite{Hubbard_2}, where $\Delta E  \!\approx\! \frac{U}{2}$. 
Here we assumed the positions of lower and upper Hubbard bands to be symmetric for simplicity, but generalization to asymmetric case is direct. 
We write the total $A_f(\omega)$ as a mixing of $A_f^{(\rm at)}(\omega)$ and the $A_f^{(1,2,3)}(\omega)$ components defined above. 
To guarantee proper normalization, we also add smooth high-energy cutoffs to $A_f^{(1,2,3)}(\omega)$. 
We treat $\Delta E$ and the mixing amplitudes of each component as tuning parameters in our ans\"atze. 
We then obtain $G_f(\omega)$ via the Kramer-K\"onig relation, and the correlation self-energy via $\Sigma_f(\omega) \!=\! \omega - G_f^{-1}(\omega)$. 
Within the DMFT approximation that neglects spatial correlations, the lattice spectral function is given by $A(k,\omega) \!=\! -\frac{1}{\pi} \Im \frac{1}{\omega - H(k) - \Sigma_f(\omega + \ii0^+)}$, where $H(k)$ is the topological heavy fermion (THF) Hamiltonian of MATBG with hetero-strain (\cref{app:AfSig-MATBG} in SM \cite{supplement}). 

Our \textit{ans\"atze} are natural generalizations of the Hubbard-I approximation (HIA) \cite{Hubbard_2, Lichtenstein_abinitio_1998, hu_projected_2025}, where $A_f \!=\! A_f^{\rm (at)}$. 
As $A_f^{(1,2,3)}(\omega)$ are included, we are able to capture the pseudogap and in-gap excitations in the lattice spectrum of MATBG, in addition to the Hubbard bands \cite{hu_projected_2025} that are also captured by other approaches \cite{ledwith_nonlocal_2024,zhao_topological_mott_2025}. 

\begin{figure}[tb]
    \centering
    \includegraphics[width=0.8\linewidth]{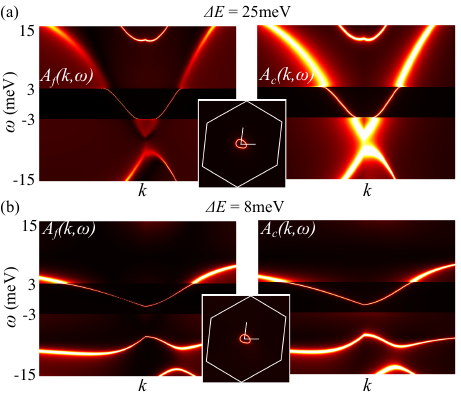}
    \caption{Lattice spectral function in the LS phase, plotted component-wise, $A(k,\omega) \!=\! A_f(k,\omega) \!+\! A_c(k,\omega)$. 
    (a) Same as \cref{fig:Af}(d), but with a different $k$-path (inset). 
    Both Hubbard peaks of $f$ lie at high energies $\pm 25$meV, thus the Fermi surface majorly comprises of $c$. 
    (b) Lowering the Hubbard peaks to $\pm 8$meV. 
    Although the Fermi volume is still given by $\frac{\nu-2}{4}$ (inset), hybridization between $c$ bands and the $f$ Hubbard bands has strongly suppressed the Fermi velocity. }
    \label{fig:Af_EM}
\end{figure}

Several remarks are in order. 
In the THF model, the hybridization strength $\Delta_0$ of the quantum impurity problem is not determined by the $f^\dagger c$ hopping strength $\gamma$, but by the hybridization function $\sim \! \gamma^2N(\omega)$, where $N(\omega)$ is the density of states. 
The AD and LS phases can exist in the $\gamma^2 N(\omega) \ll U$ limit even if $U \!\lesssim\! \gamma$. 
Moreover, in the LS phase, the poles corresponding to $c$ bands will hybridize with the $f$ Hubbard bands. 
If the $f$ Hubbard bands are at high energy, the influence of hybridization will be weak [\cref{fig:Af}(d) or \cref{fig:Af_EM}(a)]. 
If the $f$ Hubbard bands are low, however, hybridization can strongly suppress the Fermi velocity, and the quasiparticle poles near the Fermi energy will also acquire significant $f$ weights [\cref{fig:Af_EM}(b)]. 
However, since the Hubbard bands are physically broadened, it is only an approximation that they are treated as poles as in HIA. 
It thus should be studied in the future whether the excitations with strong hybridization to Hubbard bands can still remain well-defined quasiparticles.  

\paragraph{NRG calculations}
We perform NRG calculations \cite{krishna-murthy_renormalization-group_1980-1,Bulla_2008_nrg} using the MuNRG toolbox \cite{lee_adaptive_2016,lee_doublon-holon_2017} based on the QSpace tensor library \cite{10.21468/SciPostPhysCodeb.40,10.21468/SciPostPhysCodeb.40-r4.0}. 
We exploit the charge-$\Uo$, valley-$\Uo$, and spin-$\SUt$ symmetries. 
Unless otherwise specified, we keep $\sim \! 3000$ multiplets ($\sim \! 8000$ states). 
The Wilson chain is constructed with a discretization parameter $\Lambda = 3$. 
The $z$-averaging technique \cite{yoshida_renormalization-group_1990,campo_alternative_2005,zitko_energy_2009} with $n_z\!=\!2$ is employed for calculating the correlation functions.
We fix the Hubbard repulsion $U\!=\!3$, and use a box-shaped hybridization function $\Delta(\omega)\!=\!\Delta_0\theta(D \!-\! |\omega|)$ with $\Delta_0\!=\!0.2$ and $D\!=\!10$. 

\begin{figure*}[!t]
    \centering
    \includegraphics[width=0.9\linewidth]{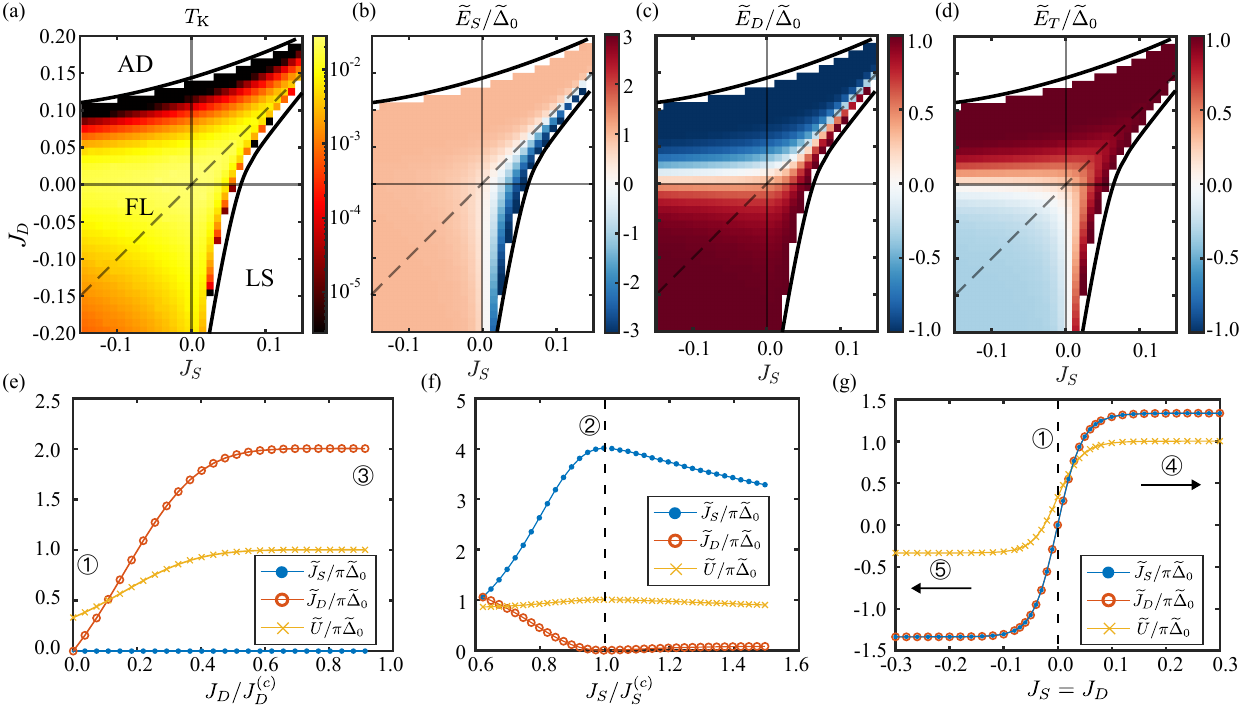}
    \caption{
    NRG results of the phase diagram, $\TK$, and effective parameters of the SVAIM. (a) Phase diagram on $(J_S,J_D)$ plane. The black solid lines sketch the phase boundary, and the color in the FL phase indicates $\TK$. The grey dashed line marks $J_S\!=\!J_D$. (b)-(d) The effective interactions in $S,D,T$ channels  $\td{E}_{S,D,T}$ compared to $\td{\Delta}_0$ as a function of $J_S,J_D$ in the FL phase. 
    (e)-(g) $\td{J}_S/\pi \td{\Delta}_0, \td{J}_D/\pi \td{\Delta}_0,\td{U}/\pi \td{\Delta}_0$ as functions of: (e) $J_D/J^{(c)}_D$ when $J_S\!=\!0$, (f) $J_S/J^{(c)}_S$ when $J_D\!=\!0.05$, (g) $J_S$ when $J_S\!=\!J_D$. The dashed lines in (f),(g) mark the FL-LS critical point and $J_S\!=\!J_D\!=\!0$, respectively. The numeric labels indicate the regions where the relations in \cref{tab:eff-int} hold, while the arrows in (g) show that these relations remain valid upon increasing $|J_S|$ along the line $J_S = J_D$. \label{fig:nrgphase}}
\end{figure*}

The NRG phase diagram is plotted in \cref{fig:nrgphase}(a). 
The three phases are distinguished by different fixed-point NRG spectra. 
FL and LS are both Fermi-liquid-like, but with opposite even-odd oscillations, corresponding to the $\frac{\pi}{2}$ and $0$ phase shifts of bath electrons, respectively. 
The AD phase simply exhibits non-universal phase shift $\rho_z$. 
$A_f(\omega)$ differs as well. 
FL, LS, and AD exhibits a sharp resonance peak, a full gap with $A_f(0) \!=\! 0$, and a gap with $A_f(0) \!\neq\! 0$, respectively, consistent with the analytical results. 
See \cref{sec:nrg-phase} in SM~\cite{supplement} for typical RG flow and impurity spectral function in these phases. 

{
\renewcommand{\arraystretch}{1.5}
\begin{table}
    \centering
    \begin{tabular}{c|c|c}\hline\hline
        Parameter region & Impurity GS & $\frac{1}{\pi \TK}(\td{E}_S,\td{E}_D,\td{E}_T)$  \\\hline
        \ding{192} $U\!\gg\!\TK;\,J_D\!=\!J_S\!=0$ &$S\!\oplus\! D \!\oplus\! T$ & $(\frac{1}{3}, \frac{1}{3}, \frac{1}{3})$ \\\hline
        \ding{193} $J_S,J_S\!-\!J_D,~U\!\gg\! \TK$ & $S$ & $(-3,1,1)$ \\\hline
        \ding{194} $J_D,U\!\gg\!\TK;\,~J_S\!=\!0$ & $D$& $(1,-1,1)$ \\\hline
        \ding{195} $J_S,U\!\gg\! \TK;\,~J_S\!=\!J_D\!>\!0$ & $S\!\oplus\!D$& $(-\frac{1}{3}, -\frac{1}{3}, 1)$ \\\hline
        \ding{196} $|J_S|,U\gg\! \TK;\,~J_S\!=\!J_D\!<0$& $T$ & $(1, 1, -\frac{1}{3})$ \\\hline\hline
    \end{tabular}
    \caption{\label{tab:eff-int}
    Renormalized FL parameters by Ward identities. 
    The five parameter regions are marked in \cref{fig:nrgphase}(e)--(g). 
    Notice that besides the attractive channel favored by the lowest-lying impurity multiplet, other channels are equally repulsive. }
\end{table}
}

In NRG, we define $\TK$ in the FL phase by the renormalized hybridization $\TK \!=\! \td{\Delta}_0 \!=\! z\Delta_0$, where $z \!=\! [1\!-\!\partial_{\omega}\Sigma_f(\omega)|_{\omega=0}]^{-1}$ is the quasiparticle weight. 
$z$ is calculated by fitting the renormalized chain parameters \cite{hewson_renormalized_2004} as detailed in \cref{sec:nrg-eff-int} in SM~\cite{supplement}. 
We find that, generically speaking, $\TK$ is higher if the impurity ground state degeneracy is higher. 
For example, a ``ridge'' of $\TK$ extends along the line $J_D\!=\!J_S\!>0$, where $S$ and $D$ are degenerate; along the line $J_S\!=\!0 \!>\! J_D$, where $S$ and $T$ are degenerate; and along the line $J_D\!=\!0 \!>\! J_S$, where $D$ and $T$ are degenerate. 

In the FL phase, we extract the renormalized parameters $\td{U},\td{J}_D,\td{J}_S$ from the NRG spectra \cite{hewson_renormalized_2004}, as explained in \cref{sec:nrg-eff-int} in SM~\cite{supplement}. 
The effective interaction strengths in $S,D,T$ channels, $\td{E}_S\!=\!\td{U}\!-\!\td{J}_S,\td{E}_D\!=\!\td{U}\!-\!\td{J}_D,\td{E}_T\!=\!\td{U}$ are shown in \cref{fig:nrgphase}(b)--(d), where the attractive regions agree with \cref{fig:intro}(b). 
Three linecuts, $J_S\!=\!0$, $J_D\!=\!0.05$, and $J_S\!=\!J_D$ are plotted in \cref{fig:nrgphase}(e)--(g), in order to better illustrate that the interaction strengths indeed asymptote to the Ward identity results as $\TK \!\to\! 0$ (\cref{tab:eff-int}). 

Renormalized parameters can also be defined in the LS phase, where the bath still forms a Fermi liquid. 
As the impurity decouples in the low-energy regime, we regard the first bath site in NRG as a new ``impurity'', following Ref.~\cite{nishikawa_convergence_2012} (\cref{sec:nrg-eff-int} in SM~\cite{supplement}). 
As shown in \cref{fig:nrgphase}(f), renormalized parameters obtained in FL and LS phases both asymptote to the Ward identity result when approaching the second-order critical point. 
Deep in the LS phase, the perturbative analysis to the effective bath interaction applies.

\begin{figure}[h]
    \centering
    \includegraphics[width=\linewidth]{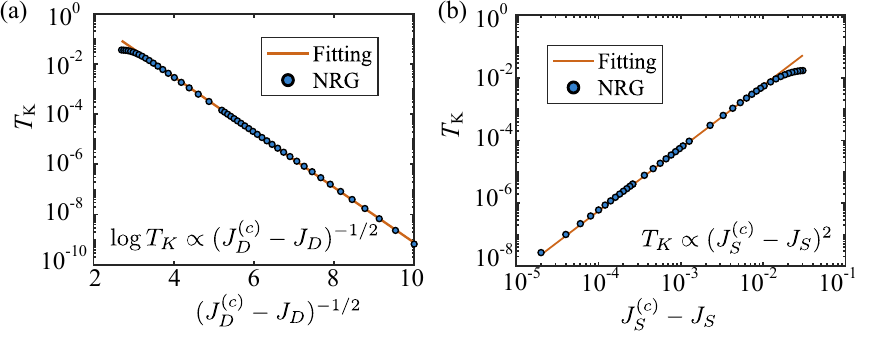}
    \caption{$\TK$ near the phase transitions. 
    (a) BKT type, obtained along $J_S\!=\!0$, with critical $J^{(c)}_D \!\approx\! 0.137$. 
    (b) Second-order, obtained along $J_D\!=\!0.05$, with critical $J^{(c)}_S \!\approx\! 0.08026$. 
    }
    \label{fig:critical}
\end{figure}

The scaling of $\TK$ near phase transitions indicates the nature of the phase transitions. 
We select two linecuts, $J_S\!=\!0$ [\cref{fig:critical}(a)] and $J_D\!=\!0.05$ [\cref{fig:critical}(b)], to illustrate the BKT and the second-order transitions, respectively. 
For BKT, $\TK$ can be fitted by $\TK \!\propto\! e^{-c\sqrt{\frac{\Delta_0}{J^{(c)}_{D}-J_D}}}$ with some constant $c$, consistent with our analytical RG calculation and previous numerical results \cite{Galpin_2005_quantumphasetransition}. 
In \cref{sec:nrg-eff-int} of SM~\cite{supplement}, we further show that from the finite-size NRG spectrum flow, we can numerically extract the flow diagram of $\lambda_x,\lambda_z$, as plotted in \cref{fig:bkt}(d), validating \cref{fig:intro}(c). 
On the other hand, $\TK$ near the second-order transition can be fitted by $\TK \!\propto\! (J^{(c)}_{S} \!-\! J_S)^2$, consistent with our analytical RG calculation and previous numerical results \cite{fabrizio_nontrivial_2003,leo_spectral_2004,nishikawa_convergence_2012}. 

\clearpage
\appendix 

\onecolumngrid
\tableofcontents

\section{The quantum impurity model}
\label{sec:Himp}

\subsection{The Kondo model}  \label{sec:Kondo-coupling-general}

% \begin{table}[tb]
%     \centering
%     \begin{tabular}{c|c|c}
%     \hline
%         irrep $[L,S]$ & $\mrm{DEG}_{[L,S]}$ & basis \\
%     \hline
%         $[A_1, 0]$ & 1 & $\sigma^0 \spin^0$ \\
%         $[A_2, 0]$ & 1 & $\sigma^z \spin^0$ \\
%         $[2, 0]$ & 2 & $\sigma^{x,y} \spin^0$ \\
%         $[A_1, 1]$ & 3 & $\sigma^0 \spin^{x,y,z}$ \\
%         $[A_2, 1]$ & 3 & $\sigma^z \spin^{x,y,z}$ \\
%         $[2, 1]$ & 6 & $\sigma^{x,y} \spin^{x,y,z}$ \\
%     \hline
%     \end{tabular}
%     \caption{Hermitian bilinear bath operators classified into irreps of $[L,S]$. }
%     \label{tab:bath_oprt}
% \end{table}

We focus the parameter regime $\epsilon_f \approx -\frac{3}{2}U$, where the low-energy impurity ($f$) configurations are dominated by two-electron states, and the multiplet splitting plays a significant role. 
To obtain the corresponding low-energy theory, we carry out a Schrieffer-Wolff (SW) transformation to integrate out the charge fluctuations on the $f$-impurity that cost $\mcl{O}(U)$ energies, resulting in an effective Kondo model (Section II in Ref.~\cite{wang_bosonization_2025})
\begin{align} \label{eq:HK}
    H_{\rm K} = H_0 + H_{\rm imp}^{\rm (K)} + H_{\rm c}^{\rm (K)} \ .
\end{align}
The impurity Hamiltonian becomes $H_{\rm imp}^{\rm (K)} = \sum_{\Gamma=S,D,T} E_\Gamma  \PP_\Gamma$, where $\PP_\Gamma$ is the projector to the $\Gamma=S,D,T$ manifolds. 
The low-energy impurity Hilbert space now includes all two-electron states, $\PP_2 = \sum_{\Gamma=S,D,T} \PP_\Gamma$. 
In general, the SW transformation may also generate corrections to the splitting of the multiplets, which can be absorbed as a re-definition to $J_S$ and $J_D$, hence we neglect them. 
$H_{\rm c}^{\rm (K)}$ is the Kondo coupling between the impurity and the bath. 
As detailed in Section~II and Appendix~B in Ref.~\cite{wang_bosonization_2025}, $U(1)_v$, $C_{2z}$, $\SUt_s$, and a $C_2T$ symmetry (see \cref{sec:basis-rotation}) in SVAIM restrict it to the form
\begin{align}   \label{eq:HK_c}
    H^{\rm (K)}_{\rm c} &= 2\pi \zeta_{0z} \sum_{\nu = x,y,z} \Theta^{0 \nu} \cdot \psi^\dagger \sigma^0 \spin^\nu \psi \Big|_{x=0} + 2\pi \zeta_{xz} \sum_{\mu=x,y} \sum_{\nu=x,y,z} \Theta^{\mu \nu} \cdot \psi^\dagger \sigma^\mu \spin^\nu \psi \Big|_{x=0} + 2\pi \zeta_{zz} \sum_{\nu = x,y,z} \Theta^{z\nu} \cdot \psi^\dagger \sigma^z \spin^\nu \psi \Big|_{x=0} \\\nonumber
    &+ 2\pi \lambda_z \cdot \Theta^{z0} \cdot \psi^\dagger \sigma^z \spin^0 \psi \Big|_{x=0} + 2\pi \zeta_{x} \sum_{\mu=x,y} \Theta^{\mu 0} \cdot \psi^\dagger \sigma^\mu \spin^0 \psi \Big|_{x=0}  \ .
\end{align}
Here, we have defined the representation of the $\SUf$ generators on the 6 two-electron states as
\begin{align}   \label{eq:Theta}
    \Theta^{\mu \nu}  = \PP_2   \frac{f^\dagger \sigma^{\mu} \spin^{\nu} f }{2} \PP_2 \qquad \qquad \mu\nu \not=00
\end{align}
where we have abbreviated $\psi^\dagger \sigma^\mu \spin^\nu \psi = \sum_{ls,l's'} \psi^\dagger_{ls}(0) [\sigma^\mu]_{ll'} [\spin^\nu]_{ss'} \psi_{l's'}(0)$. 
We neglect couplings like $\PP_{\Gamma} \cdot :\psi^\dagger\sigma^0\varsigma^0\psi:$, which break the particle-hole symmetry (PHS) and are in general not relevant in the low-energy physics \cite{wang_bosonization_2025}.

Among terms in \cref{eq:HK_c}, $\lambda_z$ acts within the $D$ states (note that $\Theta^{z0} = \Lambda_z$ defined in the main text), $\zeta_{0z}$ acts within the $T$ states. Besides, $\zeta_{zz}$ acts between $S$ and $T$ states, $\zeta_{xz}$ acts between $D$ and $T$ states, and $\zeta_x$ acts between $S$ and $D$ states ($\Theta^{x0} = \frac{1}{\sqrt{2}}(\Theta_+ +\Theta_-), \Theta^{y0} = \frac{-\ii}{\sqrt{2}}(\Theta_+ - \Theta_-)$ defined in the main text). 

Suppose we are carrying out an RG (for example, a poorman scaling) to the quantum impurity model. 
After the charge fluctuation has been integrated out and \cref{eq:HK_c} is obtained, we are at an ``initial'' energy scale $\mcl{D}_0$ satisfying $|J_{S,D}| \ll \mcl{D}_0 \ll U$. 
Here, the multiplet splitting induced by $J_{S,D}$ is not important, and the five  independent moment-moment couplings in \cref{eq:HK_c} ($\zeta_{0z}, \zeta_{zz}, \zeta_{xz}, \zeta_x, \lambda_z$) will remain approximately equal, which we denote as $\zeta$. 
$\zeta$ will grow as the energy scale $\mcl{D}$ is lowered, similar to the $\Uf$ symmetric case \cite{zhou_kondo_2024}. 
If $\zeta$ already diverges at some $\mcl{D}_{\rm K} \gg |J_{S,D}|$ (or equivalently speaking, the system flows to a strong-coupling fixed point, evidenced by \textit{e.g.} the low-energy bath phase shift saturating $\frac{\pi}{2}$), then the system should share the same universal properties as the $\Uf$ symmetric model. 
However, if $\zeta$ has not diverged when $\mcl{D} \sim |J_{S,D}|$, yet we are still interested in physics with temperature $k_BT \ll  |J_{S,D}|$, then we will have to further downfold the low-energy Hilbert space, from $\PP_2$ to some $\PP_\Gamma$ (or a summation over some $\PP_\Gamma$), where a second SW transformation that integrates out the multiplet fluctuation from $\PP_\Gamma$ to $\PP_2 - \PP_\Gamma$ will be required. 
For the doublet regime, downfolding $\PP_2$ to $\PP_D$ gives rise to the pair Kondo (PK) Hamiltonian. 
Further details can be found in Ref.~\cite{wang_bosonization_2025}.

\subsection{Quasiparticle operators and spectral functions in Kondo-type models}   \label{sec:qp} 

We are interested in the spectral function of the physical $f$-electrons in the original Anderson model, defined by 
\begin{align}
    A_f(\omega) = -\frac{1}{\pi} \Im G_f(\omega + \ii 0^+)
\end{align}
where $G_f(\omega)$ can be obtained by analytical continuing the imaginary-time Green's function, 
\begin{align}
    G_f(\tau) = -\left\langle T_\tau ~ f_{ls}(\tau) ~ f^\dagger_{ls}(0) \right\rangle \qquad \qquad 
    G_f(\ii \omega) = \int_{-\infty}^{\infty} \mrm{d} \tau ~ G_f(\tau) ~ e^{\ii \omega \tau} \ . 
\end{align}
Operators here are in the Heisenberg representation, $f_{ls}(\tau) = e^{ H\tau} f_{ls} e^{- H \tau}$. 
However, to calculate $A_f(\omega)$ in the Kondo or PK models is not so obvious, as the $f$ electron has gone through several SW transformations. 
We discuss the SW transformed $f$ operator and calculation for spectral function $A_f$ in this subsection. 

During one SW transformation, we apply a unitary transformation $e^{\ii S}$ to the Hamiltonian, $\td{H} = e^{\ii S}He^{-\ii S}$, in order to eliminate the off-diagonal element between low-energy (with projector $\PP^{\rm (L)}$) and high-energy (with projector $1-\PP^{\rm (L)}$) Hilbert spaces. 
Therefore, $e^{\ii S}$ is chosen such that $\PP^{\rm (L)} \td{H} (1-\PP^{\rm (L)})=0$. 
Correspondingly, $f_{ls}$ leaves a component of $\td{f}_{ls} = \PP^{(\rm L)} e^{\ii S} f_{ls} e^{-\ii S} \PP^{(\rm L)}$ in the low-energy space. 
Crucially, in the original Anderson model, the relation $f_{ls} \propto [H-H_0, \psi_{ls}(0)]$ holds as an identity, where $H_0$ is the Hamiltonian of the bath electrons. 
Now we show that, to the leading order of $\frac{1}{U}$, where $U$ denotes the large energy gap between $\PP^{\rm (L)}$ and $1 - \PP^{\rm (L)}$ subspaces, $\td{f}_{ls}$ in the low-energy models can be calculated as $\td{f}_{ls} \propto [H^{(\rm L)} - H^{(\rm L)}_0, \psi_{ls}(0)]$, where $H^{(\rm L)} = \PP^{(\rm L)} e^{\ii S} H e^{-\ii S} \PP^{(\rm L)}$ and $H^{(\rm L)}_0 = \PP^{(\rm L)} e^{\ii S} H_0 e^{-\ii S} \PP^{(\rm L)}$ are the low-energy effective Hamiltonians. 

By the above definition, 
\begin{align}
    \td{f}_{ls} &\propto \PP^{(\rm L)} e^{\ii S} [H-H_0, \psi_{ls}(0)] e^{-\ii S}  \PP^{(\rm L)} = \PP^{(\rm L)} [\td{H} - \td{H}_0 , \td{\psi}_{ls}(0) ] \PP^{(\rm L)} \\\nonumber
    &= \left(\PP^{(\rm L)} \td{H} \td{\psi}_{ls}(0) \PP^{(\rm L)} - \PP^{(\rm L)} \td{\psi}_{ls}(0) \td{H} \PP^{(\rm L)} \right) 
    - \left(\PP^{(\rm L)} \td{H}_0 \td{\psi}_{ls}(0) \PP^{(\rm L)} - \PP^{(\rm L)} \td{\psi}_{ls}(0) \td{H}_0 \PP^{(\rm L)} \right)
\end{align}
where we have denoted $\td{H}_0 = e^{\ii S} H_0 e^{-\ii S}$, and $\td{\psi}_{ls}(0) = e^{\ii S} \psi_{ls}(0) e^{-\ii S}$ for brevity, and expanded the commutators explicitly. 
Making use of $\PP^{(\rm L)}\td{H}(1-\PP^{(\rm L)}) = 0$ mentioned above and noticing that $\PP^{(\rm L)}\td{H}_0(1-\PP^{(\rm L)}) \neq 0$, we have
\begin{align}
    \td{f}_{ls} 
    &\propto \left(\PP^{(\rm L)} \td{H} \PP^{(\rm L)} \td{\psi}_{ls}(0) \PP^{(\rm L)} - \PP^{(\rm L)} \td{\psi}_{ls}(0) \PP^{(\rm L)} \td{H} \PP^{(\rm L)} \right) - \left(\PP^{(\rm L)} \td{H}_0 \PP^{(\rm L)} \td{\psi}_{ls}(0) \PP^{(\rm L)} - \PP^{(\rm L)} \td{\psi}_{ls}(0) \PP^{(\rm L)} \td{H}_0 \PP^{(\rm L)} \right)  \\\nonumber
    & \qquad - \left(\PP^{(\rm L)} \td{H}_0 (1-\PP^{(\rm L)}) \td{\psi}_{ls}(0) \PP^{(\rm L)} - \PP^{(\rm L)} \td{\psi}_{ls}(0) (1-\PP^{(\rm L)}) \td{H}_0 \PP^{(\rm L)} \right) \\\nonumber
    &= [H^{\rm (L)} - H^{\rm (L)}_0, \PP^{(\rm L)} \td{\psi}_{ls}(0) \PP^{(\rm L)}] - \left(\PP^{(\rm L)} \td{H}_0 (1-\PP^{(\rm L)}) \td{\psi}_{ls}(0) \PP^{(\rm L)} - \PP^{(\rm L)} \td{\psi}_{ls}(0) (1-\PP^{(\rm L)}) \td{H}_0 \PP^{(\rm L)} \right) \ .
\end{align}
In the first term, $\PP^{(\rm L)} \td{\psi}_{ls} \PP^{(\rm L)} = \psi_{ls} + \mcl{O}(\frac{1}{U^2})$, and the second term itself is of $\mcl{O}(\frac{1}{U^2})$, as both $\PP^{(\rm L)} \td{H}_0 (1 - \PP^{(\rm L)})$ and $\PP^{(\rm L)} \psi (1 - \PP^{(\rm L)})$ are of order $\mcl{O}(\frac{1}{U})$. Therefore, there is 
\begin{align}
    \td{f}_{ls} 
    &\propto  [H^{\rm (L)} - H^{\rm (L)}_0, \psi_{ls}(0) ]  
\end{align}
which by itself is of order $\mcl{O}(\frac{1}{U})$. 

In this Supplementary Material, we will refer to $\td{f}$ as a ``quasiparticle operator''. 
However, we remark that the well-defined (heavy-fermion) quasiparticle \textit{excitation} only forms when $\td{f}$ contributes a Kondo resonance peak in $A_f(\omega)$. 
In the AD or LS phases, $\td{f}$ operators only contribute to the pseudogap shoulders or in-gap excitations. 

The above result can be understood from another perspective. 
In the original Anderson model, from the viewpoint of a bath $\psi$-electron (or in the tunneling experiments, an electron on the tip), an $f$-electron is nothing but the intermediate process when $\psi$ is scattered at the origin $x=0$, hence $G_f(\omega)$ is proportional to the scattering $T(\omega)$-matrix of $\psi$ electrons. 
After the SW transformation, as the bath electrons remain largely unchanged, namely, $e^{\ii S} \psi e^{-\ii S}  = \psi + \mcl{O}(\frac{1}{U})$, one can still extract $G_f(\omega)$ by computing the $T(\omega)$-matrix in these low-energy effective models. 
The scattering $T(\omega)$-matrix will turn out to be given by the Green's function of the operator $\td{f}_{ls} = [H^{\rm (L)}-H^{\rm (L)}_0, \psi_{ls}(0)]$ \cite{Costi_2000_Kondo,Bulla_2008_nrg,Moca_2019_quantumcriticality}.

\subsection{Relation to MATBG}
\label{sec:basis-rotation}

In this section we show how to relate the actual MATBG topological heavy fermion (THF) model \cite{song_magic-angle_2022} to the SVAIM discussed in this paper. 
We discuss the origin of symmetries, and also show how to map the \textit{bare} (anti-)Hund's splitting parameters obtained in Ref. \cite{wang_2025_epc} to $J_S$ and $J_D$. 
By \textit{bare} parameters, here we mean that we have only integrated out the free propagating phonons (or for the atomic-scale Coulomb, the bare Coulomb lines), and have \textit{not} considered any renormalization effect due to on-site or spatial electronic correlations. 

In MATBG, each AA-stacking center behaves as a four-orbital (eight-flavor) quantum impurity \cite{song_magic-angle_2022}, with the electron operator dubbed as $f^\dagger_{\beta \eta s}$. 
$s = \uparrow,\downarrow$ denotes the spin, $\eta = \pm$ denotes the graphene valley, and $\beta=1,2$ distinguishes the orbital angular momentum (OAM) in each valley as $(-1)^{\beta-1} \eta$ mod 3. 
Let us dub the Pauli matrices associated with $\beta,\eta,s$ as $\sigma,\tau,\spin$, respectively. 
The symmetry group consists of charge $\Uo_c$ (generated by $\sigma^0 \tau^0 \spin^0$), spin $\SUt_s$ (generated by $\sigma^0 \tau^0 \spin^{x,y,z}$), and valley $\Uo_v$ (generated by $\sigma^0 \tau^z \spin^0$). 
At each $f$ site, per valley, there is a point group $D_3$ (generated by $C_{3z}=e^{\ii \frac{2\pi}{3} \sigma^z \tau^z \spin^0}$ and $C_{2x}= \sigma^x \tau^0 \spin^0$), and the two valleys are linked by $C_{2z} = \sigma^x \tau^x \spin^0$. 
Finally, there is the Kramer's time-reversal $\mcl{T}=\ii \sigma^0 \tau^x \spin^y K$, where $K$ is complex conjugation. 
It can be combined with a spin $\SUt$ rotation $e^{-\ii \frac{\pi}{2} \sigma^0 \tau^0 \spin^y} = -\ii \sigma^0 \tau^0 \spin^y$ to produce the spinless time-reversal symmetry $T=\sigma^0 \tau^x \spin^0 K$. 
We also have $C_{2z}T = \sigma^x \tau^0 \spin^0 K$, and $C_{2y}T = \sigma^0 \tau^0 \spin^0 K$. 

Due to the highly localized nature of the Wannier functions, it is a good approximation that the $C_{3z}$ rotation symmetry (per valley) at the $f$ site can be upgraded to a continuous rotation symmetry \cite{song_magic-angle_2022, shi_heavy-fermion_2022}, so that $\sigma^z \tau^z \spin^0$ becomes the generator of the corresponding OAM $\Uo$ charge. 
Such an upgrade also naturally occurs in the effective impurity problem during the DMFT calculations, where the hybridization function of the $f$ impurity is realized by an auxiliary bath, with the OAM turning into an internal degree of freedom. 
In that case, any bilinear or quartic Hamiltonian that conserves OAM mod 3 can only change OAM by $0$, but not $3$, $6$, etc. Therefore, OAM will be automatically conserved as a continuous rotation symmetry. 
In this work, we will also adopt this approximation, and treat OAM as a $\Uo$ charge. 

It is shown in Ref.~\cite{wang_2025_epc} that, the microscopic interactions due to vibrating phonons and the atomic-scale Coulomb repulsion (\textit{e.g.} carbon-atom Hubbard), when projected to an $f$ impurity, can lead to multiplet splittings with the form of
\begin{align}    \label{eq:H_AH_MATBG}
    H &= -\frac{1}{2} \sum_{\beta_1\beta_2\beta_1'\beta_2'} \sum_{\eta s s'} 
    \Bigg[ f^\dagger_{\beta_1 \eta s} f^\dagger_{ \beta_1'\eta s'}  \begin{pmatrix}
        J_{\rm a} & 0 & 0 & 0 \\
        0 & -J_{\rm a} & J_{\rm b} & 0 \\
        0 & J_{\rm b} & -J_{\rm a} & 0 \\
        0 & 0 & 0 & J_{\rm a} \\
    \end{pmatrix}_{\beta_1'\beta_1, \beta_2'\beta_2} f_{ \beta_2'\eta s'} f_{ \beta_2\eta s} \\\nonumber
    & ~+~ f^\dagger_{ \beta_1 \eta s} f^\dagger_{ \beta_1'\ovl{\eta} s'}  \begin{pmatrix}
        J_{\rm a} & 0 & 0 & J_{\rm b} \\
        0 & -J_{\rm a} & 0 & 0 \\
        0 & 0 & -J_{\rm a} & 0 \\
        J_{\rm b} & 0 & 0 & J_{\rm a} \\
    \end{pmatrix}_{\beta_1'\beta_1, \beta_2'\beta_2} f_{ \beta_2' \ovl{\eta} s'} f_{ \beta_2\eta s} 
    ~+~ f^\dagger_{ \beta_1 \ovl{\eta} s} f^\dagger_{ \beta_1'\eta s'}  \begin{pmatrix}
        J_{\rm e} & 0 & 0 & J_{\rm d} \\
        0 & 0 & J_{\rm d} & 0 \\
        0 & J_{\rm d} & 0 & 0 \\
        J_{\rm d} & 0 & 0 & J_{\rm e} \\
    \end{pmatrix}_{\beta_1'\beta_1, \beta_2'\beta_2} f_{ \beta_2' \ovl{\eta} s'} f_{ \beta_2\eta s} \Bigg]
\end{align}
where the index $(\beta'\beta) = (11),(12),(21),(22)$. 
For the phonon-mediated interactions, $J_{\rm a,b,d,e}>0$, implying an anti-Hund's nature, while for the atomic-scale Coulomb (\textit{i.e.} carbon-atom Hubbard), $J_{\rm a,b,d,e}<0$, implying a Hund's nature. 

In experimental samples, the degeneracy between the two OAM can be externally broken by heterostrain. On the other hand, the degeneracy between the two valleys can be spontaneously broken. 
With electron-doping or hole-doping on such a symmetry-breaking background, the remaining active flavors will form a two-orbital quantum impurity problem, which is nothing but the SVAIM we consider in this work. 
Here, the ``valley'' degree of freedom in SVAIM may either represent the original valley or represent the OAM. 

We now show in detail how heterostrain or valley order downfolds the original eight-flavor problem to an SVAIM. 
Especially, we identify how the $D_{\infty} = \Uo_v \rtimes \Zt$ valley symmetry and a $C_2T$ symmetry in SVAIM arise from the MATBG symmetries. 
The spin $\SUt_s$ and charge $\Uo_c$ symmetries are simply inherited from MATBG. 
We then downfold the bare anti-Hund's splitting parameters in MATBG to SVAIM. 

\paragraph{Heterostrain}
Heterostrain of various strengths is inevitable in experiments. It explicitly breaks $C_{3z}$, and leads to a Zeeman splitting on the $f$ site as \cite{herzog_2025_kekule} 
\begin{align}    \label{eq:H_strain}
    m_x \cdot \left[ \sigma^{x} \tau^0 \spin^0 \cos\varphi_0 + \sigma^y \tau^z \spin^0 \sin\varphi_0 \right]\ .
\end{align}
Here, $\varphi_0$ denotes the azimuthal angle of the heterostrain axis. 
For a typical heterostrain $\sim 0.2\%$ in experiments, $m_x \approx 10$meV. 
The active flavors are the eigen-states of \cref{eq:H_strain}, which can be parameterized as 
\begin{align}    \label{eq:f_ls_strain}
    f_{ls} =  \frac{1}{\sqrt{2}} \left( e^{\frac{\ii}{2} \eta \vartheta_0} f_{1 \eta s} + e^{- \frac{\ii}{2} \eta \vartheta_0} f_{2 \eta s}  \right) \qquad \qquad \mrm{where}~ l=\eta
\end{align}
where $\vartheta_0 = \varphi_0$ or $\varphi_0+\pi$ for electron or hole doping, respectively, but the two cases do not need to be distinguished for our purpose. 

To make connection with the symmetry of the SVAIM, we check the origin of the $D_{\infty} = \Uo_v \rtimes \Zt$ symmetry and the $C_2T$ symmetry. 
Here, $\Uo_v$ is given by the unbroken valley $\Uo$, and the $\Zt$ factor is generated by $C_{2z}$ that anti-commutes with the valley $\Uo$ generator. 
For the gauge choice of \cref{eq:f_ls_strain}, $C_{2z} f_{ls} C_{2z} = f_{\ovl{l} s}$. 
Finally, $C_{2z}T$ acts as $(C_{2z}T) f_{ls} (C_{2z}T)^{-1} = f_{ls}$, which serves as the $C_2T$ symmetry in SVAIM. 

Next, we project the full multiplet splitting in MATBG \cref{eq:H_AH_MATBG} to the active flavors. 
According to \cref{eq:f_ls_strain}, such a projection amounts to replacing $f_{1 \eta s} \to \frac{e^{-\frac{\ii}{2} \eta \vartheta_0}}{\sqrt{2}} f_{ls}$ and $f_{2 \eta s} \to \frac{e^{\frac{\ii}{2} \eta \vartheta_0}}{\sqrt{2}} f_{ls} $, where $l = \eta$. 
A crucial observation that simplifies the calculation is that, these complex phases $e^{\pm\frac{\ii}{2} \eta \vartheta_0}$ are proportional to the OAM of the $f$ operator that they are replacing, while for all non-vanishing matrix elements in \cref{eq:H_AH_MATBG}, the OAM adds to 0. Therefore, all complex phases multiply to $1$. 
The final result reads, 
\begin{align}\label{eq:H_AH-strain}
    H = -\frac{1}{2} \sum_{l_1l_1'l_2l_2'} \sum_{ss'} f^\dagger_{l_1 s} f^\dagger_{l_1' s'} \begin{pmatrix}
        \frac{J_{\rm b}}{2} &  &  &  \\
         & \frac{J_{\rm b}}{2} & J_{\rm d} + \frac{J_{\rm e}}{2} &  \\
         & J_{\rm d} + \frac{J_{\rm e}}{2} & \frac{J_{\rm b}}{2}  &  \\
         &  &  & \frac{J_{\rm b}}{2} \\
    \end{pmatrix}_{l_1'l_1, l_2'l_2} f_{l_2' s'} f_{l_2 s}\ .
\end{align}
Note that the identity component of the above matrix simply contributes to the Hubbard $U$ and does not affect $J_{S,D}$. 
For later convenience, we rewrite $H_{\rm AH}$ [\cref{eq:H_AH}] as
\begin{align} \label{eq:Himp_1}
    H_{\rm AH} &= - \frac{J_S}{4} \sum_{ll's} f_{ls}^\dagger f_{\bar l \bar s}^\dagger f_{\bar l' \bar s} f_{l's}   - \frac{J_D}{2} \sum_{ls} f_{ls}^\dagger f_{l\bar s}^\dagger f_{l\bar s} f_{ls} \\\nonumber
    &= -\frac{1}{2} \sum_{ss'} \sum_{l_1l_1'l_2l_2'} f^\dagger_{l_1 s} f^\dagger_{l_1' s'}  \begin{bmatrix}
        J_D & 0 & 0 & 0 \\
        0 & \frac{J_S}{2} & \frac{J_S}{2} & 0 \\
        0 & \frac{J_S}{2} & \frac{J_S}{2} & 0 \\
        0 & 0 & 0 & J_D \\
    \end{bmatrix}_{l_1'l_1,l_2'l_2} f_{l_2's'} f_{l_2 s}\ .
\end{align}

Comparing \cref{eq:H_AH-strain} with this, one concludes that $J_D = \frac{J_S}{2} = J_{\rm d} + \frac{J_{\rm e}}{2}$, as summarized in \cref{tab:downfold}. 

\paragraph{Spontaneous valley orders}
A variety of valley orders have been proposed in MATBG, including the valley-polarized order (VP), the Kramer's inter-valley coherent order (KIVC), the spinless-$T$ symmetric inter-valley coherent order (TIVC), and the incommensurate Kekul\'e spiral order (IKS). Their corresponding order parameters are given below as
\begin{equation}  \label{eq:IVC_order}
    \sigma^0 \tau^z \spin^0 \qquad  \sigma^y (\tau^{x} \cos \varphi_0 + \tau^{y} \sin\varphi_0) \spin^0 \qquad \sigma^x (\tau^{x} \cos \varphi_0 + \tau^{y} \sin\varphi_0) \spin^0 \qquad \sigma^x (\tau^{x} \cos(\qq\cdot\RR) + \tau^{y} \sin(\qq\cdot\RR) \spin^0 \qquad 
\end{equation}
respectively. 
$\varphi_0$ characterizes the IVC angles, while in IKS, such IVC angle is ``spiraling'' across different moir\'e unit cells $\RR$ with some wave-vector $\qq$. 
Viewed locally from one $f$ impurity, IKS is barely distinguishable from TIVC, hence we treat them identically below. 
In this work, we do not intend to discuss which order is more likely to appear in MATBG; instead, we only discuss that if any of the above order forms, what two-valley impurity problem they will give rise to. 

The active flavors are also given by eigenstates of the order parameters \cref{eq:IVC_order}, which we tabulate in \cref{tab:downfold}. 
As valley degeneracy is broken, $l$ among the active flavors labels OAM, and hence the effective $\Uo_v$ symmetry in the two-valley model corresponds to the OAM $\Uo$ symmetry in the original model. 
For VP and TIVC (IKS), the degeneracy of opposite OAM is protected by a $\Zt$ group generated by $C_{2x}$, which will combine with $\Uo_v$ to span the $D_{\infty}$ valley group. 
For KIVC, it is protected by $C_{2x}$ dressed by a valley $\Uo$ rotation, $C_2 = C_{2x} \cdot e^{\ii \pi \frac{\tau^z-\tau^0}{2}}$, although both $C_{2x}$ and valley $\Uo$ are individually broken. This new action  shares the same algebra as $C_{2x}$: $C_2^2 = 1$, and $C_2$ anti-commutes with the $\Uo_v$ generator $\sigma^z \tau^z \spin^0$. 
Consequently, KIVC also enjoys the $D_{\infty}$ valley group. 
As for the time-reversal $C_2T$, it is fulfilled by $C_{2y}T$, or $C_{2y}T$ dressed by some valley $\Uo$ rotations. 
One can directly verify that for the wave-functions tabulated in \cref{tab:downfold}, and the corresponding definition of $C_2$ and $C_2T$ actions, there are $C_2 f_{ls} C_2^{-1} = f_{\ovl{l}s}$, and $(C_2T) f_{ls} (C_2T)^{-1} = f_{ls}$. 

Finally, we also project the multiplet splitting interaction \cref{eq:H_AH_MATBG} to the active flavors. For VP, such projection is very straightforward, by simply keeping the first line of \cref{eq:H_AH_MATBG}. 
For KIVC, we replace $f_{\beta\eta s} \to \frac{e^{-\frac{\ii}{2} \eta \vartheta_0}}{\sqrt{2}} e^{\ii \frac{\pi}{2} l \frac{1-\eta}{2}} f_{ls}$ where $l = \beta \eta$ mod 3. Calculation shows that
\begin{align}
    H = -\frac{1}{2} \sum_{l_1l_1'l_2l_2'} \sum_{ss'} f^\dagger_{l_1 s} f^\dagger_{l_1' s'} \begin{pmatrix}
        \frac{J_{\rm d}}{2} &  &  &  \\
         & -\frac{J_{\rm d}}{2} & \frac{J_{\rm e}}{2} &  \\
         & \frac{J_{\rm e}}{2} & -\frac{J_{\rm d}}{2}  &  \\
         &  &  & \frac{J_{\rm d}}{2} \\
    \end{pmatrix}_{l_1'l_1, l_2'l_2} f_{l_2' s'} f_{l_2 s}\ .
\end{align}
By comparison with \cref{eq:H_AH}, we find $J_S = J_{\rm e}$, and $J_D - \frac{J_{S}}{2} = J_{\rm d}$ hence $J_D = J_{\rm d} + \frac{J_{\rm e}}{2}$. 
For TIVC, we replace $f_{\beta \eta s} \to \frac{e^{+\ii \eta \frac{\vartheta_0}{2}}}{\sqrt{2}} f_{ls}$, where $l=\beta \eta$ mod 3. Calculation shows that
\begin{align}
    H = -\frac{1}{2} \sum_{l_1l_1'l_2l_2'} \sum_{ss'} f^\dagger_{l_1 s} f^\dagger_{l_1' s'} \begin{pmatrix}
        \frac{J_{\rm d}}{2} &  &  &  \\
         & \frac{J_{\rm d}}{2} & J_{\rm b} + \frac{J_{\rm e}}{2} &  \\
         & J_{\rm b} + \frac{J_{\rm e}}{2} & \frac{J_{\rm d}}{2}  &  \\
         &  &  & \frac{J_{\rm d}}{2} \\
    \end{pmatrix}_{l_1'l_1, l_2'l_2} f_{l_2' s'} f_{l_2 s}\ .
\end{align}
By comparison with \cref{eq:Himp_1}, we read off $J_D = \frac{J_S}{2} = J_{\rm b} + \frac{J_{\rm e}}{2}$.

\begin{table}[tb]
    \centering
    \begin{tabular}{l|l|l|l|l|l|l}
    \hline
        Order & Definition of $f_{l s}$ for $l=\pm$ & $J_S$ & $J_D$ & \multicolumn{2}{l|}{Origin of $D_{\infty} = \Uo_v \rtimes \Zt$} & Origin of $C_2T$ \\
    \cline{5-6}
        &  &  &  & $\Uo_v$ & $\Zt$ &  \\
    \hline
        Strain & $f_{+s} = \frac{1}{\sqrt{2}} \Big( e^{\ii \frac{\vartheta_0}{2}} f_{1 + s} + e^{-\ii \frac{\vartheta_0}{2}} f_{2 + s} \Big) $  & $2J_{\rm d} + J_{\rm e}$ & $J_{\rm d} + \frac{J_{\rm e}}{2}$ & Valley $\Uo$ & $C_{2z}$ & $C_{2z}T$ \\
        & $f_{-s} = \frac{1}{\sqrt{2}} \Big( e^{- \ii \frac{\vartheta_0}{2}} f_{1 - s} + e^{\ii \frac{\vartheta_0}{2}} f_{2 - s} \Big) $  &  &  &  &  &  \\
    \hline
        VP & $f_{+s} = f_{1 \eta s} $ & $2 J_{\rm b}$ & $ 2 J_{\rm a} + J_{\rm b} $ & OAM $\Uo$ & $C_{2x}$ & $C_{2y}T$ \\
        & $f_{-s} = f_{2 \eta s} $ &  &  &  &  & \\
    \hline
        KIVC & $f_{+s} = \frac{1}{\sqrt{2}} \Big( e^{\ii \frac{\vartheta_0}{2}} f_{1 + s} - \ii \cdot e^{-\ii \frac{\vartheta_0}{2}} f_{2 - s} \Big) $ & $J_{\rm e}$ & $J_{\rm d} + \frac{J_{\rm e}}{2}$ & OAM $\Uo$ & $C_{2x} \cdot e^{\ii \pi \frac{\tau^0 - \tau^z}{2}}$ & $C_{2y}T \cdot e^{\ii \vartheta_0 \tau^z} \cdot e^{\ii \pi \frac{\tau^0 - \tau^z}{2}}$ \\
        & $f_{-s} = \frac{1}{\sqrt{2}} \Big( e^{\ii \frac{\vartheta_0}{2} } f_{2 + s} + \ii \cdot e^{-\ii \frac{\vartheta_0}{2}} f_{1 - s} \Big) $ & & & & &  \\
    \hline
        TIVC (IKS) & $f_{+ s} = \frac{1}{\sqrt{2}} \Big( e^{\ii \frac{\vartheta_0}{2}} f_{1 + s} + e^{-\ii \frac{\vartheta_0}{2}} f_{2 - s} \Big) $ & $2J_{\rm b} + J_{\rm e}$ & $J_{\rm b} + \frac{J_{\rm e}}{2}$ & OAM $\Uo$ & $C_{2x}$ & $C_{2y}T \cdot e^{\ii \vartheta_0 \sigma^0 \tau^z \spin^0}$  \\
         & $f_{- s} = \frac{1}{\sqrt{2}} \Big( e^{\ii \frac{\vartheta_0}{2}} f_{2 + s} + e^{-\ii \frac{\vartheta_0}{2}} f_{1 - s} \Big) $ & & & & & \\
    \hline
    \end{tabular}
    \caption{Downfolding the four-orbital (eight-flavor) quantum impurity in MATBG to the two-valley (four-flavor) one. }
    \label{tab:downfold}
\end{table}

This subsection demonstrates that projecting onto different active flavors yields the same form of SVAIM.
In addition to the $J_{S,D}$ values obtained from the projection of bare parameters (\cref{tab:downfold}), several other effects may affect the competition between $S$, $D$, $T$ states.
For example, as discussed at the end of Sec.~VII of Ref.~\cite{wang_bosonization_2025} , the $\rho_z$ coupling further lowers the energy of $D$ states. 
Other factors including fluctuations involving the inactive orbitals and the deviations of the actual active Wannier functions from the simply projected ones. 
But these factors will not change the form of the two-valley Hamiltonian, which is restricted by symmetry. 
Therefore, in this work we will not specify the values of $J_{S,D}$ in \cref{tab:downfold} but treat them as free parameters. 

\clearpage

\section{Spectral function and correlation self-energy ansatz in the AD and LS phases}     \label{app:AfSig}

In this section, we analytically calculate the spectral function $A_f(\omega) = -\frac{1}{\pi} \Im G_f(\omega+\ii0^+)$ in the AD and LS phases. 
From the analytic form of $A_f$ and $G_f$, we also construct an ansatz for the correlation self-energy $\Sigma_f(\omega)$, which will be an extension of the Hubbard-I approximation (HIA) \cite{Lichtenstein_abinitio_1998,hu_projected_2025} to capture the low-energy spectral function at and below $\mcl{O}(J_{S,D})$. 
We finally apply this ansatz to MATBG with heterostrain to obtain the lattice spectral function. 

\subsection{Spectral function \texorpdfstring{$A_f(\omega)$}{Af}}   \label{app:AfSig-Af}

Following \cref{sec:qp}, we will first analyze the form of the ``quasiparticle operator'' $\td{f}$, which is the SW transformation of the physical $f$ electron operator, projected to the low-energy effective models.  
Then, we analytically calculate $A_f(\omega)$ for $\omega \ll \mcl{O}(J_{S,D})$ using bosonization, where the physics is governed by the fixed point Hamiltonian. 
By re-introducing a high-energy multiplet to the fixed-point Hamiltonian, we also demonstrate the formation of the pseudogap shoulders at $\omega \sim \mcl{O}(J_{S,D})$. 
Since the irrelevant terms dropped from the fixed point Hamiltonian are not negligible at this energy scale, this latter calculation only serves as a qualitative demonstration. 

The FL phase is not discussed here, and there will be a quasiparticle peak at zero energy. 

\paragraph{AD phase}
In the original Anderson model, which includes the charge fluctuation, the $f$-electron is by definition created by $f^\dagger_{ls}$. 
Adding or removing one $f$-electron costs an energy of $\mcl{O}(U)$, and leads to the upper and lower Hubbard peaks in the spectral function $A_f(\omega)$ at $\omega \approx \pm\frac{U}{2}$, respectively. 
After lowering the energy scale to $\omega \sim |J_{D}| \ll U$, we apply the first SW transformation $e^{\ii S}$ that integrates out the charge fluctuation, obtaining the (anisotropic) $\Uf$ Kondo model (see \cref{sec:Kondo-coupling-general}). 
At this stage, following \cref{sec:qp}, we identify the low-energy component of the $f$-electron as $\td{f}^\dagger_{+\uparrow} = \PP_2 \Big( e^{\ii S_1} f^\dagger_{+\uparrow} e^{-\ii S_1} \Big) \PP_2 \propto [H_{\rm K} - H_0, \psi_{+\uparrow}^\dagger(0)]$, where $H_{\rm K}$ is the Kondo Hamiltonian and $H_0$ is the Hamiltonian of bath electrons. 
Considering the terms acting in $S,D$ manifold in \cref{eq:HK_c}, we obtain 
\begin{align}  \label{eq:tdf_K}
    \td{f}^\dagger_{+\uparrow} &\propto 
    \lambda_z \cdot \Lambda_z \cdot \psi_{+\uparrow}^\dagger 
    + \zeta_x \sqrt{2} \cdot \Theta_+ \cdot \psi_{-\uparrow}^\dagger \\\nonumber
    &+ \zeta_{0z} \cdot \Bigg( \Theta^{0z} \cdot \psi_{+\uparrow}^\dagger + \left( \Theta^{0x} + \ii \Theta^{0y} \right) \cdot \psi^\dagger_{+\downarrow} \Bigg)
    + \zeta_{zz} \cdot \Bigg( \Theta^{zz} \cdot \psi_{+\uparrow}^\dagger + \left( \Theta^{zx} + \ii \Theta^{zy} \right) \cdot \psi^\dagger_{+\downarrow} \Bigg) \\\nonumber
    &+ \zeta_{xz} \cdot \Bigg( \left(\Theta^{xz} + \ii \Theta^{yz} \right) \psi^\dagger_{-\uparrow} +  \left(\Theta^{xx} + \ii \Theta^{xy} + \ii \Theta^{yx} - \Theta^{yy} \right) \cdot \psi^\dagger_{-\downarrow} \Bigg)\ . 
\end{align}
We have omitted the particle-hole breaking couplings $\gamma_{S,D,T}$ as they are irrelevant in the low-energy physics as discussed after \cref{eq:HK_c}.  
The $\Theta^{\mu\nu}$ operators are defined in \cref{sec:Kondo-coupling-general} and $\Lambda_z=\ket{D,2}\bra{D,2}-\ket{D,\bar 2}\bra{D,\bar 2},\Theta_+ = \ket{S}\bra{D,\bar2} + \ket{D,2}\bra{S}$. 
$\zeta_x$ term in \cref{eq:HK_c} represents the multiplet fluctuation from $D$ to $S$, and the $\zeta_{xz}$ term represents the multiplet fluctuation from $D$ to $T$, hence correspond to excitations upon the ground states that involve the $D$ manifold. 
They will lead to the shoulders of the pseudogap in $A_f(\omega)$ at the energy scale of $\omega \approx \pm |E_S-E_D|, \pm|E_T-E_D|$, respectively. 
$\zeta_{0z}$ and $\zeta_{zz}$, on the other hand, act within the $S \oplus T$ manifold, and will annihilate the ground $D$ states. 

To further integrate out the multiplet fluctuation away from the $D$ manifold, we need to apply a second SW transformation $e^{\ii S'}$, and arrive at the final low-energy theory described by the pair-Kondo model $H_{\rm PK}$ (see the discussion in the main text around \cref{eq:Hx-maintext}). 
At this stage, we identify the low-energy $f$-component as $\td{f}^\dagger_{+\uparrow} = \PP_D \big( e^{\ii S'} \PP_2 \big(e^{\ii S} f^\dagger_{+\uparrow} e^{-\ii S} \big) \PP_2 e^{-\ii S'} \big) \PP_D \propto [H_{\rm PK} - H_{\rm PK,0}, \psi^\dagger_{+\uparrow}(0)]$, where $H_{\rm PK,0}$ is the bath Hamiltonian in the pair-Kondo model. 
We find 
\begin{align}    \label{eq:tdf_PK}
    \td{f}^\dagger_{+\uparrow} \propto \lambda_z \cdot \Lambda_z \cdot \psi^\dagger_{+\uparrow} + (2\pi \lambda_x x_c) \cdot \Lambda_+ \cdot \psi^\dagger_{-\uparrow} \psi^\dagger_{-\downarrow} \psi_{+\downarrow}\ .
\end{align}
Notice that besides $H_{\rm PK}$, the second SW transformation may generate terms of the form of $\PP_D \cdot \psi^\dagger \psi^\dagger \psi \psi$, $\Lambda_z \cdot \psi^\dagger \psi^\dagger \psi \psi$, which leads to components like $\PP_D \cdot \psi^\dagger \psi^\dagger \psi$ and $\Lambda_z \cdot \psi^\dagger \psi^\dagger \psi$ in $\td{f}^\dagger_{+\uparrow}$. 
Nevertheless, as irrelevant perturbations, they only lead to a smooth $\omega^2$ correction in the low-energy regime besides the major contributions from \cref{eq:tdf_PK}, as will be clear soon. 

Next, we compute the Green's function of $\td{f}^\dagger_{+\uparrow}$ in the AD phase using the fixed point Hamiltonian, 
\begin{align} \label{eq:H_AD_DS} 
    H = \int \frac{\mrm{d}x}{4\pi} \sum_{\chi=c,s,v,vs} :(\partial_x \phi_\chi(x))^2: + 2\rho_z \Lambda_z \cdot \partial_x \phi_{v}(x) - J_D \cdot \PP_D - J_S \cdot \PP_S\ .
\end{align}
As we are in the AD phase, the off-diagonal terms $\zeta_x,\lambda_x$ vanish in the fixed point Hamiltonian. (See also Ref.~\cite{wang_bosonization_2025}.) To demonstrate how the multiplet fluctuation to $S$ can lead to the shoulders of the pseudogap, we re-introduced the singlet state $|S\rangle = |0\rangle$ to the model, which has a large energy gap of $J_S-J_D$ above the $D$ manifold. 
The transverse couplings (PK coupling) within the $D$ manifold $\lambda_x$ has flowed to 0 at the fixed point, and the transverse coupling between the $S$ and $D$ manifolds $\zeta_x$ is also assumed to be 0 at the fixed point. 
With PHS, $\PP_S$ cannot couple to any fermion bilinear terms, while the quartic couplings are ignored. 

To solve \cref{eq:H_AD_DS}, as we have done in main text around \cref{eq:ovlHx} and in Sec III.B in \cite{wang_bosonization_2025}, we apply a gauge transformation $U = e^{\ii 2 \rho_z \Lambda_z \phi_{v}(0)}$, after which the bath and the impurity completely decouple 
\begin{align}   \label{eq:H_AD_DS_2}
    \ovl{H} = U H U^\dagger 
    = \int \frac{\mrm{d}x}{4\pi} \sum_{\chi=c,s,v,vs} :(\partial_x \phi_\chi(x))^2: - J_D' \PP_D - J_S \PP_S\ .
\end{align}
Here, $J_D' = J_D + \frac{4 \rho_z^2}{x_c}$ absorbs the energy correction due to the coupling in the $D$ sector $-\frac{4\rho_z^2}{x_c}\Lambda_z^2 = - \frac{4\rho_z^2}{x_c} \PP_D$. 
We now define $J = J_D' - J_S$ as the multiplet excitation energy to the $S$ manifold. 

\cref{eq:tdf_PK} describes the $\td{f}$-operator at the PK energy scale $\omega \ll \mcl{O}(J)$. 
We dub its two components as $\td{f}^{(1)\dagger}_{+\uparrow} \propto \Lambda_z \cdot \psi^\dagger_{+\uparrow}$, $\td{f}^{(2)\dagger}_{+\uparrow} \propto  \Lambda_+ \cdot \psi^\dagger_{-\uparrow} \psi^\dagger_{-\downarrow} \psi_{+\downarrow}$. 
According to \cref{eq:tdf_K}, the $\td f$-operator should also incorporate an component $\td{f}^{(3)\dagger}_{+\uparrow} \propto  \Theta_+ \cdot \psi^\dagger_{-\uparrow}$ at the energy scale $\mcl{O}(J)$, which excites the $D$ states to the $S$ state. 
At the fixed point Hamiltonian, the impurity $\Uo_v$ charge, which distinguishes between the three impurity states, $|D,2\rangle$, $|D,\ovl{2}\rangle$, and $|S\rangle$, and the bath charges of each flavor $ls$ are separately conserved. 
Thus, there will be no cross terms between the correlation functions of the three components above as they carry different charges. 
We now compute them individually. 

After $U = e^{\ii 2 \rho_z \Lambda_z \phi_{v}(0)}$, the three components are transformed into
\begin{align}
    U \left( \td{f}^{(1)\dagger}_{+\uparrow} \right) U^\dagger &\propto \Lambda_z \cdot F_{+\uparrow}^\dagger \cdot e^{\ii \left( \frac{\phi_c}{2}(0) + \frac{\phi_s}{2}(0) + \frac{\phi_v}{2}(0) + \frac{\phi_{vs}}{2}(0) \right)} \ ,\\
    U \Big( \td{f}^{(2)\dagger}_{+\uparrow} \Big) U^\dagger &\propto \Lambda_+ \cdot F_{-\uparrow}^\dagger F_{-\downarrow}^\dagger F_{+\downarrow} \cdot e^{\frac{\ii}{2} \phi_c(0)} e^{\frac{\ii}{2} \phi_s(0)} e^{\frac{\ii}{2} \phi_{vs}(0)} e^{-\ii(\frac{3}{2} - 4\rho_z)\phi_v(0)} \ , \\  \label{eq:U_tdf3}
    U \Big( \td{f}^{(3)\dagger}_{+\uparrow} \Big) U^\dagger &\propto \Theta_+ \cdot F_{-\uparrow}^\dagger \cdot e^{\frac{\ii}{2} \phi_c(0)} e^{\frac{\ii}{2} \phi_s(0)} e^{-\frac{\ii}{2} \phi_{vs}(0)} e^{-\ii(\frac{1}{2} - 2\rho_z)\phi_v(0)} \ ,
\end{align}
where we have exploited $[\Lambda_z, \Theta_+] = \Theta_+$, $[\Lambda_z, \Lambda_+] = 2 \Lambda_+$ and hence $U \Theta_+ U^\dagger = \Theta_+ e^{\ii 2\rho_z\phi_v(0)}$, $U \Lambda_+ U^\dagger = \Lambda_+ e^{\ii 4 \rho_z\phi_v(0)}$. 
Then, according to the correlation functions in Eqs.~(A61) and (A62) in \cite{wang_bosonization_2025}, the imaginary-time Green's function for $\td f^{(1)\dagger}$ in the $T\to 0^+$ limit reads
\begin{align}
    G_f^{(1)}(\tau) &= -\Big\langle T_\tau ~ e^{\tau H} \Big( \td{f}^{(1)}_{+\uparrow} \Big) e^{-\tau H} \cdot \Big( \td{f}^{(1) \dagger}_{+\uparrow} \Big) \Big\rangle_{0} \sim - \left\langle \Lambda_z^2 \right\rangle \left[ \theta(\tau) \frac{x_c}{|\tau|} - \theta(-\tau) \frac{x_c}{|\tau|} \right] \ .
\end{align}
As $\Lambda_z$ commutes with the Hamiltonian, it produces a factor $\Lambda_z^2$, whose average can be factored out, and produces $\langle\Lambda_z^2\rangle = 1$ in the $D$ manifold. 
The remaining correlation function is then identical to the correlation function of a bath electron $\psi_{+\uparrow}^\dagger(0)$, which decays as $\frac{1}{\tau}$, and corresponds to a constant density of states across all $\omega$. 
Correspondingly, $G_f^{(1)}(\omega)$ contributes a constant background in $A_f(\omega)$, 
\begin{align}
    A_f^{(1)}(\omega) \propto \mrm{const}\ .
\end{align}

For $\td{f}^{(2)\dagger}$, as $\Lambda_\pm$ commutes with $\ovl{H}$ (because $\Lambda_\pm$ commutes with both $\PP_D = \Lambda_z^2$ and $\PP_S$), the Green's function is simply determined by the remaining vertex operators of bath fields.
According to Eqs.~(A59) to (A62) in \cite{wang_bosonization_2025}, we obtain
% \cref{eq:vertex_corr_2,eq:vertex_corr_2n,eq:free-fermion-propagator}
\begin{align}
    G_f^{(2)}(\tau) &= -\Big\langle e^{\tau H} \Big( \td{f}^{(2)}_{+\uparrow} \Big)  e^{-\tau H} \cdot \td{f}^{(2) \dagger}_{+ \uparrow} \Big\rangle_{0} \sim -\left[ \theta(\tau) \left( \frac{x_c}{|\tau|} \right)^{\alpha_2} -  \theta(-\tau) \left( \frac{x_c}{|\tau|} \right)^{\alpha_2} \right]
\end{align}
where the power $\alpha_2 = \frac{3}{4} + (\frac{3}{2} - 4\rho_z)^2$. 
Following the same trick of contour integral of irrational functions used in App.~C in Ref.~\cite{wang_bosonization_2025}, the corresponding spectral function should be 
\begin{align}
    A_f^{(2)}(\omega) \sim x_c^{\alpha_2} |\omega|^{\alpha_2-1}\ .
\end{align}
Importantly, for $0 < \rho_z < \rho_z^c = \frac{1}{2} - \frac{1}{2\sqrt{2}} \approx 0.1464$, $2 > \alpha_2-1 > 0.5858$, so $A_f^{(2)}$ can either behave as a smooth dip (if $\rho_z$ is small, so that $\alpha_2-1>1$), or a kink downward (if $\rho_z$ is large and approaches the BKT critical value, so that $\alpha_2-1<1$). 
Added up, $A_f^{(1)} + A_f^{(2)}$ determines the spectral features at low frequency $\omega \ll \mcl{O}(J)$.

Finally, we compute the Green's funciton for $\td{f}^{(3)}$, which contains a multiplet excitation to the $S$ manifold, 
{\small
\begin{align}  \label{eq:G_tdf3}
    G_f^{(3)}(\tau) &= -\Big\langle T_\tau \cdot e^{\tau H} \Big( \td{f}^{(3)}_{+\uparrow} \Big)  e^{-\tau H} \cdot \td{f}^{(3) \dagger}_{+\uparrow} \Big\rangle_{0}  \\\nonumber
    &\propto -\Big\langle T_\tau \cdot e^{\tau \ovl{H}} \Big( \Theta_- \cdot e^{-\ii \frac{\phi_c(0)}{2}} e^{-\ii \frac{\phi_s(0)}{2}} e^{\ii \frac{\phi_{vs}(0)}{2}} e^{\ii (\frac{1}{2} - 2\rho_z) \phi_v(0)} \Big)  e^{-\tau \ovl{H}} \cdot \Big( \Theta_+ \cdot e^{\ii \frac{\phi_c(0)}{2}} e^{\ii \frac{\phi_s(0)}{2}} e^{-\ii \frac{\phi_{vs}(0)}{2}} e^{-\ii (\frac{1}{2} - 2\rho_z) \phi_v(0)} \Big) \Big\rangle_{\ovl{0}}  \ . 
\end{align}}%%
Crucially, the $S$ sector is higher by an energy of $J$ than the $D$ sector. 
Therefore, if $\tau>0$, the time-evolution operator $e^{\tau \ovl{H}}$ lives in the $D$ sector, while $e^{-\tau \ovl{H}}$ lives in the $S$ sector, as it is sandwiched by $\Theta_-$ and $\Theta_+$, leading to a $e^{-\tau J}$ factor. 
On the other hand, if $\tau<0$, $e^{-\tau \ovl{H}}$ will live in the $D$ sector, while $e^{\tau \ovl{H}}$ will live in the $S$ sector, leading to a $e^{\tau J}$ factor. 
By also computing the bath correlations, which is directly determined by the total scaling dimension $\alpha_3 = \frac{3}{4} + \left( \frac{1}{2} - 2\rho_z \right)^2$, we obtain
\begin{align}
    G_f^{(3)}(\tau) &\sim -\left[ \theta(\tau) \left( \frac{x_c}{|\tau|} \right)^{\alpha_3} e^{- \tau |J|} - \theta(-\tau) \left( \frac{x_c}{|\tau|} \right)^{\alpha_3} e^{\tau |J|}  \right] \ .
\end{align}
The Mastubara Green's function is given by  $G(\ii \omega) = \int_{-\infty}^{\infty} \mrm{d} \tau ~ G(\tau) ~ e^{\ii \omega \tau}$.
Following the same trick of contour integral of irrational functions used in App.~C in Ref.~\cite{wang_bosonization_2025}, where $f(z)$ should be chosen as $(J-\ii\omega)^{\alpha_3-1}$ and it has a branch-cut at $z=-\ii y$ ($y\ge J$) (Fig.~5 in Ref.~\cite{wang_bosonization_2025}). 
We obtain 
\begin{align}
    G_f^{(3)}(\ii \omega) &= - x_c^{\alpha_3} \cdot \Gamma(1 \!-\!\alpha_3 ) \cdot \Big( (J - \ii \omega)^{\alpha_3 - 1} - (J + \ii \omega)^{\alpha_3 - 1} \Big) \ ,\\
    G_f^{(3)}(\omega +\ii 0^+) &= - x_c^{\alpha_3} \cdot \Gamma(1 \!-\!\alpha_3 ) \cdot \Big( (J - \omega - \ii 0^+)^{\alpha_3 - 1} - (J + \omega +\ii 0^+)^{\alpha_3 - 1} \Big) \ .
\end{align}
The $(J - \omega - \ii 0^+)^{\alpha_3 - 1} $ and $(J + \omega + \ii 0^+)^{\alpha_3 - 1} $ factors in the retarded Green's functions should be interpreted as $f(z=-\ii\omega + 0^+)$ and $f(z=\ii\omega - 0^+)$, respectively. 
According to the branch-cut shown in Fig.~5(b), there are 
\begin{equation}
\Im[f(z=-\ii\omega + 0^+)] = - \sin((\alpha_3-1)\pi) \cdot |\omega-J|^{\alpha_3-1} \cdot \theta(\omega-J)\ ,
\end{equation}
\begin{equation}
\Im[f(z=\ii\omega - 0^+)] = \sin((\alpha_3-1)\pi) \cdot |\omega-J|^{\alpha_3-1} \cdot \theta(-\omega-J)\ . 
\end{equation}
Thus, the corresponding spectral function is 
\begin{align}   \label{eq:Af3_ori}
    A_f^{(3)}(\omega) &\sim  x_c^{\alpha_3} \cdot \frac{\pi}{\Gamma(\alpha_3) } \cdot \bigg( \theta(\omega-J) \Big| \omega-J \Big|^{\alpha_3-1} +\theta(-\omega-J) \Big| \omega+J \Big|^{\alpha_3-1} \bigg) \ ,
\end{align}
where the relation $\Gamma(1-\alpha_3) \cdot \sin(\pi(1-\alpha_3)) = \frac{\pi}{\Gamma(\alpha_3) }$ is used. 
For $0 < \rho_z < \rho_z^c = \frac{1}{2} - \frac{1}{2\sqrt{2}}$, $0 > \alpha_3-1 > -0.2071$. 

We also remark on the `irrelevant' components in $\td{f}$, with the form of $\Lambda_z \cdot \psi^\dagger \psi^\dagger \psi$ or $\PP_D \cdot \psi^\dagger \psi^\dagger \psi$. 
As the gauge transformation commutes with $\Lambda_z$ and $\PP_D=\Lambda_z^2$, it does not alter the scaling dimension of these components, hence the time-decaying power $\alpha$ is completely determined by the bath fields, which will be $\alpha=3$. 
The corresponding spectral function must be proportional to $\omega^2$, i.e.,
\begin{equation}\label{eq:Af_45}
    A^{(4)}_f(\omega) \sim x_c^3 \omega^2 \ ,
\end{equation}
which is negligible compared to $A_f^{(1)} + A_f^{(2)}$ in the low-energy regime. 

$A^{(1)}_f(\omega) + A^{(2)}_f(\omega) + A^{(3)}_f(\omega) + A^{(4)}_f(\omega)$ sketches the basic features of the spectral function in the AD phase:
$A^{(1)}_f(\omega)$ gives a constant spectral weight around $\omega=0$, $A^{(2)}_f(\omega)$ gives a kink at $\omega=0$ when the system is close to the BKT transition point, 
and $A^{(3)}_f(\omega)$ qualitatively reproduces the pseudogap shoulders at the multiplet excitation energy. 
As has been remarked, since the fixed point Hamiltonian is only valid at energy scales far below $\mcl{O}(J_{S,D})$, only $A_f^{(1,2)}(\omega)$ are quantitatively reliable at $\omega \ll J_{S,D}$. 
$A_f^{(3)}(\omega)$, on the other hand, corresponds to features at $\omega \sim J_{S,D}$, and only qualitatively demonstrates that the shoulder peaks are contributed by excitations like $\Theta_+ \cdot \psi^\dagger$.

\paragraph{LS phase} 
To understand the LS phase, similarly, we can carry out another second SW transformation $e^{\ii S''}$ that integrates out the multiplet fluctuation away from the $S$ manifold. 
In the final low-energy theory, the only impurity operator that can be written is $\PP_S$, which cannot couple to any bilinear bath operator, if assuming PHS (see \cref{sec:Kondo-coupling-general}). 
However, the second SW transformation can lead to quartic couplings of the form $\PP_S \cdot \psi^\dagger \psi^\dagger \psi \psi$, namely, an effective interaction of the bath electrons at the spatial origin $x=0$. 
The form of this effective interaction will be calculated in \cref{sec:effective-interaction-2}, and such a quartic terms will be verified as irrelevant at the LS fixed point. 
Nevertheless, it brings about the following components of the quasiparticle operator $\td{f}^{(4) \dagger}_{+\uparrow} \propto \PP_S \cdot \psi^\dagger_{+\uparrow} \psi^\dagger_{ls} \psi_{ls}$, $\td{f}^{(5) \dagger}_{+\uparrow} \propto \PP_S \cdot \psi^\dagger_{-\uparrow} \psi^\dagger_{+\downarrow} \psi_{-\downarrow}$. 
Such operators all possess scaling dimension $\alpha=3$, hence contribute a quadratic term in the spectral function (\cref{eq:Af_45}).

In order to demonstrate the formation of pseudogap, we also re-introduce the $D$ manifold, and consider the fixed point Hamiltonian which has the same form as \cref{eq:H_AD_DS} (See Sec.~VII in Ref.~\cite{wang_bosonization_2025} for details) 
\begin{equation}
H = \int \frac{\dd x}{4\pi} 
    :(\partial_{x} \phi_{v})^2 :
    + \frac{\varepsilon_{D}}{x_c} \cdot \PP_D 
    + 2 \rho_z \Lambda_z \cdot \partial_x \phi_v(x) \Big|_{x=0} \ .
\end{equation}
Here we already dropped $\chi=c,s,vs$ components that decouple with the impurity.
By similarly applying the gauge transformation $U = e^{\ii 2\rho_z \phi_v(0)}$, the $D$ sector also decouples from the bath, 
\begin{equation}   \label{eq:H_LS_fp_U}
\ovl{H} = \int \frac{\dd x}{4\pi} 
    :(\partial_{x} \phi_{v})^2 :
    + \frac{\varepsilon_{D}'}{x_c} \cdot \PP_D 
\end{equation}
where $\varepsilon_D' = \varepsilon_D - 4\rho_z^2 > 0$. 
\cref{eq:H_LS_fp_U} is identical to \cref{eq:H_AD_DS_2}, except with the sign of $\varepsilon_D$ reversed, so that the $S$ multiplet becomes the ground state. 
We can now compute the spectral function due to excitations like $\td{f}^{(3)\dagger}_{+\uparrow} \propto \Theta_+ \cdot \psi_{-\uparrow}^\dagger$. 
$U \td{f}^{(3)\dagger} U^\dagger$ follows identically as \cref{eq:U_tdf3} after the gauge transformation, and hence in the LS phase, the Green's function of $\td{f}^{(3)\dagger}$ has the same expression as \cref{eq:G_tdf3}. 
The only difference is that $\ovl{0}$ is in the $S$ manifold, instead of $D$, but all the other derivation follows identically.  
Finally, we arrive at $A_f^{(3)}$ in \cref{eq:Af3_ori}, with $J$ given by $\frac{\varepsilon_D'}{x_c}$. 

$A^{(3)}(\omega) + A^{(4)}_f(\omega)$ sketches the basic features of the spectral function in the LS phase.

We finally remark that, without PHS, there can be a term of $\PP_S \cdot :\psi^\dagger \sigma^0 \spin^0 \psi :$ in $[H-H_0]$, hence there will be a component proportional to $\PP_S \cdot \psi^\dagger$ in the definition of $\td{f}^\dagger$. This term has scaling $\alpha=1$, and will lead to a finite constant background in $A_f(\omega)$ at $\omega=0$.

\subsection{Ansatz for correlation self-energy \texorpdfstring{$\Sigma_f(\omega)$}{Sig}}   \label{app:AfSig-Sig}

\begin{figure}[tb]
    \centering
    \includegraphics[width=0.6\linewidth]{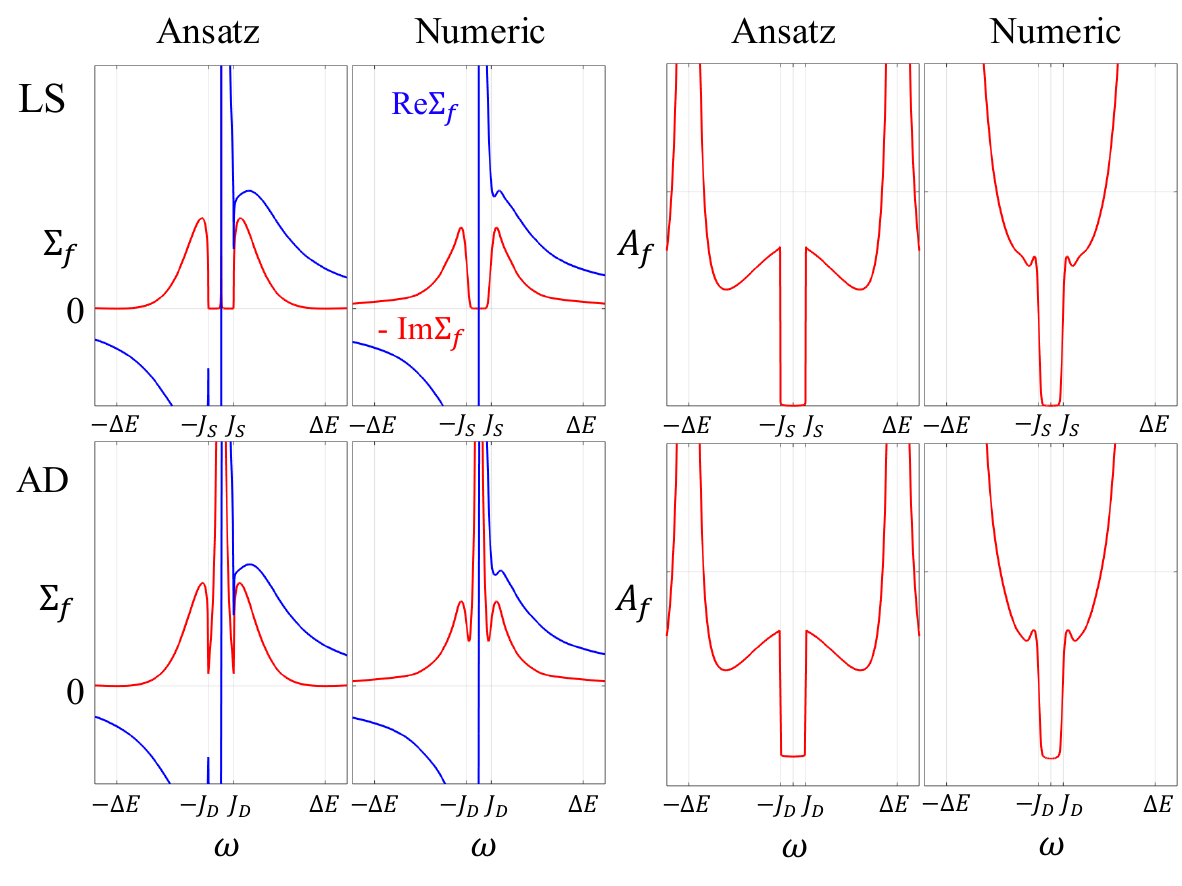}
    \caption{Ansatz for correlation self-energy $\Sigma_f(\omega+ \ii 0^+)$, compared with NRG result. 
    We have absorbed the on-site potential $\epsilon_f$ to cancel the Hartree-Fock value of $\Sigma_f(\omega +\ii 0^+)$. 
    We choose $U=3$ and $\Delta_0=0.04$ for the Anderson model parameters. 
    (Upper) For the LS phase, the anti-Hund's couplings are chosen as $J_S=0.2,J_D=0$. 
    The parameters in the analytical ansatz are $\beta_3 = 0.15$, $D_3=7J_S$. 
    (Lower) For the AD phase, $J_S=0$, $J_D=0.2$. 
    The tuning parameters adopted in the analytical ansatz are $\beta_1 = \beta_2 = 0.03$, $\beta_3 = 0.12$, $D_1=D_2=D_3=7J_D$, and $\alpha_2=1.9$. 
    In the atomic limit, the peaks in the spectral density experience weak hybridization-induced broadening, so we keep more multiplets (~8000) and use a larger $n_z=8$ than the default choice to better resolve the spectral function.}
    \label{fig:ansatz}
\end{figure}

The correlation self-energy $\Sigma_f(\omega)$ and the impurity Green's function is related via the following relation, 
\begin{align}   \label{eq:Gf}
    G_f(\omega) = \frac{1}{\omega - \Sigma_f(\omega) - \Delta(\omega)}
\end{align}
where we have chosen the hybridization self-energy as constant $\Delta(\omega+\ii 0^+) = -\ii \Delta_0$. 
By our convention, the on-site potential $\epsilon_f$ is absorbed into $\Sigma_f$. 
In this section, we construct an ansatz for $\Sigma_f(\omega)$, with the aim to reproduce $A_f(\omega) = -\frac{1}{\pi} \Im G_f(\omega+\ii 0^+)$ in the full energy range. 
We follow the procedures below. 
We will write an ansatz $A_f(\omega)$ as a mixing of the Hubbard peaks $A_f^{\rm(at)}(\omega)$ and the $A_f^{(1,2,3)}(\omega)$ components defined above. 
We ignore $A_f^{(4)}(\omega)$, as the $\omega^2$ dependence will naturally arise when the hybridization self-energy $\Delta(\omega)$ is also present. 
To guarantee proper normalization of $A_f^{(1,2,3)}(\omega)$, we add smooth high-energy cutoffs, because they are only well-defined in the low-energy end. 
We treat the mixing amplitudes of each component as tuning parameters. 
We then obtain an ansatz $G_f(\omega)$ via the Kramer-K\"onig relation, and define the correlation self-energy ansatz via $\Sigma_f(\omega) \!=\! \omega - G_f^{-1}(\omega)$. 
The physical impurity spectral function is then computed by inserting $\Sigma_f$ into \cref{eq:Gf} and $A_f(\omega) = -\frac{1}{\pi} \Im G_f(\omega+\ii 0^+)$.

Our \textit{ans\"atze} are natural generalizations of the Hubbard-I approximation (HIA) \cite{Hubbard_2, Lichtenstein_abinitio_1998, hu_projected_2025}, where one writes the ansatz $A_f \!=\! A_f^{\rm (at)}$. 
As $A_f^{(1,2,3)}(\omega)$ are included, we are able to capture the pseudogap and in-gap excitations, in addition to the Hubbard bands. 
These Hubbard bands in MATBG \cite{hu_projected_2025} are also captured by other approaches \cite{ledwith_nonlocal_2024,zhao_topological_mott_2025}. 

Let us start with the high-energy end $\omega \sim \mcl{O}(U)$, where the main feature of $G_f$, the Hubbard peaks, is already captured by HIA. 
HIA approximates the correlation self-energy as the ``atomic'' one $\Sigma^{(\rm at)}(\omega+\ii 0^+) = \omega - \left( G^{\rm (at)}_f(\omega+ \ii 0^+) \right)^{-1}$, where $G^{\rm (at)}_f$ is the Green's function of $f$ electron computed for an isolated impurity (an ``atom'') \cite{Lichtenstein_abinitio_1998}, 
\begin{align}   \label{eq:Gf_at}
    G^{\rm (at)}_{f}(\omega+\ii 0^+) = \frac{1}{Z} \sum_{\Xi,\Xi_0} \frac{|\langle \Xi| f_{l s} |\Xi_0\rangle|^2}{\omega + E_{\Xi} - E_{\Xi_0} +\ii 0^+} + \frac{1}{Z} \sum_{\Xi} \frac{|\langle \Xi| f^\dagger_{l s} |\Xi_0\rangle|^2}{\omega + E_{\Xi_0} - E_{\Xi} +\ii 0^+} \ .
\end{align}
The above expression is at low-temperature limit $T \ll J_{S,D}$. 
Here, $\Xi_0$ runs over the impurity ground states, and $Z$ is the impurity ground state degeneracy - in AD, $\Xi_0 \in D$ and $Z=2$, and in LS, $\Xi_0 \in S$ and $Z=1$. 
$\Xi$ represents impurity excited states, with $E_{\Xi} - E_{\Xi_0} \ge 0$. 
From \cref{eq:Gf_at}, one can easily read off that, removing one $f$ electron contributes a pole at negative frequency $\omega$, while adding one electron contributes a pole at positive frequency $\omega$. 
Since all flavors $ls$ are degenerate, the result will be equal for all $ls$. 
For the current model at PHS, for both AD and LS, there will be
\begin{align}
    G_f^{(\rm at)}(\omega +\ii 0^+) = \frac{1}{2} \frac{1}{\omega + \Delta E +\ii 0^+} + \frac{1}{2} \frac{1}{\omega - \Delta E +\ii 0^+} = \frac{\omega+\ii 0^+}{(\omega +\ii 0^+)^2 - (\Delta E)^2}
\end{align}
where $\Delta E \approx \frac{U}{2}$. 
Directly inverting this Green's function, one obtains the HIA ansatz of correlation self-energy, 
\begin{align}   \label{eq:Sig_at}
    \Sigma^{(\rm at)}_{f}(\omega+\ii 0^+) = \frac{(\Delta E)^2}{\omega +\ii 0^+}\ .
\end{align}
For this PHS result, $\Sigma_f^{(\rm at)}$ has a pole at $\omega=0$. 
Notice that the spectral function of $A_f^{(\rm at)}(\omega) = -\frac{1}{\pi} \Im G_f^{(\rm at)}(\omega + \ii \eta)$ is already normalized, namely, $1 = \int\dd\omega ~ A_f^{(\rm at)}(\omega)$. 

Besides the poles at $\pm\Delta E$, the asymptotic behaviors of $G_f^{(\rm at)}$ include
\begin{align}
    G_f^{(\rm at)}(\omega+ \ii 0^+) \stackrel{\omega\to 0}{=} -\frac{\omega}{(\Delta E)^2} + \cdots \ ,
    \qquad \qquad G_f^{(\rm at)}(\omega+ \ii 0^+) \stackrel{\omega\to \infty}{=} \frac{1}{\omega} + \cdots\ . 
\end{align}

\paragraph{LS phase}
Next we add the pseudogap at $\mcl{O}(J)$. 
As discussed in the previous section, the non-universal power-law singularities at $\omega \sim \mcl{O}(J)$ need not be treated as quantitively valid features. 
Also, in the true `atomic' limit, we expect $\alpha_3 \to 1$ as $\rho_z \to 0$, where the singularity becomes rather weak.  
Therefore, for simplicity and for practical convenience, we simply set $\alpha_3=1$. 
Meanwhile, we impose a smooth cutoff with width $D_3\sim J$, in order to describe the fact that the $\td{f}^{(3)}$ component is not well-defined at arbitrary energy scale, but only within some range near $\mcl{O}(J)$. Therefore,
\begin{align}
    A_f^{(3)}(\omega) = \frac{1}{2 \arctan\frac{D_3}{J}} \frac{D_3}{\omega^2 + D_3^2} \Big[ \theta(-J-\omega) + \theta(\omega-J)  \Big] \ .
\end{align}
Here, we have attached a constant to guarantee the normalization that $1 = \int\dd\omega A_f^{(3)}(\omega)$, which can be quickly verified from
\begin{align}
    \int_{-\infty}^{-J}\dd\omega \frac{D_3}{\omega^2 + D_3^2} + \int_{J}^{\infty}\dd\omega \frac{D_3}{\omega^2 + D_3^2}  = 2 \arctan\frac{D_3}{J}\ .
\end{align}
Next, we compute the real-part of the Green's function corresponding to $A_f^{(3)}(\omega)$, from the Kramer-K\"onig relation, 
{\small
\begin{align}  \label{eq:Gf3}
    G_f^{(3)}(\omega+\ii 0^+) &= \int_{-\infty}^{\infty} \dd \epsilon \frac{A_f^{(3)}(\epsilon)}{\omega +\ii 0^+ - \epsilon} 
    = \frac{1}{2\arctan\frac{D_3}{J}} \Bigg[ \int_{-\infty}^{-J} \dd \epsilon + \int_{J}^{\infty} \dd \epsilon \Bigg]  \frac{1}{\omega +\ii 0^+ - \epsilon} \frac{D_3}{\epsilon^2 + D_3^2}  \\\nonumber
    &=  \frac{1}{2 \arctan\frac{D_3}{J}} \Bigg( - \frac{D_3}{\omega^2+D_3^2} \ln\frac{\epsilon-(\omega+\ii 0^+)}{D} + \frac{\ii}{2} \frac{1}{\omega + \ii D_3}\ln(\frac{\epsilon+\ii D_3}{D_3}) - \frac{\ii}{2}\frac{1}{\omega - \ii D_3}\ln(\frac{\epsilon-\ii D_3}{D_3})  \Bigg) (\Bigg|_{-\infty}^{-J} + \Bigg|_J^{\infty} ) \\\nonumber
    &= \frac{1}{2 \arctan\frac{D_3}{J}} \frac{D_3}{\omega^2 + D_3^2} \ln\frac{J-(\omega+\ii 0^+)}{J+(\omega+\ii 0^+)} 
    +  \frac{\omega }{\omega^2+D_3^2} \ .
\end{align}}%%
Notice that, with the second term, the poles at $\omega=\pm\ii D_3$ introduced by the artificial Lorentzian envelope have been canceled. 

We choose the correlation self-energy ansatz as
\begin{align}
	\Sigma_f(\omega+\ii 0^+) = \omega - \left( \beta_{\rm at} G^{(\rm at)}_f(\omega+\ii 0^+) + \beta_3 G^{(3)}_f(\omega+\ii 0^+)  \right)^{-1}
\end{align}
where $\beta_{\rm at} + \beta_3 = 1$ are tuning parameters. 
Notice that, in presence of hybridization $\Delta$, a $A_f \sim \omega^2$ component in the spectral function will be automatically generated using the above ansatz of correlation self-energy (\cref{fig:ansatz}), which further justifies omitting  $A^{(4)}_f$ (\cref{eq:Af_45}) in the construction. 

The Lorentz truncation function  $ \frac{D_3^2}{\omega^2 + D_3^2}$ leads to an artifact of the correlation self-energy at $\omega \to \infty$.
According to the Lehmann spectral representation, the correlation self-energy must vanish in the $\omega\to \infty$ limit.
However, since $G_f = \frac{1}{\omega} - \ii \frac{D}{\omega^2} + \cdots$, the correlation self-energy has a finite imaginary part in the $\omega\to \infty$ limit: 
\begin{align}
    \Sigma_f(\omega +\ii 0^+) = \omega - (G_f)^{-1} \sim \omega - \frac{1}{\frac{1}{\omega} - \ii\frac{D_3}{\omega^2}} \sim \omega - \frac{\omega}{1 - \ii\frac{D_3}{\omega}} \sim -\ii D_3\ .
\end{align}
Such an artifact can be avoided using a faster-decaying truncation function, {\it e.g.,} $ \frac{D_3^3}{\omega^4 + D_3^4}$. 
Nonetheless, for the purpose of this work, this artifact does not lead to any physical error, hence we adopt the current truncation choice.

\paragraph{AD phase}
We also consider 
\begin{align}
    A_f^{(1)}(\omega) &= \frac{1}{\pi} \frac{D_1}{\omega^2 + D_1^2}  \ ,\\
    A_f^{(2)}(\omega) &= \frac{\sin\frac{\pi \alpha_2}{2}}{\pi D_2^{\alpha_2-2}} \frac{1}{\omega^2 + D_2^2} |\omega|^{\alpha_2 - 1} \ .
\end{align}
$D_1$ and $D_2$ are parameters representing the energy scale where $\td f^{(1)}$ and $\td f^{(2)}$ are justified. 
The normalization of $A_f^{(1)}$ is obvious.
For $\alpha_2< 2$, with which $A_f^{(2)}(\omega)$ exhibits a kink at $\omega=0$, $A_f^{(2)}$ is also normalized because $\int_{-\infty}^{\infty} \dd\omega \frac{|\omega|^{\alpha_2-1}}{\omega^2 + D^2} = 2 D_2^{\alpha_2-2} \int_{0}^{\infty} \dd x \frac{x^{\alpha_2-1}}{x^2 + 1} = \frac{\pi D_2^{\alpha_2-2}}{\sin\frac{\pi \alpha_2}{2}}$ (see calculations around \cref{eq:csccot}). 
For $\alpha_2 > 2$, $ A_f^{(2)}(\omega) $ is not normalized. Nevertheless, $A_f^{(2)}$ in this case is featureless because it is smooth (in the sense that the first-order derivative is zero, although the second-order derivative still diverges) and small around $\omega=0$, and one may omit it. 
If one were to keep $A^{(2)}_f(\omega)$ with $\alpha_2>2$ in the low-energy physics, one may choose a faster-decaying truncation function, {\it e.g.,} $\frac{D_2^3}{\omega^4 + D_2^4}$, instead of the Lorentz function.

$G_f^{(1)}$ can be obtained from $G_f^{(3)}$ (\cref{eq:Gf3}) by setting $J\to 0^+$, 
\begin{align}
    G_f^{(1)}(\omega+\ii 0^+) = \frac{1}{\omega + \ii D_1}
\end{align}
where the pole at $\omega = \ii D_1$ is canceled. 
$G_f^{(2)}$ is given by 
\begin{align}
    G_f^{(2)}(\omega+ \ii 0^+) &= \int_{-\infty}^{\infty}\dd\epsilon \frac{A_f^{(2)}(\epsilon)}{\omega+ \ii 0^+ - \epsilon} 
    = \frac{\sin\frac{\pi \alpha_2}{2}}{\pi D_2^{\alpha_2-2}} \int_{-\infty}^{\infty} \dd\epsilon \frac{1}{\omega + \ii0^+ - \epsilon} \frac{|\epsilon|^{\alpha_2 - 1} }{\epsilon^2 + D_2^2} 
    = \frac{\sin\frac{\pi \alpha_2}{2}}{\pi D_2^{\alpha_2-2}} \int_{0}^{\infty} \dd\epsilon \frac{2 \omega}{(\omega+ \ii 0^+)^2 - \epsilon^2} \frac{\epsilon^{\alpha_2 - 1} }{\epsilon^2 + D_2^2} \\\nonumber
    &= \frac{\sin\frac{\pi \alpha_2}{2}}{\pi D_2^{\alpha_2-2}} \cdot \frac{2\omega}{\omega^2 + D^2} \int_{0}^{\infty} \dd\epsilon \Big( \frac{\epsilon^{\alpha_2 - 1} }{\epsilon^2 + D^2} + \frac{\epsilon^{\alpha_2 - 1} }{(\omega+\ii 0^+)^2 - \epsilon^2} \Big) \ .
\end{align}
For $0<\nu<\mu$, there are
\begin{align}   \label{eq:csccot}
    \int_0^{\infty}\dd x \frac{x^{\mu-1}}{x^\nu+1} = \frac{\pi}{\nu} \frac{1}{\sin\frac{\pi \mu}{\nu}}\ , \qquad \mcl{P} \int_0^{\infty} \dd x \frac{x^{\mu-1}}{1 - x^\nu} = \frac{\pi}{\nu} \cot\frac{\pi\mu}{\nu}\ . 
\end{align}
Carrying out the principal value integral, we obtain 
\begin{align}
    \Re G_f^{(2)}(\omega+ \ii 0^+) &= \frac{\omega}{\omega^2 + D_2^2} + \frac{\sgn(\omega)}{\omega^2 + D_2^2} \frac{|\omega|^{\alpha_2-1}}{D_2^{\alpha_2-2}} \cos\frac{\pi\alpha_2}{2}\ .
\end{align}
By also matching the imaginary part $\Im G_f^{(2)}(\omega + \ii0^+) = -\pi A_f^{(2)}(\omega)$, we obtain
\begin{equation}
G_f^{(2)}(\omega+ \ii 0^+) =
\frac{\omega}{\omega^2 + D_2^2} 
+ \frac{1}{\omega^2 + D_2^2} \cdot \frac{|\omega|^{\alpha_2-1}}{D_2^{\alpha-2}}
\cdot \pare{ \cos\frac{\pi\alpha_2}2 \cdot \sgn(\omega) - \ii \cdot \sin\frac{\pi \alpha_2}2  }\ . 
\end{equation}
We can rewrite the factor $|\omega|^{\alpha_2-1} \pare{ \cos\frac{\pi\alpha_2}2 \cdot \sgn(\omega) - \ii \cdot \sin\frac{\pi \alpha_2}2  }$ as 
\begin{align}
    -\ii (-\ii \omega + 0^+)^{\alpha_2-1} &= -\ii |\omega|^{\alpha_2-1} e^{-\ii \frac{\pi}{2}(\alpha_2-1) \sgn(\omega)} = -\ii |\omega|^{\alpha_2-1} \left( \cos\frac{\pi (\alpha_2-1)}{2} - \ii \cdot \sgn(\omega) \sin\frac{\pi(\alpha_2-1)}{2} \right) \\\nonumber
    &= -\ii  |\omega|^{\alpha_2-1} \left( \sin\frac{\pi \alpha_2}{2} + \ii \cdot \sgn(\omega) \cos\frac{\pi \alpha_2}{2} \right) \ . 
\end{align}
It is direct to verify that the pole of the first term in $G^{(2)}_f$ at $\omega = \ii D_2$ is canceled by the second term. 

We choose the correlation self-energy ansatz as
\begin{align}
	\Sigma_f(\omega + \ii 0^+) = \omega - \left( \beta_{\rm at} G^{(\rm at)}_f(\omega+\ii 0^+) + \beta_1 G^{(1)}_f(\omega+\ii 0^+) + \beta_2 G^{(2)}_f(\omega+\ii 0^+) + \beta_3 G^{(3)}_f(\omega+\ii 0^+)  \right)^{-1}
\end{align}
where $\beta_{\rm at,1,2,3}$ are tuning parameters and satisfy $\beta_{\rm at} + \beta_1 + \beta_2 + \beta_3 = 1$. 

Let us check the asymptotic behavior of $\Sigma_f$ in AD at $\omega \to 0$. The ansatz 
\begin{align}
    G_f(\omega + \ii 0^+) &= \beta_{\rm at} G^{(\rm at)}_f(\omega+\ii 0^+) + \beta_1 G^{(1)}_f(\omega+\ii 0^+) + \beta_2 G^{(2)}_f(\omega+\ii 0^+) + \beta_3 G^{(3)}_f(\omega+\ii 0^+)  \\\nonumber
    \Re G_f(\omega + \ii 0^+)  &= \left( \frac{-\beta_{\rm at} }{(\Delta E)^2} + \frac{\beta_1 }{D_1^2} + \frac{\beta_2}{D_2^2} + \frac{\beta_3}{D_3^2} \right) \omega  + \mcl{O}(\omega^3) + \beta_2  \frac{|\omega|^{\alpha_2-1}}{D_2^{\alpha_2}} \cos\frac{\pi \alpha_2}{2} \sgn(\omega) + \mcl{O}(\omega^{\alpha_2+1}) \\\nonumber
    \Im G_f(\omega + \ii 0^+)  &= - \left(\frac{\beta_1}{D_1}  + \mcl{O}(\omega^2) + \frac{\beta_2 |\omega|^{\alpha_2-1}}{D_2^{\alpha_2}} \sin\frac{\pi\alpha_2}{2} + \mcl{O}(\omega^{\alpha_2+1}) \right)
\end{align}
Let us focus on the most singular case, namely, $0.5858<\alpha_2-1<1$. Then $\frac{\Re G_f}{\Im G_f} = \mcl{O}(\omega^{\alpha_2-1})$. 
Then, 
\begin{align}
    \Im \Sigma_f(\omega+\ii 0^+) &= -\Im [G_f(\omega+\ii 0^+)]^{-1} = \frac{\Im G_f}{(\Re G_f)^2 + (\Im G_f)^2} 
    = \frac{1}{\Im G_f} + \mcl{O}(\omega^{2(\alpha_2-1)}) \\\nonumber
    &= - \frac{1}{\frac{\beta_1}{D_1} + \frac{\beta_2 |\omega|^{\alpha_2-1}}{D_2^{\alpha_2}} \sin\frac{\pi\alpha_2}{2} + \mcl{O}(\omega^2)} = - \frac{D_1}{\beta_1} \left( 1 - \frac{\beta_2}{\beta_1} \frac{D_1 |\omega|^{\alpha_2-1}}{D_2^{\alpha_2}} \right) + \mcl{O}(\omega^{2(\alpha_2-1)})\ .
\end{align}

In \cref{fig:ansatz}, we compare the ansatz self-energy with the numeric ones. 
Using the ansatz $\Sigma_f(\omega+\ii0^+)$, we also re-compute the spectral function in presence of the constant hybridization $\ii \Delta_0$, $A_f(\omega) = -\frac{1}{\pi} \Im G_f(\omega+\ii \eta) = -\frac{1}{\pi}\Im \frac{1}{\omega - \Sigma_f(\omega+\ii\eta) + \ii \Delta_0}$, and compare with the NRG result.

\subsection{Application to MATBG with heterostrain}  \label{app:AfSig-MATBG}

% \begin{figure}[tb]
%     \centering
%     \includegraphics[width=0.9\linewidth]{Af_EM.pdf}
%     \caption{Lattice spectral function in the LS phase, plotted component-wise, $A(k,\omega) \!=\! A_f(k,\omega) \!+\! A_c(k,\omega)$. 
%     (a) Parameters are the same as \cref{fig:Af}(d) of main text, but with a different $k$-path (inset). 
%     Both Hubbard peaks of $f$ lie at high energies $\Delta E = \pm 25$meV, thus the Fermi surface majorly comprises of $c$. 
%     (b) Lowering the Hubbard peaks to $\Delta E = \pm 8$meV. 
%     Although the Fermi volume is still given by $\frac{\nu-2}{4}$ (inset), hybridization between $c$ bands and the $f$ Hubbard bands has strongly suppressed the Fermi velocity. }
%     \label{fig:Af_EM}
% \end{figure}

With an ansatz of $\Sigma_f$, we can calculate the lattice spectral function via $A(\kk,\omega) = -\frac{1}{\pi} \Im \frac{1}{\omega - H(\kk) - \Sigma_f(\omega+\ii 0^+)}$, where $H(\kk)$ is the lattice Bloch Hamiltonian, within the framework of dynamical mean-field theory (DMFT)~\cite{Georges1996}. 
To be concrete, we study MATBG with heterostrain, and exploit the topological heavy fermion (THF) basis \cite{song_magic-angle_2022, herzog_2025_efficient, herzog_2025_kekule}. 
The heterostrain tensor (namely, the difference of the strain tensors in the two graphene layers) is given by 
\begin{align}
    \mcl{E} = \begin{pmatrix}
        \epsilon_+ + \epsilon_- & \epsilon_{xy} \\
        \epsilon_{xy} & \epsilon_+ - \epsilon_- \\
    \end{pmatrix} \ . 
\end{align}
$(\epsilon_{xy}, \epsilon_{-}) = -\frac{\nu_G+1}{2} \epsilon(\cos(2\varphi), \sin(2\varphi))$ describes the orientation of the strain field, which stretches in one direction and squeezes in another. 
$\epsilon_+ = \frac{\nu_G - 1}{2} \epsilon$ describes an isotropic expansion. 
$\nu_G = 0.16$ is the Poisson ratio, linking the two effects. 
We take $(\epsilon_{xy}, \epsilon_{-}) = (0,1)$, and $\epsilon = 0.2\%$ for concreteness, which are typical values in experiments. 

The heterostrain shears the moir\'e Brillouin zone, which is characterized by three vectors, 
\begin{align}
    \qq_j = \theta\frac{4\pi}{3 a_G} \left(\sin\frac{2\pi(j-1)}{3}, -\cos\frac{2\pi(j-1)}{3} \right) + \frac{4\pi}{3 a_G} \left(\cos\frac{2\pi(j-1)}{3}, \sin\frac{2\pi(j-1)}{3} \right) \cdot \mcl{E}   \qquad j = 1,2,3
\end{align}
where $\theta = 1.05^\circ$ denoting the twist angle, and $a_G = 0.246$nm denoting the graphene lattice constant. 

Due to the valley and spin degeneracies, we only write down the lattice Green's function in one flavor, the $\eta = +$ valley and $s = \uparrow$ spin.  
The kinetic Hamiltonian on the lattice consists of 
\begin{align}
    H(\kk) = H_0(\kk) + \delta H_{\epsilon}(\kk) + \delta H_{\rm mf}(\kk)\ .
\end{align}
$H_0(\kk)$ is the THF Hamiltonian with no heterostrain, following Ref. \cite{song_magic-angle_2022}, 
\begin{align}
    H_0(\kk) = \left(
    \begin{array}{c|cc|c}
         0 & h.c. & 0 & \cdots \\
    \hline
        \big(\gamma \sigma^0 + v_\star' (k_x \sigma^x + k_y \sigma^y) \big) e^{-\frac{\lambda^2|\kk|^2}{2}} & 0 & h.c. & \cdots \\
        0 & v_\star (k_x \sigma^0 - \ii k_y \sigma^z) & M \sigma_x &  \\
    \hline
        \cdots & \cdots &  & \cdots \\
    \end{array} \right)
\end{align}
where the columns are $(f_{\kk 1}, f_{\kk 2}, c_{\kk 1}, c_{\kk 2}, c_{\kk 3}, c_{\kk 4}, c_{(\kk+\mbf{G}) 1}, c_{(\kk+\mbf{G}) 2}, c_{(\kk+\mbf{G}) 3}, c_{(\kk+\mbf{G}) 4},  \cdots )^T$. 
Omitted blocks follow by replacing $c_{(\kk+\mbf{G})b}$ to $c_{(\kk+\mbf{G}')b}$ with $b=1,2,3,4$. 
$\mbf{G},\mbf{G}'$ run over moir\'e reciprocal lattice vectors, spanned by $\mbf{G}_1 = \qq_2 - \qq_1$ and $\mbf{G}_2 = \qq_3 - \qq_1$. 
$\gamma= -24.75$meV, $M = 3.697$meV, $v_\star = -430.3 $meV$ \cdot $nm, and $v_\star' = 162.2$meV$\cdot$nm \cite{song_magic-angle_2022}.

$\delta H_{\epsilon}(\kk)$ is the couplings induced by heterostrain, 
\begin{align}
    \delta H_\epsilon(\kk) = \left(
    \begin{array}{c|cc|c}
        M_f (\epsilon_{xy} \sigma^x + \epsilon_{-} \sigma^y ) & h.c. & h.c. & \cdots \\
    \hline
        \ii \gamma' \epsilon_+ \sigma^z & c (\epsilon_{xy} \sigma^x + \epsilon_{-} \sigma^y) & h.c. & \cdots \\
        c'' (\epsilon_{xy} \sigma^0 - \ii  \epsilon_{-} \sigma^z) & c' (\epsilon_{xy} \sigma^x - \epsilon_{-} \sigma^y) & M' \epsilon_+ \sigma^y &  \\
    \hline
        \cdots & \cdots &  & \cdots \\
    \end{array} \right)
\end{align}
where $c = -8750$meV, $c' = 2050$meV, $c'' = -3362$meV, $M_f = 4380$meV, $\gamma' = -3352$meV, and $M' = -4580$meV \cite{herzog_2025_efficient}. 

In Refs. \cite{herzog_2025_kekule, crippa_2025_dynamicalcorrelation}, it is found that a typical heterostrain at charge-neutrality point (CNP) of MATBG ($\nu=0$) fully polarizes the $f$ flavors along the ``Zeeman'' splitting of $M_f (\epsilon_{xy} \sigma^x + \epsilon_{-} \sigma^y)$. 
Doped to $\nu > 0$, only the flavors that were empty at CNP remain active, while the occupied flavors remain frozen. 
We make the same assumption here. 
From this frozen background, there can be a Fock exchange term at the mean-field level of the form $\Delta (\epsilon_{xy} \sigma^x + \epsilon_{-} \sigma^y)$, where $\Delta$ is of $\mcl{O}(U)$. 
Besides, the main effect of the other Coulomb interactions between $f$ and $c$ (terms $U_2, W, J, V$ in Ref. \cite{song_magic-angle_2022}) is to adjust the chemical potential for $f$ and $c$ electrons separately, which we define as $\epsilon_f, \epsilon_{c,1}, \epsilon_{c,3}$, following Ref. \cite{zhou_kondo_2024}. 
In accordance with the rest of this paper, the chemical potential $\epsilon_f$ should be adjusted so that the active flavors lie at the Fermi surface. 
In sum, 
\begin{align}
    \delta H_{\rm mf}(\kk) = \left(\begin{array}{c|cc|c}
        \epsilon_f  \sigma^0 + \Delta (\epsilon_{xy} \sigma^x + \epsilon_{-} \sigma^y) &  & & \cdots \\
    \hline
         & \epsilon_{c,1} &  & \cdots \\
         &  & \epsilon_{c,2}  &  \\
    \hline
        \cdots & \cdots &  & \cdots \\
    \end{array} \right)\ .
\end{align}
We choose $(\Delta+M_f) \frac{\nu_G+1}{2} \epsilon + \epsilon_f = 0$, in order to align the active $f$ flavors with the Fermi energy. 
For calculations in \cref{fig:Af} of main text and in \cref{fig:Af_EM}, we choose $\epsilon_{c,1} = -8$meV and $\epsilon_{c,2} = -12$meV. 
In \cref{fig:Af}(c,d) of main text and \cref{fig:Af_EM}(a), we choose $U \approx 50$meV, so that Hubbard peaks of active $f$ levels lie at $\Delta E = \pm 25$meV, and the frozen $f$ levels lie at $\epsilon_f - (\Delta+M_f) \frac{\nu_G+1}{2} \epsilon = -50$meV. 
In \cref{fig:Af_EM}(b), we choose $U \approx 16$meV, so that Hubbard peaks of active $f$ levels at $\Delta E = \pm 8$meV, and we choose the frozen $f$ levels to lie at $\epsilon_f - (\Delta+M_f) \frac{\nu_G+1}{2} \epsilon = -16$meV. 

For the anti-Hund's parameters, we use $J_S=3$meV and $J_D=0$ for the LS phase, and use $J_D=3$meV and $J_S=0$ for the AD phase. 
For the self-energy ansatz, for the LS phases in \cref{fig:Af}(d) of main text and \cref{fig:Af_EM}, we use $\beta_3=0.1$, $D_3=7 J_S$; for the AD phase in \cref{fig:Af}(d) of main text, we use $\beta_1=0.04$, $\beta_2=0.01$, $\beta_3 = 0.1$, and $D_1=D_2=4J_D$, $D_3=7J_D$. 

Finally, the lattice Green's function is given by 
\begin{align}
    A(\kk,\omega) &= -\frac{1}{\pi} \Im \pare{\Tr\brak{\frac{1}{(\omega + \ii0^+) - H(\kk) - \Sigma_f(\omega+\ii0^+)} } }\ . 
\end{align}

\clearpage
\section{Effective interactions}
\subsection{Fully anti-symmetrized form of the local interaction}
For later convenience, we fully anti-symmetrize $H_{\rm imp}$ as 
\begin{equation}
    H_{\rm imp} = \epsilon_f  \hat{N} + \frac14 \sum_{1234}  \Gamma^0_{1234} f_{1}^\dagger f_{2}^\dagger f_3 f_4
\end{equation}
where the Arabic numbers are composite indices, i.e., $1 \equiv (l_1,s_1)$, $2 \equiv (l_2,s_2)$, {\it etc.}. 
We can read the (not fully anti-symmetrized yet) vertex function from the definition of $H_{\rm imp}$ and \cref{eq:H_AH} in the main text as
% (more concretely, the $U$ term can be re-written as $\frac{U}{2} \hat{N}(\hat{N}-1) = \frac{U}{2} \sum_{12} f^\dagger_{1} f^\dagger_{2} f_{2} f_{1}$, while the $J_S$ and $J_D$ terms can be more conveniently read from \cref{eq:Himp_1})
\begin{equation}
2U \cdot \delta_{l_1 l_4} \delta_{l_2 l_3} \delta_{s_1 s_4}  \delta_{s_2 s_3} 
- 2J_D \cdot \delta_{l_1 l_2} \delta_{l_2 l_3} \delta_{l_3 l_4}  \delta_{s_1 s_4} \delta_{s_2 s_3} 
- J_S \cdot \delta_{l_1 \bar l_2} \delta_{l_3 \bar l_4}  \delta_{s_1 s_4} \delta_{s_2 s_3}  \ .
\end{equation}
The fully anti-symmetrized vertex is given by 
\begin{align} \label{eq:Gamma0}
\Gamma^0_{1234} =& 
\Gamma^0_U \cdot \pare{ \delta_{l_1 l_4} \delta_{l_2 l_3} \delta_{s_1 s_4}  \delta_{s_2 s_3}
    - \delta_{l_2 l_4} \delta_{l_1 l_3} \delta_{s_2 s_4}  \delta_{s_1 s_3}} 
+ \Gamma^0_D \cdot  \delta_{l_1 l_2} \delta_{l_2 l_3} \delta_{l_3 l_4} 
    \pare{  \delta_{s_1 s_3} \delta_{s_2 s_4}  -  \delta_{s_1 s_4} \delta_{s_2 s_3}  } \nonumber\\
& + \frac{\Gamma^0_S}2 \cdot \delta_{l_1 \bar l_2} \delta_{l_3 \bar l_4} 
    \pare{  \delta_{s_1 s_3} \delta_{s_2 s_4}   -  \delta_{s_1 s_4} \delta_{s_2 s_3}   }\ .
\end{align}
The bare parameters are given by $\Gamma^0_U = U$, $\Gamma^0_D = J_D$, $\Gamma^0_S = J_S$. 
These parameters may flow under renormalization, but the form of $\Gamma^0$ will remain unchanged, as it is already the most general form allowed by the symmetry group.

There are several special limits of $J_S, J_D$, where the symmetry group $\Ut_{c,s} \times D_{\infty}$ is further enlarged. 
Here, $\Ut_{c,s} = \big(\Uo_c \times \SUt_s \big) / \{\sigma^0 \spin^0, -\sigma^0 \spin^0 \}$ denotes the charge-spin actions. 

\paragraph{The $\rm U(4)$ limit}
When $J_D = J_S = 0$, the Anderson model is fully $\rm U(4)$ symmetric, with generators given by $\sigma^{\mu} \spin^{\nu}$ for $\mu, \nu = 0,x,y,z$. 
No multiplet splitting is allowed to occur. 
Accordingly, in the fully anti-symmetric vertex, only $\Gamma^0_U$ survives, while $\Gamma^0_{S} = \Gamma^0_{D}$ remains 0. 

\paragraph{The $\Ut_{c,s} \times \SUt_v$ limit}
When $J_D = J_S \neq 0$, the doublet and singlet become degenerate, and the valley symmetry group will be promoted to an $\SUt_v$ group, generated by $\sigma^{x,y,z} \spin^0$. 
In particular, the original $C_2 = \sigma^x \spin^0$ action can be understood as $e^{-\ii \frac{\pi}{2} \sigma^0 \spin^0} \cdot e^{\ii \frac{\pi}{2} \sigma^x \spin^0}$, a product of a $\Uo_c$ rotation and an $\SUt_v$ rotation. 

We now derive the vertex function in this limit.  
We denote $\Gamma^0_J \equiv \Gamma^0_S=\Gamma^0_D$ and split $\Gamma_U^0 = (\Gamma_U^0 - \frac12 \Gamma_J^0) + \frac12\Gamma_J^0$.
Then the vertex function can be written as 
{\small
\begin{align} 
& \Gamma^0_{1234} =
\pare{\Gamma^0_U - \frac12 \Gamma^0_J }\cdot \pare{ \delta_{l_1 l_4} \delta_{l_2 l_3} \delta_{s_1 s_4}  \delta_{s_2 s_3} 
   - \delta_{l_2 l_4} \delta_{l_1 l_3} \delta_{s_2 s_4}  \delta_{s_1 s_3}}  \nonumber\\
&+ \frac{\Gamma^0_J}2 \cdot \delta_{s_1 s_4} \delta_{s_2 s_3}
    \pare{  \delta_{l_1 l_4 } \delta_{l_2 l_3} - \delta_{l_1 \bar l_2} \delta_{l_3\bar l_4} - 2 \delta_{l_1 l_2} \delta_{l_1 l_3} \delta_{l_2l_4} }
- \frac{\Gamma^0_J}2 \cdot \delta_{s_1 s_3} \delta_{s_2 s_4}
    \pare{  \delta_{l_1 l_3 } \delta_{l_2 l_4} - \delta_{l_1 \bar l_2} \delta_{l_3\bar l_4} - 2 \delta_{l_1 l_2} \delta_{l_1 l_3} \delta_{l_2l_4} }\ .
\end{align}}%%
To simplify the first term in the second row, we rewrite 
$\delta_{l_1 l_4} \delta_{l_2 l_3} = \delta_{l_1 l_2} \delta_{l_1 l_4} \delta_{l_2 l_3} + \delta_{l_1\bar l_2} \delta_{l_1 l_4} \delta_{l_2 l_3} 
=  \delta_{l_1 l_2} \delta_{l_1 l_3} \delta_{l_2 l_4} + \delta_{l_1\bar l_2} \delta_{l_3 \bar l_4} \delta_{l_2 l_3}$.
Then, using $1-\delta_{l_2 l_3} = \delta_{l_2 \bar l_3}$, the Kronecker delta functions involving $l$-indices become
$-\delta_{l_1 l_2} \delta_{l_1 l_3} \delta_{l_2 l_4} - \delta_{l_1 \bar l_2} \delta_{l_3 \bar l_4} \delta_{l_2 \bar l_3}
=-\delta_{l_1 l_2} \delta_{l_1 l_3} \delta_{l_2 l_4} - \delta_{l_1 \bar l_2} \delta_{l_2 l_4} \delta_{l_1 l_3} = -\delta_{l_1 l_3} \delta_{l_2 l_4}$. 
Hence, the first term in the second row is proportional to
$- \delta_{l_1 l_3} \delta_{l_2 l_4} \delta_{s_1 s_4} \delta_{s_2 s_3}$. 
The second term in the second row is obtained by permuting the indices 3 and 4. 
Therefore, the vertex equals to  
{\small
\begin{align} 
\Gamma^0_{1234} = & 
\pare{\Gamma^0_U - \frac12 \Gamma^0_J }\cdot \pare{ \delta_{l_1 l_4} \delta_{l_2 l_3} \delta_{s_1 s_4}  \delta_{s_2 s_3} 
   - \delta_{l_2 l_4} \delta_{l_1 l_3} \delta_{s_2 s_4}  \delta_{s_1 s_3}}  
+
   \frac{\Gamma^0_J}2 \cdot \pare{ 
    \delta_{l_1 l_4} \delta_{l_2 l_3} \delta_{s_1 s_3} \delta_{s_2 s_4} 
- \delta_{l_1 l_3} \delta_{l_2 l_4} \delta_{s_1 s_4} \delta_{s_2 s_3}  }\ ,
\end{align}}%%
which has the form of the models in Refs.~\cite{yoshimori_perturbation_1976,nishikawa_renormalized_2010}. 
Comparing the above equation to Eq.~(4.1) of Ref.~\cite{yoshimori_perturbation_1976}, we identify our $\Gamma^0_U - \frac12 \Gamma^0_J$ and $\frac12 \Gamma_J^0$ as $\Gamma_C$ and $-\Gamma_e$ of Ref.~\cite{yoshimori_perturbation_1976}, respectively.

\paragraph{The $\rm U(2)_+ \times U(2)_- \rtimes Z_2$ limit}
When $J_S = 0$, the spins in the $l=\pm$ valleys are conserved independently. 
Since $\Uo_c$ and $\Uo_v$ are also preserved, the charges in the $l=\pm$ valleys are conserved independently as well. 
We dub the continuous group generated by $\frac{\sigma^0 + l\cdot \sigma^z}2 \spin^{0,x,y,z}$ as ${\rm U(2)}_l$ for $l=\pm$, which is the charge-spin rotation group per valley-$l$. 
Note the valley-flipping $\rm Z_{2}$ factor (generated by $\sigma_x$) is not promoted to a continuous symmetry in this case. 
We can use the valley quantum number $L$ and two spin quantum numbers $S_l$ for $l = \pm$ to label the irreps of scattering channels or two-electron states. 
The doublet states ($[L,S]=[2,0]$) are now denoted as $[L,S_+,S_-]=[2,0,0]$. 
The singlet ($[L,S]=[0,0]$) and triplet ($[L,S]=[0,1]$) states now become degenerate, as they can be related by an independent spin rotation in $l=+$ and/or $l=-$. They together form a four-fold degenerate irrep $[L,S_+,S_-]=[0,\frac12,\frac12]$. 
\subsection{Exact asymptotic vertex functions in the FL phase}
\label{sec:effective-interaction-1}

In this section, we briefly summarize the exact asymptotic relations of the renormalized interactions in the FL phase when $\TK \to 0^+$. For details about the renormalized perturbation theory, calculation of susceptibilities and Ward identities, we refer the reader to supplementary section B of Ref.~\cite{wang_molecular_2024} and other previous works \cite{yoshimori_perturbation_1976,hewson_renormalized_1993,nishikawa_renormalized_2010,nishikawa_convergence_2012}. 

In the FL phase, the local Green's function of the $f$-electron has a quasiparticle peak $\frac{z}{\ii \omega-\td{\epsilon}_f+\ii \td{\Delta}_0(\ii\omega)}$ contributed by quasiparticle $\td{f}\approx z^{-\frac{1}{2}}f$ and an incoherent part, where $z = [1-\partial_{\omega}\Sigma_f(\omega)|_{\omega=0}]^{-1}$ is the quasiparticle weight, and $\td{\Delta}_0=z\Delta_0$ is the renormalized hybridization function. The renormalized interactions on the quasiparticles are defined as the zero-frequency value of the full vertex function, scaled by quasiparticle weight  
\begin{align}
    \td{U},\td{J}_D,\td{J}_S=z^2\Gamma_{U,D,S}(0,0;0,0)\ .
\end{align}
We have defined the fully anti-symmetrized vertex $\Gamma$ by 
\begin{equation}
\includegraphics[width=0.7\linewidth]{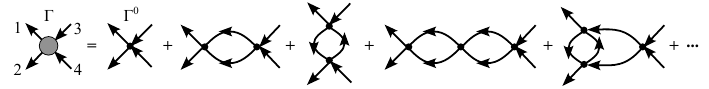}\ ,
\end{equation}
where the black dots are the bare interaction $\Gamma^0$ and the solid lines are the full Green function. $\Gamma$ is then separated into different channels by 
\begin{align} \label{eq:Gamma}
\Gamma_{1234} =& 
\Gamma_U \cdot \pare{ \delta_{l_1 l_4} \delta_{l_2 l_3} \delta_{s_1 s_4}  \delta_{s_2 s_3}
    - \delta_{l_2 l_4} \delta_{l_1 l_3} \delta_{s_2 s_4}  \delta_{s_1 s_3}} 
+ \Gamma_D \cdot  \delta_{l_1 l_2} \delta_{l_2 l_3} \delta_{l_3 l_4} 
    \pare{  \delta_{s_1 s_3} \delta_{s_2 s_4}  -  \delta_{s_1 s_4} \delta_{s_2 s_3}  } \nonumber\\
& + \frac{\Gamma_S}2 \cdot \delta_{l_1 \bar l_2} \delta_{l_3 \bar l_4} 
    \pare{  \delta_{s_1 s_3} \delta_{s_2 s_4}   -  \delta_{s_1 s_4} \delta_{s_2 s_3}   }\ ,
\end{align}
similar to the bare one (\cref{eq:Gamma0}).

In general, it is difficult to evaluate the vertex function non-perturbatively. However, in the $T_{\rm K}\to 0^+$ limit, when some degrees of freedom of impurity are frozen and symmetry is high enough, we can obtain exact relations about these renormalized parameters. 
Consider a symmetry generator $\hat{O} = \sum_{ls} f_{ls}^\dagger O_{ls} f_{ls}$ of the system, where the matrix $O$ is assumed diagonal for simplicity. 
The exact static susceptibility of $\hat{O}$ is related to the vertex function by the Ward identity (supplementary Eq.~(B88) of Ref.~\cite{wang_molecular_2024}) 
\begin{equation}
\chi^{O} = \frac{\sin \delta_f}{\pi \td \Delta_0} \cdot 
\pare{ 
    \sum_{ls} O_{ls}^2 
    - \frac{\sin \delta_f}{\pi \td \Delta_0} \sum_{l_1 s_1 l_2 s_2}
        z^2 \Gamma_{l_1 s_1,\ l_2 s_2 \ ;\ l_2 s_2,\ l_1 s_1}(0,0;0,0) 
    \cdot O_{l_1 s_1} \cdot O_{l_2 s_2} 
} \ .
\end{equation}
Here $\delta_f= \frac{n_f}{4}\pi$ and $n_f$ is the filling of the impurity. As we are only interested in the half-filling case ($n_f=2$), hereafter we set $\sin\delta_f=1$. 
We calculate the susceptibilities of charge, spin, and valley  $\chi_{c,s,v}$.
The corresponding $O$ matrices are $\sigma^0\varsigma^0$, $\sigma^0\varsigma^s$, $\sigma^z\varsigma^0$, respectively, and they are indeed generators of the symmetry group $[\Ut_{c,s}\times D_{\infty}]/\mathbb{Z}_2$.
We obtain
\begin{align}
    \chi_c =& 4\frac{1}{\pi\td{\Delta}_0}\left[1-\frac{1}{\pi\td{\Delta}_0}\left(3\td{U}-\td{J}_D-\frac{1}{2}\td{J}_S\right)\right] \label{eq:chic}\\
    \chi_s =& 4\frac{1}{\pi\td{\Delta}_0}\left[1-\frac{1}{\pi\td{\Delta}_0}\left(-\td{U}+\td{J}_D+\frac{1}{2}\td{J}_S\right)\right] \label{eq:chis}\\
    \chi_v =&  4\frac{1}{\pi\td{\Delta}_0}\left[1-\frac{1}{\pi\td{\Delta}_0}\left(-\td{U}-\td{J}_D+\frac{1}{2}\td{J}_S\right)\right]  \label{eq:chiv}\ . 
\end{align}

We define the Kondo temperature $\TK$ by $\TK=\td{\Delta}_0$ and the above equations yield \cref{eq:Ward} in the main text. This definition just differs by an order 1 constant from some other definition of Kondo temperature; for example, Refs.~\cite{costi_transport_1994,nishikawa_convergence_2012} defined $\TK=\frac{\pi}{4}\td{\Delta}_0$.

We then consider several limits. For all cases, we let $U\gg \TK$ and $n_f$ is fixed to an integer. We also calculate the effective interactions in $S,T,D$ channel $\td{E}_S=\td{U}-\td{J}_S,\td{E}_T=\td{U},\td{E}_D=\td{U}-\td{J}_D$ (\ie the two-particle energies, also listed in Table.~I in Ref.~\cite{wang_bosonization_2025}). They are related to the pairing susceptibility (supplementary section B.5 of Ref.~\cite{wang_molecular_2024}), and a negative two-particle energy indicates an attractive interaction in that channel.

\paragraph{$J_D=J_S=0,U\gg \TK$ limit}
This limit enjoys an SU(4) symmetry, which enforces $\td {J}_D = \td{J}_S=0$. 
The charge degree of freedom is frozen at the Kondo energy scale,  hence $\chi_c$ is not contributed by the quasiparticles, implying $\chi_c\ll \td{\Delta}_0^{-1}$ and $3\td{U} = \pi\td{\Delta}_0$. 
Therefore,
\begin{align}\label{eq:eff-int-U4}
(\td{U},\td{J}_D,\td{J}_S)=\pi\td{\Delta}_0 \pare{\frac{1}{3},0,0},\quad 
(\td{E}_S,\td{E}_T,\td{E}_D) = \pi\td{\Delta}_0 \pare{\frac{1}{3},\frac{1}{3},\frac{1}{3}}\ .
\end{align}

\paragraph{$J_S,J_S-J_D,U\gg \TK$ limit}
The atomic ground state is the singlet state. 
As the splittings between the singlet state and other atomic levels are much larger than $\TK$, in addition to the charge degree of freedom, the spin and valley degrees of freedom are also frozen at the Kondo energy scale, implying $\chi_{c,s,v} \ll \td{\Delta}_0^{-1}$.
We obtain 
\begin{align}\label{eq:eff-int-singlet-GS}
\left(\td{U},\td{J}_D,\td{J}_S\right)=\pi\td{\Delta}_0(1,0,4),\quad (\td{E}_S,\td{E}_T,\td{E}_D) = \pi\td{\Delta}_0 (-3,1,1)\ .
\end{align}
Notably, the singlet channel in the renormalized interaction becomes attractive (negative), favoring the singlet pairing. 

\paragraph{$J_D,U\gg \TK,J_S=0$ limit}
This limit enjoys the $\rm U(2)_+\times U(2)_-$ symmetry, which enforces $\td{J}_S=0$. 
The atomic ground states are the doublet states, which are spin-singlet, so both $\chi_{c,s}$ are frozen, implying
\begin{align}\label{eq:eff-int-doublet-GS}
\left(\td{U},\td{J}_D,\td{J}_S\right)=\pi\td{\Delta}_0(1,2,0),\quad (\td{E}_S,\td{E}_T,\td{E}_D) = \pi\td{\Delta}_0 (1,1,-1)\ .
\end{align}
Notably, the doublet channel in the renormalized interaction becomes attractive (negative), favoring the doublet pairing. 

\paragraph{$J_S,U\gg \TK,J_S=J_D>0$ limit}
This limit enjoys an additional valley $\SUt_v$ symmetry, which enforces $\td{J}_S = \td{J}_D$. 
The atomic ground states are the singlet and doublet states, which are valley-triplet and spin-singlet, so we have both $\chi_{c,s}$ frozen, implying 
\begin{align}\label{eq:eff-int-singlet-doublet-GS}
\left(\td{U},\td{J}_D,\td{J}_S\right)=\pi\td{\Delta}_0 \pare{ 1,\frac{4}3{},\frac{4}{3}},\quad 
(\td{E}_S,\td{E}_T,\td{E}_D) = \pi\td{\Delta}_0 \pare{-\frac{1}{3},1,-\frac{1}{3} }\ .
\end{align}
Notably, the doublet and singlet channels in the renormalized interaction become attractive (negative), favoring the valley-triplet pairing. 

\paragraph{$|J_S|,U\gg \TK,J_S=J_D<0$ limit}
The $\SUt_v$ symmetry enforces $\td{J}_S=\td{J}_D$. 
The atomic ground states are the triplet states, which are valley-singlet, so we have both $\chi_{c,v}$ frozen, implying
\begin{align}\label{eq:eff-int-triplet-GS} 
\left(\td{U},\td{J}_D,\td{J}_S\right)=\pi\td{\Delta}_0\pare{-\frac{1}{3},-\frac{4}{3},-\frac{4}{3}},\quad 
(\td{E}_S,\td{E}_T,\td{E}_D) = \pi\td{\Delta}_0 \pare{1,-\frac{1}{3},1} \ .
\end{align}
Notably, the spin-triplet channel in the renormalized interaction becomes attractive (negative), favoring the spin-triplet pairing. 

We have sketched those regions with attractive interaction in \cref{fig:intro}(b) of main text.
The relations above are verified by the NRG calculation as shown in \cref{fig:nrgphase}(b-g). 
\cref{eq:eff-int-U4,eq:eff-int-triplet-GS} were also obtained in Ref.~\cite{nishikawa_renormalized_2010}, and \cref{eq:eff-int-singlet-GS,eq:eff-int-doublet-GS} were also obtained in Ref.~\cite{nishikawa_convergence_2012}. 
Moreover, \cref{eq:eff-int-singlet-doublet-GS} is equivalent to \cref{eq:eff-int-triplet-GS} upon interchanging valley and spin.

\subsection{Effective interactions in the LS and AD phases}
\label{sec:effective-interaction-2}

In the LS and AD phases, the $f$-quasiparticle has zero quasiparticle weight. 
Nevertheless, we can still extract the effective interaction by examining energies of two-particle excitations perturbatively.

\paragraph{LS phase}
In the LS phase, $\zeta_x$ runs towards $0$ and $\varepsilon_D$ runs towards infinity under the RG.
When parameters are close to the fixed point, we can integrate out the high-energy $|\pm 2\rangle$ states to obtain an effective interaction induced by $\zeta_x$. 
The effective Hamiltonian near fixed point is given by Eqs.~(135) and (136) in Ref.~\cite{wang_bosonization_2025} with the renormalized $\zeta_x,\varepsilon_D$, and we find it convenient to reverse the gauge transformations and rewrite it in the fermion Hamiltonian (Eqs.~(131) in Ref.~\cite{wang_bosonization_2025}): 
\begin{align}\label{eq:HSD-fermion}
    H^{(S,D)} &= \sum_{ls} k : d^\dagger_{ls}(k) d_{ls}(k) : + J \cdot \Lambda_z^2 + (2\pi \lambda_z) \Lambda_z \sum_{ls} l \cdot \psi^\dagger_{ls}(0) \psi_{ls}(0) + (2\pi \zeta_{x}) \left( \Theta_+ \cdot \sum_{s} \psi^\dagger_{-s}(0) \psi_{+s}(0) + h.c.  \right)\, .
\end{align}
$\rho_z$ as well as $\lambda_z$ are unchanged under RG. $J=\varepsilon_D+\frac{4\rho_z^2}{x_c}$ is large and we treat $\zeta_x$ as perturbatively. 
Applying the second-order perturbation theory, the correction from the $\zeta_x$ term is 
\begin{align}
    H^{(S)}_{\rm int} &= (4\pi\zeta_x)^2\sum_{L_z=2,\bar2}\frac{|\langle L^{z} |\left( \Theta_+ \cdot \sum_{s} \psi^\dagger_{-s}(0) \psi_{+s}(0) + h.c.  \right) | 0\rangle|^2}{-J} \nonumber \\
    &= -\frac{32\pi^2\zeta_x^2}{J}\sum_{i=x,y}\psi^\dagger(0)\sigma^i\varsigma^0\psi(0) \cdot \psi^\dagger(0)\sigma^i\varsigma^0\psi(0) \ . 
\end{align}
We define 
$\hat{T}^i=\frac{1}{2}\psi^\dagger\sigma^i\varsigma^0\psi$, 
$\hat{S}^{i}=\frac{1}{2}\psi^\dagger\sigma^0\varsigma^i\psi$, 
$\hat{S}^{i}_l=\frac{1}{2}\psi^\dagger \frac{\sigma^0 + l\cdot\sigma^z}{2} \varsigma^i\psi$, 
$\hat{N}=\psi^\dagger\sigma^0\varsigma^0\psi$,
$\hat{\bf T}=(\hat{T}^x,\hat{T}^y,\hat{T}^z)$,
$\hat{\bf S}=(\hat{S}^x,\hat{S}^y,\hat{S}^z)$,
$\hat{\bf S}_l = (\hat{S}^x_l,\hat{S}^y_l,\hat{S}^z_l)$ in this section. Making use of the operator identities:
\begin{align}
    \hat{\bf T}^2+\hat{\mathbf{S}}^2+\frac{1}{2}(\hat{N}-2)^2 = 2, \qquad 
    \hat{\mathbf{S}}_{l}^2 = \frac{3}{4}\hat{N}_l (2-\hat{N}_l)\ ,
\end{align}
we have
\begin{align}
    \hat{T}^x\hat{T}^x+\hat{T}^y \hat{T}^y &= \hat{\bf T}^2 - \hat{T}^z \hat{T}^z = -\frac{1}{4}\hat{N}(\hat{N}-1)+\frac{1}{2}\hat{N}+ \frac{1}{2}\sum_{l}\hat{N}_{l\uparrow}\hat{N}_{l\downarrow} - 2\hat{\mathbf{S}}_{+}\cdot\hat{\mathbf{S}}_{-} \ .
\end{align}
The effective interaction can be rewritten as
\begin{align} \label{eq:Heff-bath-singlet}
    H^{(S)}_{\rm int} = & \epsilon'_f \hat{N} 
    +  \pare{ U' - \frac14 J'_S} \frac{\hat{N}(\hat{N}-1)}2
    +  J'_S \cdot \hat{\mathbf{S}}_+ \cdot \hat{\mathbf{S}}_-
    - \pare{J'_D - \frac14 J'_S} \sum_l \hat{N}_{l\uparrow}\hat{N}_{l\downarrow}
    \end{align}
where $(\epsilon'_f,U',J_D',J_S') =\frac{32\pi^2\zeta_x^2}{J}\left(-\frac{1}{2},1,1,2\right) $. 

The two-particle eigenstates of $H^{(S)}_{\rm  int}$ are the singlet, doublet and triplet states, same as the two-particle eigenstates of the impurity Hamiltonian except that they are formed by $\psi(0)$-particles. 
They have energies
\begin{align}
    (E_S,E_D,E_T) & = \frac{32\pi^2\zeta_x^2}{J} (-2,-1,0)= 2\epsilon_f' +\frac{32\pi^2\zeta_x^2}{J}(-1,0,1) \ .
\end{align}
The singlet state has the lowest energy, which is also less than twice the single-particle energy. 
Therefore, the interaction is attractive in this channel.

\paragraph{AD phase}
In this phase, the degenerate doublet $|2\rangle,|\bar 2\rangle$ always remains in Hilbert space, and we cannot integrate out the impurity. 
To see the effective interaction, we diagonalize the part of the pair-Kondo Hamiltonian that contains only the impurity and $\psi(0)$. 
The remaining part only adds kinetic energy to the electrons but does not affect the interaction. 
We start with the Hamiltonian $\ovl{H}_{\rm PK}$ near the fixed point (\ie $\lambda_x\sim 0$) and reverse the gauge transformation and bosonization procedure to the original fermion form $H_{\rm PK}$. 
Notice that $\psi$ in $H_{\rm PK}$ now is not the same as the original $\psi$. This is because $\rho_z$ flows under RG. The gauge transformation to absorb $\rho_z\Lambda_z\partial_x\phi_v(x)|_{x=0}$ before RG uses bare $\rho_z$, but the inverse gauge transformation to rewrite the Hamiltonian in original form after RG uses the renormalized $\rho_z$. 

The impurity and impurity-bath coupling part of the pair-Kondo Hamiltonian is
\begin{align} \label{eq:HPK-impurity}
2\pi \lambda_z \cdot \Lambda_z \sum_{ls } l \cdot 
   \psi_{l s}^\dagger(0) \psi_{ls} (0) 
 + (2\pi)^2 x_c \lambda_x  
    \pare{ \Lambda_+ \cdot \psi_{-\downarrow}^\dagger (0) \psi_{-\uparrow}^\dagger (0) \psi_{+\uparrow}(0) \psi_{+\downarrow}(0) 
    +  h.c.
    } \ .
\end{align}
where $\lambda_x$ takes the renormalized value and $\lambda_z$ is related to the renormalized $\rho_z$ by $\rho_z = \frac{\arctan(\pi\lambda_z)}{\pi}\in (0,\frac{1}{2})$.
The eigenstates and energies are shown in \cref{tab:LM-twosite}, where we  denote $\psi_{ls}\equiv\psi_{ls}(0)$ for simplicity. 
Notice that we have used the convention $\delta(0)=\frac{1}{\pi x_c}$ consistent with Ref.~\cite{wang_bosonization_2025}.  The lowest two-particle state also has an energy less than twice the single-particle energy, indicating an attractive interaction.

\begin{table}[ht]
    \centering
    \begin{tabular}{c|c|c|c}\hline
       $N_{ \psi}$ &  $\mrm{DEG}$ & wave-function &  $E\cdot x_c $ \\ \hline
       0  &  2 & $|2\rangle,|\bar2\rangle$ & 0 \\ \hline
       1 &  4& $\psi^\dagger_{-s}|2\rangle,\psi^\dagger_{+s}|\bar2\rangle,\,\forall s$ &  $-2\lambda_z$ \\
       & 4 & $\psi^\dagger_{+s}|2\rangle,\psi^\dagger_{-s}|\bar2\rangle,\,\forall s$ &  $2\lambda_z$\\\hline
       2 & 1 & $\frac{1}{\sqrt{2}}(\psi^\dagger_{-\uparrow}\psi^\dagger_{-\uparrow}|2\rangle- \psi^\dagger_{+\uparrow}\psi^\dagger_{+\uparrow}|\bar2\rangle) \quad $  &   $-4\lambda_z - 4\lambda_x$\\
    & 1& $\frac{1}{\sqrt{2}}(\psi^\dagger_{-\uparrow}\psi^\dagger_{-\uparrow}|2\rangle+ \psi^\dagger_{+\uparrow}\psi^\dagger_{+\uparrow}|\bar2\rangle) $ & $-4\lambda_z + 4\lambda_x$ \\
    & 8 & $\psi^\dagger_{+s}\psi^\dagger_{-s'}|L^z\rangle,\,\forall s,s',L^z$  & 0 \\
    & 2 &$\psi^\dagger_{+\uparrow}\psi^\dagger_{+\downarrow}|2\rangle,\psi^\dagger_{-\uparrow}\psi^\dagger_{-\downarrow}|\bar2\rangle$ & $4\lambda_z$ \\\hline
    3 &4& $\psi^\dagger_{+s}\psi^\dagger_{-\uparrow}\psi^\dagger_{-\downarrow}|2\rangle,\psi^\dagger_{-s}\psi^\dagger_{+\uparrow}\psi^\dagger_{+\downarrow}|\bar2\rangle,\,\forall s$ &$-2\lambda_z$ \\
    &4& $\psi^\dagger_{-s}\psi^\dagger_{+\uparrow}\psi^\dagger_{+\downarrow}|2\rangle,\psi^\dagger_{+s}\psi^\dagger_{-\uparrow}\psi^\dagger_{-\downarrow}|\bar2\rangle,\,\forall s$ &$2\lambda_z$ \\\hline
    4  &2& $\psi^{\dagger}_{+\uparrow}\psi^{\dagger}_{+\downarrow}\psi^{\dagger}_{-\uparrow}\psi^{\dagger}_{-\downarrow}|2\rangle,\psi^{\dagger}_{+\uparrow}\psi^{\dagger}_{+\downarrow}\psi^{\dagger}_{-\uparrow}\psi^{\dagger}_{-\downarrow}|\bar2\rangle$  & 0 \\\hline
    \end{tabular}
    \caption{The eigenstates of the pair-Kondo model without kinetic energy term. }
    \label{tab:LM-twosite}
\end{table}

To close this subsection, we add two remarks. 
First, the transverse couplings ($\zeta_x$ in the LS phase and $\lambda_x$ in the AD phase) flow to zero only at asymptotically low energies but remain finite at intermediate scales. 
While they eventually vanish in the single-impurity model, in the lattice model, they may trigger pairing instabilities before vanishing through the attractive interaction they mediate.
Second, the $\psi$ electron in the Kondo-type model also contains components of the original $f$ electron in the Anderson model, as discussed in \cref{app:AfSig-Af}. 
Consequently, if superconductivity could arise in the LS/AD phases due to the effective attraction acting on the $\psi$ electron derived in this section, the pairing would involve both $c$- and $f$-electrons.

\clearpage
\section{NRG verification}   \label{app:NRG}

% \begin{figure*}[h]
%     \centering
%     \includegraphics[width=\linewidth]{nrgphase.pdf}
%     \caption{
%     NRG results of the phase diagram, $\TK$, and effective parameters of the SVAIM. (a) Phase diagram on $(J_S,J_D)$ plane. The black solid lines sketch the phase boundary, and the color in the FL phase indicates $\TK$. The grey dashed line marks $J_S\!=\!J_D$. (b)-(d) The effective interactions in $S,D,T$ channels  $\td{E}_{S,D,T}$ compared to $\td{\Delta}_0$ as a function of $J_S,J_D$ in the FL phase. 
%     (e)-(g) $\td{J}_S/\pi \td{\Delta}_0, \td{J}_D/\pi \td{\Delta}_0,\td{U}/\pi \td{\Delta}_0$ as functions of: (e) $J_D/J^{(c)}_D$ when $J_S\!=\!0$, (f) $J_S/J^{(c)}_S$ when $J_D\!=\!0.05$, (g) $J_S$ when $J_S\!=\!J_D$. The dashed lines in (f),(g) mark the FL-LS critical point and $J_S\!=\!J_D\!=\!0$, respectively. The numeric labels \ding{192}-\ding{196} correspond to \cref{eq:eff-int-U4} to \cref{eq:eff-int-triplet-GS}, respectively, while the arrows in (g) show that these relations remain valid upon increasing $|J_S|$ along the line $J_S = J_D$. \label{fig:nrgphase}}
% \end{figure*}

% \begin{figure}[!t]
%     \centering
%     \includegraphics[width=0.7\linewidth]{critical.pdf}
%     \caption{$\TK$ near the phase transitions. 
%     (a) BKT type, obtained along $J_S\!=\!0$, with critical $J^{(c)}_D \!\approx\! 0.137$. 
%     (b) Second-order, obtained along $J_D\!=\!0.05$, with critical $J^{(c)}_S \!\approx\! 0.08026$. 
%     }
%     \label{fig:critical}
% \end{figure}

\subsection{Phase \label{sec:nrg-phase}}

Here, we describe how we distinguish the three phases by the fixed-point spectra and the spectral functions to obtain the phase diagram in the main text and End Matter.

\begin{figure}[!h]
    \centering
    \includegraphics[width=1.0\linewidth]{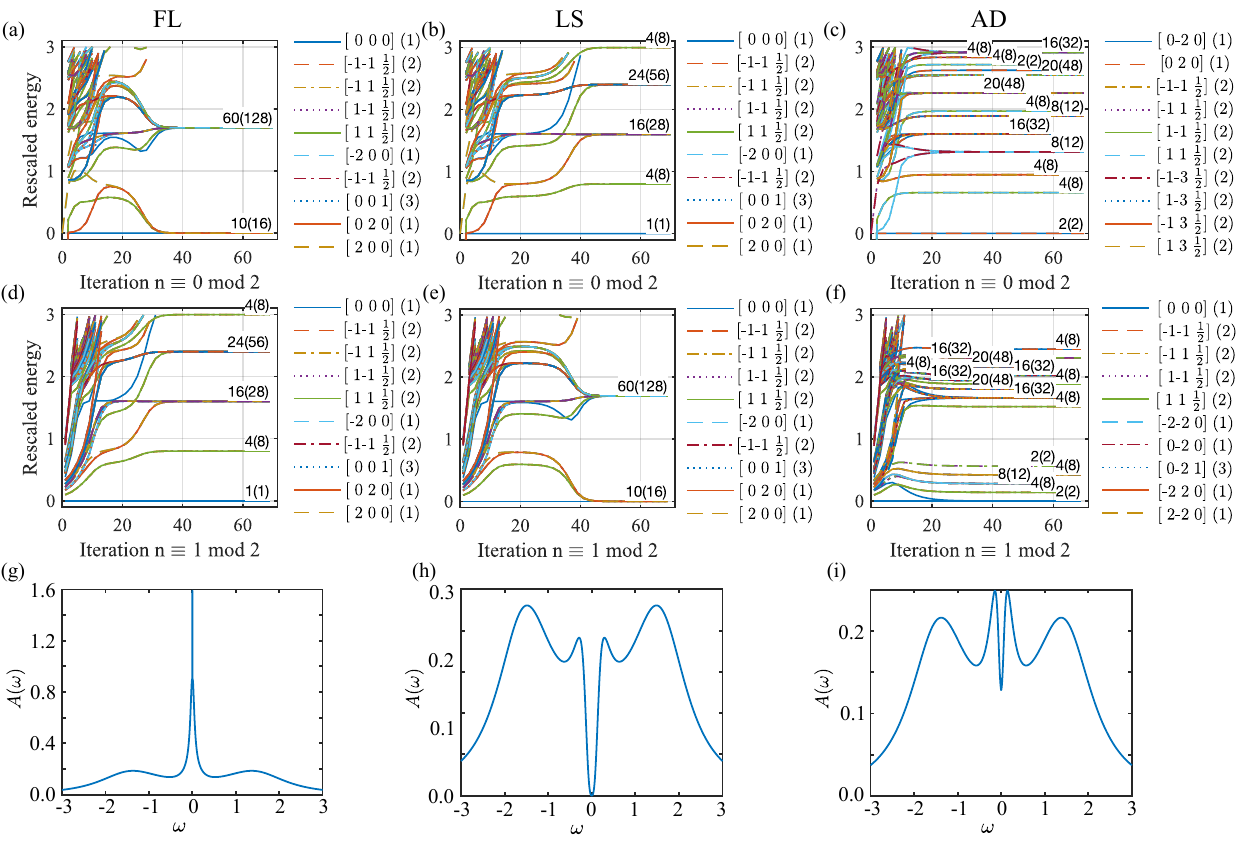}
    \caption{Typical NRG spectrum and spectral density in three phases. Top and middle panels: NRG spectrum for even and odd numbers of bath sites, respectively. Bottom panel: The spectral density. From left to right are figures for the FL phase, the LS phase, and the AD phase. The numbers $N_{mul}(N_{state})$ next to the line indicate that the line contains $N_{mul}$ multiplets and $N_{state}$ states in total. $[N,L_z,S](D)$ labels the quantum number of the multiplets where $N$ are the total particle numbers relative to half-filling, $L_z$ is the angular momentum, $S$ is the total spin and $D$ is the degeneracy. We do not utilize the $Z_2$ valley symmetry in numerical calculations, so $\pm L_z$ are labeled differently, but one can find that the $Z_2$-related states are degenerate. The parameters are chosen as follows: $J_S=0.054, J_D=0$ in the FL phase; $J_S=0.0548, J_D=0$ for the NRG spectrum, and $J_S=0.2, J_D=0$ for the spectral density in the LS phase; $J_S=0, J_D=0.2$ in the AD phase. For the local singlet phase, we use two different $J_S$. We choose a smaller $J_S$ when plotting NRG spectrum so that $J_S$ is closer to the critical point and we can illustrate the unstable fixed point in the first few iterations. We use a larger $J_S$ when plotting spectral density because though spectral density always has a full gap but the gap shrinks when approaching the critical point. } 
    \label{fig:spec}
\end{figure}

In the FL and LS phases, the NRG spectra converge to Fermi-liquid-like fixed points. The difference between these two phases is that the ground state is non-degenerate at odd iterations in the FL phase and at even iterations in the LS phase. This is because in the FL phase, impurity electrons can hybridize with the bath electrons, whereas in the LS phase, they form a singlet and are effectively decoupled. In the AD phase, the NRG spectra converge to a family of fixed-point spectra that can be interpreted as the paired Kondo model with $\lambda_x=0$ and different effective $\lambda_z$. We will construct effective Hamiltonians to capture these fixed-point spectra and perturbations around them in the next section. We notice that near the critical point of the FL-to-LS transition, there is an unstable fixed point (around $N=20$ in \cref{fig:spec} (a)(b)(d)(e) where $N$ is the number of the iteration steps), which is consistent with the results in Ref.~\cite{fabrizio_nontrivial_2003,leo_spectral_2004}. In contrast, in the FL-to-AD transition, no new type of fixed point occurs near the critical point, as expected from the RG analysis, which shows that the critical point of the pair-Kondo model lies at the end of the fixed line. 

We also plot the spectral density in \cref{fig:spec} (g)-(i), where a sharp resonance peak, a full gap, or a dip that does not touch zero appears at zero frequency in the FL, LS, and AD phases, respectively, thereby further characterizing the three phases. 

\subsection{Effective interactions\label{sec:nrg-eff-int}}

As the RG steps increase, the NRG spectrum converges to a fixed point. Once the spectrum is close to this point, an effective Hamiltonian can be constructed by adding perturbative terms to the fixed-point Hamiltonian, thereby reproducing the small deviations of the low-energy spectrum \cite{krishna-murthy_renormalization-group_1980-1,krishna-murthy_renormalization-group_1980}. This yields an estimate of the effective interactions.

\paragraph{FL phase}
In this phase, the fixed-point Hamiltonian is a free-fermion chain, and the leading-order correction terms are interactions at the first few sites \cite{krishna-murthy_renormalization-group_1980-1,hewson_renormalized_2004}. In the original NRG paper \cite{krishna-murthy_renormalization-group_1980-1}, Wilson et al. showed that the impurity site and the first bath site form a Kondo singlet and decouple from the rest of the bath. The effective Hamiltonian can then be the bath Hamiltonian without the first bath site, together with an interaction acting on the second bath site. Alternatively, Hewson {\it et al.} \cite{hewson_renormalized_2004} proposed another effective Hamiltonian consisting of the original bath Hamiltonian and renormalized impurity interactions and impurity-bath hybridizations. Hewson's method provides an estimate of the renormalized interaction $z^2\Gamma$ and the quasiparticle weight \cite{hewson_renormalized_1993,hewson_fermi_1993}, which we prefer here. 

Detailedly speaking, in NRG, the bath electrons are mapped to the Wilson chain, which is a free fermion chain with exponentially decaying energy scales \cite{krishna-murthy_renormalization-group_1980-1,krishna-murthy_renormalization-group_1980} 
\begin{align}
    H^{(N)} &= H_{\rm imp} + \sum_{ls}(t_0f^\dagger_{ls}\psi_{1ls}+h.c.)+ H^{(N)}_{\rm  bath}\ ,\\
    H^{(N)}_{\rm bath} &=\sum_{n=1}^{N} \epsilon_{n} \psi^\dagger_{nls} \psi_{nls} + \sum_{n=1}^{N-1} \left(  t_n\psi^\dagger_{nls} \psi_{n+1ls} + h.c. \right)\ ,
\end{align}
where $t_N \propto \Lambda^{-\frac{N-1}{2}}$ decides the energy scale at iteration $N$ and $\Lambda$ is the discretization parameter. The $H_{\rm imp}$ we used is just the $H_{\rm imp}$ in the main text. The transformation from $\Lambda^{\frac{N-1}{2}}H_N$ to $\Lambda^{\frac{N}{2}}H_{N+1}$ defines an RG transformation and the low-energy spectrum of $\Lambda^{\frac{N-1}{2}}H_N$ converges when $N\to \infty$, clarifying the fixed point.

Within $\TK$, the low-energy physics exhibits a Fermi-liquid feature and the effective degree of freedom is the quasiparticle $\td{f}\approx z^{-\frac{1}{2}}f$.  Correspondingly, the low-energy NRG spectrum can be fitted by a weakly interacting model with renormalized parameters:
\begin{align}
    H^{(N)} = \td{H}_{\rm imp} + \sum_{ls}(\td{t}_0\td{f}^\dagger_{ls}\psi_{1ls}+h.c.) + H^{(N)}_{\rm bath} 
\end{align}
where $\td{t}_0 = z^{1/2}t_0$ and $\td{H}_{\rm imp}$ takes the same form as $H_{\rm imp}$ except that the parameters $\epsilon,U,J_D,J_S$ are replaced by the effective values $\td{\epsilon},\td{U},\td{J}_D,\td{J}_S$. They are all symmetry-allowed terms in the impurity up to two-body interactions.

To obtain the renormalized parameters, we first adjust $\td{t}_0,\td{\epsilon}$ to fit the single-particle/single-hole excitation energy of $H^N$, by which $z$ is also obtained. $z$ can be alternatively obtained from the self-energy via its definition, but this approach has the drawback that the calculated self-energy depends on the chosen broadening parameters. In contrast, here $z$ is determined solely by the NRG spectrum.

To further obtain $\td{U},\td{J}_D,\td{J}_S$, we calculate their first-order corrections to the spectrum by perturbation theory, and then match the perturbed spectrum with the one obtained by NRG. Here, we take the case where the number of bath sites is odd (\ie the total number of sites is even) as an example. In this case, the bilinear part of the Hamiltonian can be diagonalized by 
\begin{align}
    H^{(N)}_{\rm fixed} = \Lambda^{-\frac{N-1}{2}} \sum_{j=1}^{\frac{N+1}{2}}\sum_{ls}\left(E^{(p)}_{j} c^{(p)\dagger}_{jls} c^{(p)}_{jls} - E^{(h)}_{j} c^{(h)\dagger}_{jls} c^{(h)}_{jls}\right) \label{eq:NRG-free-diag}
\end{align}
where 
\begin{itemize}
    \item $c^{(p)\dagger}_{jls}=\alpha^{(p)}_{0j}\cdot f^\dagger_{ls} + \sum_{i=1}^{N}\alpha^{(p)}_{ij}\psi^\dagger_{ils}$, $c^{(h)\dagger}_{jls}=\alpha^{(h)}_{0j}\cdot f^\dagger_{ls} + \sum_{i=1}^{N}\alpha^{(h)}_{ij}\psi^\dagger_{ils}$ are the  single-particle and single-hole eigenstates. Due to the exponentially decaying energy scale $|\alpha_{01}|\sim \Lambda^{-\frac{N}{4}}$.
    \item $E^{(p)}_1<E^{(p)}_2<\cdots E^{(p)}_{\frac{N+1}{2}}$ and $E^{(h)}_1<E^{(h)}_2<\cdots<E^{(h)}_{\frac{N+1}{2}}$ are the single-particle/hole eigenenergies of the rescaled Hamiltonian $\Lambda^{\frac{N-1}{2}}H^{(N)}_{\rm fixed}$, which converge to an order 1 value for fixed $j$ and $N\to\infty$. In particle-hole symmetric case $E^{(p)}_{j}=E^{(h)}_{j}$. 
\end{itemize}

We focus on the lowest single-particle state $c_{1ls}$ and the corresponding two-particle states. To first-order perturbation, $\td{U},\td{J}_D,\td{J}_S$ leave the single-particle levels unchanged. The two-particle states form $S,D,T$ multiplets, the same as those two particle, and the eigenenergies $E_{S,D,T}$ satisfy
\begin{equation}\label{eq:NRGspecturm-sdt} 
    \begin{aligned}
    E_S - 2E^{(p)}_{1}&= \Lambda^{\frac{N-1}{2}}\cdot |\alpha_{01}|^4 \cdot (\td{U}-\td{J}_S) \ , \\
    E_D - 2E^{(p)}_{1}&= \Lambda^{\frac{N-1}{2}}\cdot |\alpha_{01}|^4 \cdot (\td{U}-\td{J}_D) \ , \\
    E_T - 2E^{(p)}_{1}&= \Lambda^{\frac{N-1}{2}}\cdot |\alpha_{01}|^4 \cdot \td{U} \ .
\end{aligned}
\end{equation}

We can identify the single-particle states and $S,D,T$ multiplets by their corresponding quantum numbers in the NRG spectrum and then obtain the eigenenergies. 
Numerically, one will find that the left-hand sides of \cref{eq:NRGspecturm-sdt} scale as $\Lambda^{-\frac{N}{2}}$ for large enough $N$, and $\Lambda^{\frac{N-1}2} |\alpha_{01}|^4$ on the right-hand side scales as $\Lambda^{-\frac{N}2}$ as $ |\alpha_{01}| \sim \Lambda^{-\frac{N}4}$, leading to finite values of  $\td{U},\td{J}_D,\td{J}_S$ for large $N$. 

We plot the renormalized two-particle energies calculated from the obtained $\td{U},\td{J}_S,\td{J}_D$ in \cref{fig:nrgphase}(b-d), which are consistent with the Ward identities in \cref{sec:effective-interaction-1} and exhibit regions with local attractive interactions, supporting \cref{fig:intro}(b) of main text. To further illustrate the consistency with the Ward identities, several line cuts of the renormalized parameters are shown in \cref{fig:nrgphase}(e–g). 

In contrast to the works using Wilson's definition of effective interaction, which find them diverging near the critical point like Ref.~\cite{fabrizio_nontrivial_2003}, the effective interactions here tend to zero together with $\TK$, similar to Refs.~\cite{hewson_fermi_1993,nishikawa_renormalized_2010,nishikawa_convergence_2012}. The difference arises because Hewson's definition corresponds to the quasiparticle vertices $z^2\Gamma$, whereas Wilson’s does not involve adjusting the hopping and therefore produces the bare vertices $\Gamma$, without the $z^2$ factor.

\paragraph{LS phase}
Here, the fixed point is also a Fermi liquid, but the $f$-electron now has zero quasiparticle weight. The impurity itself forms a singlet and is decoupled from the bath; therefore, we treat the first bath as the impurity and repeat Hewson's procedure mentioned above again. In this case, we obtain the effective interaction and quasiparticle weight for the first bath site. 
Notably, its bare hybridization function, which will be used to fit quasiparticle $z$, is obtained by integrating out the other bath sites, different from the one for the impurity site that is obtained by integrating out all bath sites. 
As shown in \cref{fig:nrgphase}(f), with this definition, the effective parameters obtained still satisfy the prediction of Ward identities near the critical point, implying that the spin and valley moments of the first bath site are also quenched here. 
Ref.~\cite{nishikawa_convergence_2012} has verified the ratio $\td{J}_S/\td{U}$ (our definition of $\td{J}_S$ is twice theirs). We highlight that we further find the correct definition of $\td{\Delta}_0$ in the LS phase and confirm that $\td{J}_S/\td{\Delta}_0,\td{J}_D/\td{\Delta}_0$ and $\td{U}/\td{\Delta}_0$ are also consistent with the Ward identities.

\paragraph{AD phase}
For simplicity, we consider $J_S=0$ here. Consistent with the RG analysis of the pair-Kondo model, the fixed-point Hamiltonian in the AD phase is 
\begin{align}\label{eq:Hfix-lambz}
    H^{(N)}_{\rm fixed} =  \td{\lambda}_z\Lambda_z\cdot\sum_{ls}l\psi^\dagger_{1ls}\psi_{1ls} + H^{(N)}_{\rm bath} 
\end{align}
where the $f$-impurity is left with two states $|2\rangle,|\bar2\rangle$ with  $L_z=\pm 2$ ( also $\Lambda_z=\pm1)$.
The spectrum consists of two groups of free-fermion spectra. We consider even bath sites for simplicity here, which have a non-degenerate ground state before being coupled to the local moment. As \cref{eq:Hfix-lambz} commutes with $\Lambda_z$, it can be diagonalized within each $\Lambda_z=\pm 1$ sector, where it is reduced to a free-fermion Hamiltonian. We then obtain

\begin{align}
    H^{(N)}_{\rm fixed} = \Lambda^{-\frac{N-1}{2}} \sum_{j=1}^{N/2}\sum_{ls}\left[\left(E^{(p)}_{j} c^{(p)\dagger}_{jls} c^{(p)}_{jls} - E^{(h)}_{j} c^{(h)\dagger}_{jls} c^{(h)}_{jls}\right) + \Lambda_z\cdot l\cdot\left(\lambda^{(p)}_{z,j}c^{(p)\dagger}_{jls} c^{(p)}_{jls}+ \lambda^{(h)}_{z,j}c^{(h)\dagger}_{jls} c^{(h)}_{jls}\right)  \right]\label{eq:NRG-LM-diag} \, .
\end{align}
Now $c^{(p)\dagger}_{jls}=\sum_{i=1}^{N}\alpha^{(p)}_{ij}\psi^\dagger_{ils},c^{(h)\dagger}_{jls}=\sum_{i=1}^{N}\alpha^{(h)}_{ij}\psi^\dagger_{ils}$.
$E^{(p)}_{j}\pm 2\lambda^{(p)}_{z,j},(E^{(h)}_{j}\pm 2\lambda^{(h)}_{z,j})$ are single-particle/single-hole eigenenergies of the rescaled Hamiltonian $\Lambda^{\frac{N-1}{2}}H_N$, where $+(-)$ corresponds to those states with bath valley charge of same (opposite) sign with the impurity. Also, $E^{(p)}_{j}=E^{(h)}_{j},\lambda^{(p)}_{z,j}=\lambda^{(h)}_{z,j}$ in particle-hole symmetric case. Besides,  $E^{(p,h)}_{j}>2\lambda^{(p,h)}_{z,j}$, and the ground states are two-fold degenerate states $|{\rm GS}\rangle_{2}\equiv|2\rangle\otimes|{\rm GS}\rangle_0,|{\rm GS}\rangle_{\bar2}\equiv|\bar2\rangle\otimes|{\rm GS}\rangle_0$ where $|\mrm{GS}\rangle_0=\prod_{jls}c^{(h)\dagger}_{jls}|\mrm{vac}\rangle$ is the filled fermi sea of bath electrons.

Following Wilson \cite{krishna-murthy_renormalization-group_1980-1,krishna-murthy_renormalization-group_1980}, the leading correction terms are symmetric-allowed interaction terms on the first few bath sites, and we find that the following term can account for the deviation of the NRG spectrum from the fixed point:
\begin{align}
    \td{\lambda}_x 
    \pare{ \Lambda_+ \cdot \psi_{1,-\downarrow}^\dagger \psi_{1-\uparrow}^\dagger\psi_{1+\uparrow} \psi_{1+\downarrow}
    +  h.c.
    } \ .
\end{align}
We still focus on the single- and two-article states of $c^{(p)}_{1ls}$, and drop the band index $1$ hereafter for simplicity. The single particle states are also not affected by $\td{\lambda}_x$, while the two-particle states are

\begin{align}
\begin{array}{cc}
   c_{+s}^\dagger c_{-s'}^\dagger|\mrm{GS}\rangle_{L^z}, L^z=\pm2, s=\uparrow\downarrow,s'=\uparrow\downarrow, & E=2E^{(p)} \\
   c_{+\uparrow}^\dagger c_{+\downarrow}^\dagger|\mrm{GS}\rangle_2,|\bar2\rangle\otimes c_{-\uparrow}^\dagger c_{-\downarrow}^\dagger|\mrm{GS}\rangle_{\bar2}  & E=2E^{(p)}+4\lambda^{(p)}_z \\
   \frac{1}{\sqrt{2}}\left(c_{-\uparrow}^\dagger c_{-\downarrow}^\dagger|\mrm{GS}\rangle_2- c_{+\uparrow}^\dagger c_{+\downarrow}^\dagger|\mrm{GS}\rangle_{\bar2}\right) & E=2E^{(p)}-4\lambda^{(p)}_z- \Lambda^{(N-1)/2}\cdot|\alpha^{(p)}_{11}|^4\cdot \td{\lambda}_x \\
   \frac{1}{\sqrt{2}}\left(c_{-\uparrow}^\dagger c_{-\downarrow}^\dagger|\mrm{GS}\rangle_2+c_{+\uparrow}^\dagger c_{+\downarrow}^\dagger|\mrm{GS}\rangle_{\bar2}\right) &E=2E^{(p)}-4\lambda^{(p)}_z+ \Lambda^{(N-1)/2}\cdot|\alpha^{(p)}_{11}|^4\cdot \td{\lambda}_x\ .
\end{array}
\end{align}

\begin{figure}
    \centering
    \includegraphics[width=1.0\textwidth]{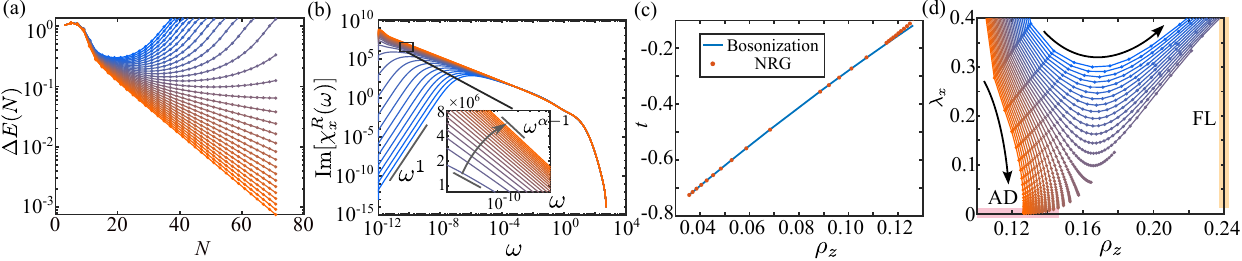}
    
    \caption{ NRG results near the BKT critical point. We fix $J_S=0$ and the colors in (a),(b),(d) represent the $J_D$, ranging from 0.12 to 0.15 every 0.01 from blue to orange. The BKT critical point is at $J^{(c)}_D\approx 0.137$. 
    (a) The binding energy $\Delta E(N)$ in NRG as a function of iteration step $N$, which corresponds to renormalized $\lambda_x(l)$ at energy scale $e^{-l}\sim \Lambda^{-N/2}$. Only even numbers of bath sites are shown.  
    (b) Log-log plot of the imaginary part of transverse valley susceptibility $\Im[\chi^R_{x}(\omega)]$ as a function of frequency $\omega$ near the BKT critical point. The inset shows a magnified view of the boxed region, revealing the non-universal power-law behavior $\omega^{\alpha-1}$ with $0<\alpha<1$ in the AD region. For those parameters in the Fermi liquid regime, $\Im[\chi^R_{x}(\omega)]\propto \omega$ below the Kondo temperature. 
    (c) The power $t$ extracted from the NRG spectrum versus $\rho_z$ extracted from the correlation function, compared with the bosonization result $t=-1+8\rho_z-8\rho_z^2$. 
    (d) RG flow of $\lambda_x,\rho_z$ extracted from NRG spectrum. Each line represents $\lambda_x,\rho_z$ obtained in fixed parameters and different NRG iterations.  The arrow indicates the direction of RG flow when the NRG iteration step increases. Here $\rho_z$ is calculated from $\lambda_z$.\label{fig:bkt}}
\end{figure}

$|\alpha^{(p)}_{11}|$ here also decays as $\Lambda^{-N/4}$ as $N$ increases. 
Unlike the cases in Fermi liquid fixed points, $\td{\lambda}_x$ does not converge to a fixed value when $N\to\infty$. To proceed, we define the binding energy $\Delta E(N)=2E_p - E_{2p}\sim \Lambda^{-N/2}\td{\lambda}_x $ where $E_p,E_{2p}$ are the rescaled lowest single- and two-particle energies at iteration $N$. Indeed, since both $\Delta E(N)$ in NRG and the running coupling $\lambda_x(l)$ in RG are rescaled under the RG flow, it is $\Delta E(N)$ instead of $\td\lambda_x$ that corresponds to $\lambda_x$ in the analytical RG calculation. We identify $\lambda_x(l)=\Delta E(N)$ with $e^{-l}\!\sim\!\Lambda^{-N/2}$. 
As shown in \cref{fig:bkt}(a), for those parameters in the AD phase, $\Delta E(N)$ shows a non-universal power law behavior $\Delta E(N)\sim \Lambda^{-tN/2},0<t<1$, which agrees with the analytical RG analysis that $\lambda_x \sim e^{-tl}$ near the critical point (Eq.~(61) in Ref.~\cite{wang_bosonization_2025}). 

To further elucidate this, we also numerically evaluate $\rho_z$ and compare the obtained $(t,\rho_z)$ to the bosonization prediction $t=-1 + 8\rho_z - 8 \rho_z^2$. (See Sec V.C in \cite{wang_bosonization_2025} for details)
We plot the correlation function $\Im[\chi^R_x(\omega)]$ as shown in \cref{fig:bkt}(b), which shows an ordinary linear in $\omega$ dependence in the FL regime and a non-universal power law $\sim |\omega|^{\alpha-1}\sgn(\omega)$ where $\alpha=16\rho_z^2$ as proposed by Eq.~(C9) in \cite{wang_bosonization_2025}. We extract $\rho_z$ from this, and $t$ from the scaling of $\Delta E(N)$. As shown in \cref{fig:bkt}(c), the relation between $t$ and $\rho_z$ agrees well with the bosonization prediction.

We also plot the extracted $\lambda_x,\rho_z$ at each iteration in \cref{fig:bkt}(d), which forms a renormalization flow as the iteration step increases. To obtain $\rho_z$ in each iteration step, we compute it using $\rho_z = \arctan (\pi \lambda_z^{(p)})/{\pi}$ (\ie the definition of $\rho_z$ in main text) where $\lambda_z^{(p)}$ at iteration $N$ is regarded as the renormalized $\lambda_z$ at this scale. When $N$ increases, this $\rho_z$ converges to a fixed value which is approximately equal to the one obtained by fitting the low-energy power-law behavior of $\Im[\chi^R_x(\omega)]$ mentioned above.
\cref{fig:bkt}(d) qualitatively reproduces the analytical RG flow from bosonization as shown in \cref{fig:intro}(c). Notably, the BKT critical point is close to the analytical value $\rho^c_z=\frac{1}{2}-\frac{1}{2\sqrt{2}}\approx 0.1464$.

One last concern is whether other interaction terms, such as $\td{U},\td{J}_S,\td{J}_D$, will also affect the low-energy spectrum. They also contain four fermi creation/annihilation operators, and the effects on the low-energy spectrum are scaled by $\Lambda^{(N-1)/2}\cdot|\alpha^{(p)}_{11}|^4$, similar to that of $\td{\lambda}_x$. However, numerically, we find that the splittings due to these interaction terms are negligible compared to those of $\td{\lambda}_x$.


\begin{thebibliography}{146}%
\makeatletter
\providecommand \@ifxundefined [1]{%
 \@ifx{#1\undefined}
}%
\providecommand \@ifnum [1]{%
 \ifnum #1\expandafter \@firstoftwo
 \else \expandafter \@secondoftwo
 \fi
}%
\providecommand \@ifx [1]{%
 \ifx #1\expandafter \@firstoftwo
 \else \expandafter \@secondoftwo
 \fi
}%
\providecommand \natexlab [1]{#1}%
\providecommand \enquote  [1]{``#1''}%
\providecommand \bibnamefont  [1]{#1}%
\providecommand \bibfnamefont [1]{#1}%
\providecommand \citenamefont [1]{#1}%
\providecommand \href@noop [0]{\@secondoftwo}%
\providecommand \href [0]{\begingroup \@sanitize@url \@href}%
\providecommand \@href[1]{\@@startlink{#1}\@@href}%
\providecommand \@@href[1]{\endgroup#1\@@endlink}%
\providecommand \@sanitize@url [0]{\catcode `\\12\catcode `\$12\catcode
  `\&12\catcode `\#12\catcode `\^12\catcode `\_12\catcode `\%12\relax}%
\providecommand \@@startlink[1]{}%
\providecommand \@@endlink[0]{}%
\providecommand \url  [0]{\begingroup\@sanitize@url \@url }%
\providecommand \@url [1]{\endgroup\@href {#1}{\urlprefix }}%
\providecommand \urlprefix  [0]{URL }%
\providecommand \Eprint [0]{\href }%
\providecommand \doibase [0]{http://dx.doi.org/}%
\providecommand \selectlanguage [0]{\@gobble}%
\providecommand \bibinfo  [0]{\@secondoftwo}%
\providecommand \bibfield  [0]{\@secondoftwo}%
\providecommand \translation [1]{[#1]}%
\providecommand \BibitemOpen [0]{}%
\providecommand \bibitemStop [0]{}%
\providecommand \bibitemNoStop [0]{.\EOS\space}%
\providecommand \EOS [0]{\spacefactor3000\relax}%
\providecommand \BibitemShut  [1]{\csname bibitem#1\endcsname}%
\let\auto@bib@innerbib\@empty
%</preamble>
\bibitem [{\citenamefont {Andrei}\ and\ \citenamefont
  {MacDonald}(2020)}]{andrei_graphene_2020}%
  \BibitemOpen
  \bibfield  {author} {\bibinfo {author} {\bibfnamefont {E.~Y.}\ \bibnamefont
  {Andrei}}\ and\ \bibinfo {author} {\bibfnamefont {A.~H.}\ \bibnamefont
  {MacDonald}},\ }\href {\doibase 10.1038/s41563-020-00840-0} {\bibfield
  {journal} {\bibinfo  {journal} {Nature Materials}\ }\textbf {\bibinfo
  {volume} {19}},\ \bibinfo {pages} {1265} (\bibinfo {year}
  {2020})}\BibitemShut {NoStop}%
\bibitem [{\citenamefont {Mak}\ and\ \citenamefont
  {Shan}(2022)}]{mak_semiconductor_2022}%
  \BibitemOpen
  \bibfield  {author} {\bibinfo {author} {\bibfnamefont {K.~F.}\ \bibnamefont
  {Mak}}\ and\ \bibinfo {author} {\bibfnamefont {J.}~\bibnamefont {Shan}},\
  }\href {\doibase 10.1038/s41565-022-01165-6} {\bibfield  {journal} {\bibinfo
  {journal} {Nature Nanotechnology}\ }\textbf {\bibinfo {volume} {17}},\
  \bibinfo {pages} {686} (\bibinfo {year} {2022})}\BibitemShut {NoStop}%
\bibitem [{\citenamefont {Nuckolls}\ and\ \citenamefont
  {Yazdani}(2024)}]{nuckolls_microscopic_2024}%
  \BibitemOpen
  \bibfield  {author} {\bibinfo {author} {\bibfnamefont {K.~P.}\ \bibnamefont
  {Nuckolls}}\ and\ \bibinfo {author} {\bibfnamefont {A.}~\bibnamefont
  {Yazdani}},\ }\href {\doibase 10.1038/s41578-024-00682-1} {\bibfield
  {journal} {\bibinfo  {journal} {Nature Reviews Materials}\ }\textbf {\bibinfo
  {volume} {9}},\ \bibinfo {pages} {460} (\bibinfo {year} {2024})}\BibitemShut
  {NoStop}%
\bibitem [{\citenamefont {Wu}\ \emph {et~al.}(2018{\natexlab{a}})\citenamefont
  {Wu}, \citenamefont {Lovorn}, \citenamefont {Tutuc},\ and\ \citenamefont
  {MacDonald}}]{Wu_Hubbard_2018}%
  \BibitemOpen
  \bibfield  {author} {\bibinfo {author} {\bibfnamefont {F.}~\bibnamefont
  {Wu}}, \bibinfo {author} {\bibfnamefont {T.}~\bibnamefont {Lovorn}}, \bibinfo
  {author} {\bibfnamefont {E.}~\bibnamefont {Tutuc}}, \ and\ \bibinfo {author}
  {\bibfnamefont {A.~H.}\ \bibnamefont {MacDonald}},\ }\href {\doibase
  10.1103/PhysRevLett.121.026402} {\bibfield  {journal} {\bibinfo  {journal}
  {Phys. Rev. Lett.}\ }\textbf {\bibinfo {volume} {121}},\ \bibinfo {pages}
  {026402} (\bibinfo {year} {2018}{\natexlab{a}})}\BibitemShut {NoStop}%
\bibitem [{\citenamefont {Koshino}\ \emph {et~al.}(2018)\citenamefont
  {Koshino}, \citenamefont {Yuan}, \citenamefont {Koretsune}, \citenamefont
  {Ochi}, \citenamefont {Kuroki},\ and\ \citenamefont
  {Fu}}]{Koshino_maximally_2018}%
  \BibitemOpen
  \bibfield  {author} {\bibinfo {author} {\bibfnamefont {M.}~\bibnamefont
  {Koshino}}, \bibinfo {author} {\bibfnamefont {N.~F.~Q.}\ \bibnamefont
  {Yuan}}, \bibinfo {author} {\bibfnamefont {T.}~\bibnamefont {Koretsune}},
  \bibinfo {author} {\bibfnamefont {M.}~\bibnamefont {Ochi}}, \bibinfo {author}
  {\bibfnamefont {K.}~\bibnamefont {Kuroki}}, \ and\ \bibinfo {author}
  {\bibfnamefont {L.}~\bibnamefont {Fu}},\ }\href {\doibase
  10.1103/PhysRevX.8.031087} {\bibfield  {journal} {\bibinfo  {journal} {Phys.
  Rev. X}\ }\textbf {\bibinfo {volume} {8}},\ \bibinfo {pages} {031087}
  (\bibinfo {year} {2018})}\BibitemShut {NoStop}%
\bibitem [{\citenamefont {Haule}\ \emph {et~al.}(2019)\citenamefont {Haule},
  \citenamefont {Andrei},\ and\ \citenamefont {Haule}}]{haule_2019_mott}%
  \BibitemOpen
  \bibfield  {author} {\bibinfo {author} {\bibfnamefont {M.}~\bibnamefont
  {Haule}}, \bibinfo {author} {\bibfnamefont {E.~Y.}\ \bibnamefont {Andrei}}, \
  and\ \bibinfo {author} {\bibfnamefont {K.}~\bibnamefont {Haule}},\ }\href
  {https://arxiv.org/abs/1901.09852} {\enquote {\bibinfo {title} {The
  mott-semiconducting state in the magic angle bilayer graphene},}\ } (\bibinfo
  {year} {2019}),\ \Eprint {http://arxiv.org/abs/1901.09852} {arXiv:1901.09852
  [cond-mat.str-el]} \BibitemShut {NoStop}%
\bibitem [{\citenamefont {Liu}\ \emph {et~al.}(2019)\citenamefont {Liu},
  \citenamefont {Liu},\ and\ \citenamefont {Dai}}]{Liu_2019_pseudoLL}%
  \BibitemOpen
  \bibfield  {author} {\bibinfo {author} {\bibfnamefont {J.}~\bibnamefont
  {Liu}}, \bibinfo {author} {\bibfnamefont {J.}~\bibnamefont {Liu}}, \ and\
  \bibinfo {author} {\bibfnamefont {X.}~\bibnamefont {Dai}},\ }\href {\doibase
  10.1103/PhysRevB.99.155415} {\bibfield  {journal} {\bibinfo  {journal} {Phys.
  Rev. B}\ }\textbf {\bibinfo {volume} {99}},\ \bibinfo {pages} {155415}
  (\bibinfo {year} {2019})}\BibitemShut {NoStop}%
\bibitem [{\citenamefont {Calderón}\ and\ \citenamefont
  {Bascones}(2020)}]{calderon_2020_interaction}%
  \BibitemOpen
  \bibfield  {author} {\bibinfo {author} {\bibfnamefont {M.~J.}\ \bibnamefont
  {Calderón}}\ and\ \bibinfo {author} {\bibfnamefont {E.}~\bibnamefont
  {Bascones}},\ }\href {\doibase 10.1103/PhysRevB.102.155149} {\bibfield
  {journal} {\bibinfo  {journal} {Physical Review B}\ }\textbf {\bibinfo
  {volume} {102}},\ \bibinfo {pages} {155149} (\bibinfo {year}
  {2020})}\BibitemShut {NoStop}%
\bibitem [{\citenamefont {Reddy}\ \emph {et~al.}(2023)\citenamefont {Reddy},
  \citenamefont {Devakul},\ and\ \citenamefont {Fu}}]{Reddy_2023_artificial}%
  \BibitemOpen
  \bibfield  {author} {\bibinfo {author} {\bibfnamefont {A.~P.}\ \bibnamefont
  {Reddy}}, \bibinfo {author} {\bibfnamefont {T.}~\bibnamefont {Devakul}}, \
  and\ \bibinfo {author} {\bibfnamefont {L.}~\bibnamefont {Fu}},\ }\href
  {\doibase 10.1103/PhysRevLett.131.246501} {\bibfield  {journal} {\bibinfo
  {journal} {Phys. Rev. Lett.}\ }\textbf {\bibinfo {volume} {131}},\ \bibinfo
  {pages} {246501} (\bibinfo {year} {2023})}\BibitemShut {NoStop}%
\bibitem [{\citenamefont {Liu}\ \emph {et~al.}(2025)\citenamefont {Liu},
  \citenamefont {Aryal}, \citenamefont {Calugaru}, \citenamefont {Fang},
  \citenamefont {Yang}, \citenamefont {Hu}, \citenamefont {Yan}, \citenamefont
  {Bernevig},\ and\ \citenamefont {Liu}}]{liu_ideal_2025}%
  \BibitemOpen
  \bibfield  {author} {\bibinfo {author} {\bibfnamefont {Y.}~\bibnamefont
  {Liu}}, \bibinfo {author} {\bibfnamefont {A.}~\bibnamefont {Aryal}}, \bibinfo
  {author} {\bibfnamefont {D.}~\bibnamefont {Calugaru}}, \bibinfo {author}
  {\bibfnamefont {Z.}~\bibnamefont {Fang}}, \bibinfo {author} {\bibfnamefont
  {K.}~\bibnamefont {Yang}}, \bibinfo {author} {\bibfnamefont {H.}~\bibnamefont
  {Hu}}, \bibinfo {author} {\bibfnamefont {Q.}~\bibnamefont {Yan}}, \bibinfo
  {author} {\bibfnamefont {B.~A.}\ \bibnamefont {Bernevig}}, \ and\ \bibinfo
  {author} {\bibfnamefont {C.-X.}\ \bibnamefont {Liu}},\ }\href {\doibase
  10.48550/arXiv.2507.06168} {\enquote {\bibinfo {title} {{Ideal''}
  {Topological} {Heavy} {Fermion} {Model} in {Two}-dimensional {Moiré}
  {Heterostructures} with {Type}-{II} {Band} {Alignment}},}\ } (\bibinfo {year}
  {2025}),\ \bibinfo {note} {arXiv:2507.06168 [cond-mat]}\BibitemShut {NoStop}%
\bibitem [{\citenamefont {Chen}\ \emph {et~al.}(2024)\citenamefont {Chen},
  \citenamefont {Nuckolls}, \citenamefont {Ding}, \citenamefont {Miao},
  \citenamefont {Wong}, \citenamefont {Oh}, \citenamefont {Lee}, \citenamefont
  {He}, \citenamefont {Peng}, \citenamefont {Pei}, \citenamefont {Li},
  \citenamefont {Hao}, \citenamefont {Yan}, \citenamefont {Xiao}, \citenamefont
  {Gao}, \citenamefont {Li}, \citenamefont {Zhang}, \citenamefont {Liu},
  \citenamefont {He}, \citenamefont {Watanabe}, \citenamefont {Taniguchi},
  \citenamefont {Jozwiak}, \citenamefont {Bostwick}, \citenamefont {Rotenberg},
  \citenamefont {Li}, \citenamefont {Han}, \citenamefont {Pan}, \citenamefont
  {Liu}, \citenamefont {Dai}, \citenamefont {Liu}, \citenamefont {Bernevig},
  \citenamefont {Wang}, \citenamefont {Yazdani},\ and\ \citenamefont
  {Chen}}]{chen_strong_2024}%
  \BibitemOpen
  \bibfield  {author} {\bibinfo {author} {\bibfnamefont {C.}~\bibnamefont
  {Chen}}, \bibinfo {author} {\bibfnamefont {K.~P.}\ \bibnamefont {Nuckolls}},
  \bibinfo {author} {\bibfnamefont {S.}~\bibnamefont {Ding}}, \bibinfo {author}
  {\bibfnamefont {W.}~\bibnamefont {Miao}}, \bibinfo {author} {\bibfnamefont
  {D.}~\bibnamefont {Wong}}, \bibinfo {author} {\bibfnamefont {M.}~\bibnamefont
  {Oh}}, \bibinfo {author} {\bibfnamefont {R.~L.}\ \bibnamefont {Lee}},
  \bibinfo {author} {\bibfnamefont {S.}~\bibnamefont {He}}, \bibinfo {author}
  {\bibfnamefont {C.}~\bibnamefont {Peng}}, \bibinfo {author} {\bibfnamefont
  {D.}~\bibnamefont {Pei}}, \bibinfo {author} {\bibfnamefont {Y.}~\bibnamefont
  {Li}}, \bibinfo {author} {\bibfnamefont {C.}~\bibnamefont {Hao}}, \bibinfo
  {author} {\bibfnamefont {H.}~\bibnamefont {Yan}}, \bibinfo {author}
  {\bibfnamefont {H.}~\bibnamefont {Xiao}}, \bibinfo {author} {\bibfnamefont
  {H.}~\bibnamefont {Gao}}, \bibinfo {author} {\bibfnamefont {Q.}~\bibnamefont
  {Li}}, \bibinfo {author} {\bibfnamefont {S.}~\bibnamefont {Zhang}}, \bibinfo
  {author} {\bibfnamefont {J.}~\bibnamefont {Liu}}, \bibinfo {author}
  {\bibfnamefont {L.}~\bibnamefont {He}}, \bibinfo {author} {\bibfnamefont
  {K.}~\bibnamefont {Watanabe}}, \bibinfo {author} {\bibfnamefont
  {T.}~\bibnamefont {Taniguchi}}, \bibinfo {author} {\bibfnamefont
  {C.}~\bibnamefont {Jozwiak}}, \bibinfo {author} {\bibfnamefont
  {A.}~\bibnamefont {Bostwick}}, \bibinfo {author} {\bibfnamefont
  {E.}~\bibnamefont {Rotenberg}}, \bibinfo {author} {\bibfnamefont
  {C.}~\bibnamefont {Li}}, \bibinfo {author} {\bibfnamefont {X.}~\bibnamefont
  {Han}}, \bibinfo {author} {\bibfnamefont {D.}~\bibnamefont {Pan}}, \bibinfo
  {author} {\bibfnamefont {Z.}~\bibnamefont {Liu}}, \bibinfo {author}
  {\bibfnamefont {X.}~\bibnamefont {Dai}}, \bibinfo {author} {\bibfnamefont
  {C.}~\bibnamefont {Liu}}, \bibinfo {author} {\bibfnamefont {B.~A.}\
  \bibnamefont {Bernevig}}, \bibinfo {author} {\bibfnamefont {Y.}~\bibnamefont
  {Wang}}, \bibinfo {author} {\bibfnamefont {A.}~\bibnamefont {Yazdani}}, \
  and\ \bibinfo {author} {\bibfnamefont {Y.}~\bibnamefont {Chen}},\ }\href
  {\doibase 10.1038/s41586-024-08227-w} {\bibfield  {journal} {\bibinfo
  {journal} {Nature}\ }\textbf {\bibinfo {volume} {636}},\ \bibinfo {pages}
  {342} (\bibinfo {year} {2024})}\BibitemShut {NoStop}%
\bibitem [{\citenamefont {Birkbeck}\ \emph {et~al.}(2025)\citenamefont
  {Birkbeck}, \citenamefont {Xiao}, \citenamefont {Inbar}, \citenamefont
  {Taniguchi}, \citenamefont {Watanabe}, \citenamefont {Berg}, \citenamefont
  {Glazman}, \citenamefont {Guinea}, \citenamefont {von Oppen},\ and\
  \citenamefont {Ilani}}]{birkbeck_quantum_2025}%
  \BibitemOpen
  \bibfield  {author} {\bibinfo {author} {\bibfnamefont {J.}~\bibnamefont
  {Birkbeck}}, \bibinfo {author} {\bibfnamefont {J.}~\bibnamefont {Xiao}},
  \bibinfo {author} {\bibfnamefont {A.}~\bibnamefont {Inbar}}, \bibinfo
  {author} {\bibfnamefont {T.}~\bibnamefont {Taniguchi}}, \bibinfo {author}
  {\bibfnamefont {K.}~\bibnamefont {Watanabe}}, \bibinfo {author}
  {\bibfnamefont {E.}~\bibnamefont {Berg}}, \bibinfo {author} {\bibfnamefont
  {L.}~\bibnamefont {Glazman}}, \bibinfo {author} {\bibfnamefont
  {F.}~\bibnamefont {Guinea}}, \bibinfo {author} {\bibfnamefont
  {F.}~\bibnamefont {von Oppen}}, \ and\ \bibinfo {author} {\bibfnamefont
  {S.}~\bibnamefont {Ilani}},\ }\href {\doibase 10.1038/s41586-025-08881-8}
  {\bibfield  {journal} {\bibinfo  {journal} {Nature}\ }\textbf {\bibinfo
  {volume} {641}},\ \bibinfo {pages} {345} (\bibinfo {year} {2025})},\ \bibinfo
  {note} {publisher: Nature Publishing Group}\BibitemShut {NoStop}%
\bibitem [{\citenamefont {Bistritzer}\ and\ \citenamefont
  {MacDonald}(2011)}]{BM_2011}%
  \BibitemOpen
  \bibfield  {author} {\bibinfo {author} {\bibfnamefont {R.}~\bibnamefont
  {Bistritzer}}\ and\ \bibinfo {author} {\bibfnamefont {A.~H.}\ \bibnamefont
  {MacDonald}},\ }\href {\doibase 10.1073/pnas.1108174108} {\bibfield
  {journal} {\bibinfo  {journal} {Proceedings of the National Academy of
  Sciences}\ }\textbf {\bibinfo {volume} {108}},\ \bibinfo {pages} {12233}
  (\bibinfo {year} {2011})}\BibitemShut {NoStop}%
\bibitem [{\citenamefont {Cao}\ \emph {et~al.}(2018)\citenamefont {Cao},
  \citenamefont {Fatemi}, \citenamefont {Demir}, \citenamefont {Fang},
  \citenamefont {Tomarken}, \citenamefont {Luo}, \citenamefont
  {Sanchez-Yamagishi}, \citenamefont {Watanabe}, \citenamefont {Taniguchi},
  \citenamefont {Kaxiras}, \citenamefont {Ashoori},\ and\ \citenamefont
  {Jarillo-Herrero}}]{Cao_2018_CI}%
  \BibitemOpen
  \bibfield  {author} {\bibinfo {author} {\bibfnamefont {Y.}~\bibnamefont
  {Cao}}, \bibinfo {author} {\bibfnamefont {V.}~\bibnamefont {Fatemi}},
  \bibinfo {author} {\bibfnamefont {A.}~\bibnamefont {Demir}}, \bibinfo
  {author} {\bibfnamefont {S.}~\bibnamefont {Fang}}, \bibinfo {author}
  {\bibfnamefont {S.~L.}\ \bibnamefont {Tomarken}}, \bibinfo {author}
  {\bibfnamefont {J.~Y.}\ \bibnamefont {Luo}}, \bibinfo {author} {\bibfnamefont
  {J.~D.}\ \bibnamefont {Sanchez-Yamagishi}}, \bibinfo {author} {\bibfnamefont
  {K.}~\bibnamefont {Watanabe}}, \bibinfo {author} {\bibfnamefont
  {T.}~\bibnamefont {Taniguchi}}, \bibinfo {author} {\bibfnamefont
  {E.}~\bibnamefont {Kaxiras}}, \bibinfo {author} {\bibfnamefont {R.~C.}\
  \bibnamefont {Ashoori}}, \ and\ \bibinfo {author} {\bibfnamefont
  {P.}~\bibnamefont {Jarillo-Herrero}},\ }\href {\doibase 10.1038/nature26154}
  {\bibfield  {journal} {\bibinfo  {journal} {Nature}\ }\textbf {\bibinfo
  {volume} {556}},\ \bibinfo {pages} {80} (\bibinfo {year} {2018})}\BibitemShut
  {NoStop}%
\bibitem [{\citenamefont {Lu}\ \emph {et~al.}(2019)\citenamefont {Lu},
  \citenamefont {Stepanov}, \citenamefont {Yang}, \citenamefont {Xie},
  \citenamefont {Aamir}, \citenamefont {Das}, \citenamefont {Urgell},
  \citenamefont {Watanabe}, \citenamefont {Taniguchi}, \citenamefont {Zhang},
  \citenamefont {Bachtold}, \citenamefont {MacDonald},\ and\ \citenamefont
  {Efetov}}]{Lu_2019_superconductors}%
  \BibitemOpen
  \bibfield  {author} {\bibinfo {author} {\bibfnamefont {X.}~\bibnamefont
  {Lu}}, \bibinfo {author} {\bibfnamefont {P.}~\bibnamefont {Stepanov}},
  \bibinfo {author} {\bibfnamefont {W.}~\bibnamefont {Yang}}, \bibinfo {author}
  {\bibfnamefont {M.}~\bibnamefont {Xie}}, \bibinfo {author} {\bibfnamefont
  {M.~A.}\ \bibnamefont {Aamir}}, \bibinfo {author} {\bibfnamefont
  {I.}~\bibnamefont {Das}}, \bibinfo {author} {\bibfnamefont {C.}~\bibnamefont
  {Urgell}}, \bibinfo {author} {\bibfnamefont {K.}~\bibnamefont {Watanabe}},
  \bibinfo {author} {\bibfnamefont {T.}~\bibnamefont {Taniguchi}}, \bibinfo
  {author} {\bibfnamefont {G.}~\bibnamefont {Zhang}}, \bibinfo {author}
  {\bibfnamefont {A.}~\bibnamefont {Bachtold}}, \bibinfo {author}
  {\bibfnamefont {A.~H.}\ \bibnamefont {MacDonald}}, \ and\ \bibinfo {author}
  {\bibfnamefont {D.~K.}\ \bibnamefont {Efetov}},\ }\href {\doibase
  10.1038/s41586-019-1695-0} {\bibfield  {journal} {\bibinfo  {journal}
  {Nature}\ }\textbf {\bibinfo {volume} {574}},\ \bibinfo {pages} {653}
  (\bibinfo {year} {2019})}\BibitemShut {NoStop}%
\bibitem [{\citenamefont {Choi}\ \emph {et~al.}(2019)\citenamefont {Choi},
  \citenamefont {Kemmer}, \citenamefont {Peng}, \citenamefont {Thomson},
  \citenamefont {Arora}, \citenamefont {Polski}, \citenamefont {Zhang},
  \citenamefont {Ren}, \citenamefont {Alicea}, \citenamefont {Refael},
  \citenamefont {von Oppen}, \citenamefont {Watanabe}, \citenamefont
  {Taniguchi},\ and\ \citenamefont {Nadj-Perge}}]{Choi_2019_electronic}%
  \BibitemOpen
  \bibfield  {author} {\bibinfo {author} {\bibfnamefont {Y.}~\bibnamefont
  {Choi}}, \bibinfo {author} {\bibfnamefont {J.}~\bibnamefont {Kemmer}},
  \bibinfo {author} {\bibfnamefont {Y.}~\bibnamefont {Peng}}, \bibinfo {author}
  {\bibfnamefont {A.}~\bibnamefont {Thomson}}, \bibinfo {author} {\bibfnamefont
  {H.}~\bibnamefont {Arora}}, \bibinfo {author} {\bibfnamefont
  {R.}~\bibnamefont {Polski}}, \bibinfo {author} {\bibfnamefont
  {Y.}~\bibnamefont {Zhang}}, \bibinfo {author} {\bibfnamefont
  {H.}~\bibnamefont {Ren}}, \bibinfo {author} {\bibfnamefont {J.}~\bibnamefont
  {Alicea}}, \bibinfo {author} {\bibfnamefont {G.}~\bibnamefont {Refael}},
  \bibinfo {author} {\bibfnamefont {F.}~\bibnamefont {von Oppen}}, \bibinfo
  {author} {\bibfnamefont {K.}~\bibnamefont {Watanabe}}, \bibinfo {author}
  {\bibfnamefont {T.}~\bibnamefont {Taniguchi}}, \ and\ \bibinfo {author}
  {\bibfnamefont {S.}~\bibnamefont {Nadj-Perge}},\ }\href {\doibase
  10.1038/s41567-019-0606-5} {\bibfield  {journal} {\bibinfo  {journal} {Nat.
  Phys.}\ }\textbf {\bibinfo {volume} {15}},\ \bibinfo {pages} {1174} (\bibinfo
  {year} {2019})}\BibitemShut {NoStop}%
\bibitem [{\citenamefont {Kerelsky}\ \emph {et~al.}(2019)\citenamefont
  {Kerelsky}, \citenamefont {McGilly}, \citenamefont {Kennes}, \citenamefont
  {Xian}, \citenamefont {Yankowitz}, \citenamefont {Chen}, \citenamefont
  {Watanabe}, \citenamefont {Taniguchi}, \citenamefont {Hone}, \citenamefont
  {Dean}, \citenamefont {Rubio},\ and\ \citenamefont
  {Pasupathy}}]{Kerelsky_2019_maximized}%
  \BibitemOpen
  \bibfield  {author} {\bibinfo {author} {\bibfnamefont {A.}~\bibnamefont
  {Kerelsky}}, \bibinfo {author} {\bibfnamefont {L.~J.}\ \bibnamefont
  {McGilly}}, \bibinfo {author} {\bibfnamefont {D.~M.}\ \bibnamefont {Kennes}},
  \bibinfo {author} {\bibfnamefont {L.}~\bibnamefont {Xian}}, \bibinfo {author}
  {\bibfnamefont {M.}~\bibnamefont {Yankowitz}}, \bibinfo {author}
  {\bibfnamefont {S.}~\bibnamefont {Chen}}, \bibinfo {author} {\bibfnamefont
  {K.}~\bibnamefont {Watanabe}}, \bibinfo {author} {\bibfnamefont
  {T.}~\bibnamefont {Taniguchi}}, \bibinfo {author} {\bibfnamefont
  {J.}~\bibnamefont {Hone}}, \bibinfo {author} {\bibfnamefont {C.}~\bibnamefont
  {Dean}}, \bibinfo {author} {\bibfnamefont {A.}~\bibnamefont {Rubio}}, \ and\
  \bibinfo {author} {\bibfnamefont {A.~N.}\ \bibnamefont {Pasupathy}},\ }\href
  {\doibase 10.1038/s41586-019-1431-9} {\bibfield  {journal} {\bibinfo
  {journal} {Nature}\ }\textbf {\bibinfo {volume} {572}},\ \bibinfo {pages}
  {95} (\bibinfo {year} {2019})}\BibitemShut {NoStop}%
\bibitem [{\citenamefont {Jiang}\ \emph {et~al.}(2019)\citenamefont {Jiang},
  \citenamefont {Lai}, \citenamefont {Watanabe}, \citenamefont {Taniguchi},
  \citenamefont {Haule}, \citenamefont {Mao},\ and\ \citenamefont
  {Andrei}}]{Jiang_2019_charge}%
  \BibitemOpen
  \bibfield  {author} {\bibinfo {author} {\bibfnamefont {Y.}~\bibnamefont
  {Jiang}}, \bibinfo {author} {\bibfnamefont {X.}~\bibnamefont {Lai}}, \bibinfo
  {author} {\bibfnamefont {K.}~\bibnamefont {Watanabe}}, \bibinfo {author}
  {\bibfnamefont {T.}~\bibnamefont {Taniguchi}}, \bibinfo {author}
  {\bibfnamefont {K.}~\bibnamefont {Haule}}, \bibinfo {author} {\bibfnamefont
  {J.}~\bibnamefont {Mao}}, \ and\ \bibinfo {author} {\bibfnamefont {E.~Y.}\
  \bibnamefont {Andrei}},\ }\href {\doibase 10.1038/s41586-019-1460-4}
  {\bibfield  {journal} {\bibinfo  {journal} {Nature}\ }\textbf {\bibinfo
  {volume} {573}},\ \bibinfo {pages} {91} (\bibinfo {year} {2019})}\BibitemShut
  {NoStop}%
\bibitem [{\citenamefont {Xie}\ \emph {et~al.}(2019)\citenamefont {Xie},
  \citenamefont {Lian}, \citenamefont {Jäck}, \citenamefont {Liu},
  \citenamefont {Chiu}, \citenamefont {Watanabe}, \citenamefont {Taniguchi},
  \citenamefont {Bernevig},\ and\ \citenamefont
  {Yazdani}}]{Xie_2019_spectroscopic}%
  \BibitemOpen
  \bibfield  {author} {\bibinfo {author} {\bibfnamefont {Y.}~\bibnamefont
  {Xie}}, \bibinfo {author} {\bibfnamefont {B.}~\bibnamefont {Lian}}, \bibinfo
  {author} {\bibfnamefont {B.}~\bibnamefont {Jäck}}, \bibinfo {author}
  {\bibfnamefont {X.}~\bibnamefont {Liu}}, \bibinfo {author} {\bibfnamefont
  {C.-L.}\ \bibnamefont {Chiu}}, \bibinfo {author} {\bibfnamefont
  {K.}~\bibnamefont {Watanabe}}, \bibinfo {author} {\bibfnamefont
  {T.}~\bibnamefont {Taniguchi}}, \bibinfo {author} {\bibfnamefont {B.~A.}\
  \bibnamefont {Bernevig}}, \ and\ \bibinfo {author} {\bibfnamefont
  {A.}~\bibnamefont {Yazdani}},\ }\href {\doibase 10.1038/s41586-019-1422-x}
  {\bibfield  {journal} {\bibinfo  {journal} {Nature}\ }\textbf {\bibinfo
  {volume} {572}},\ \bibinfo {pages} {101} (\bibinfo {year}
  {2019})}\BibitemShut {NoStop}%
\bibitem [{\citenamefont {Nuckolls}\ \emph {et~al.}(2020)\citenamefont
  {Nuckolls}, \citenamefont {Oh}, \citenamefont {Wong}, \citenamefont {Lian},
  \citenamefont {Watanabe}, \citenamefont {Taniguchi}, \citenamefont
  {Bernevig},\ and\ \citenamefont {Yazdani}}]{Nuckolls_2020_strongly}%
  \BibitemOpen
  \bibfield  {author} {\bibinfo {author} {\bibfnamefont {K.~P.}\ \bibnamefont
  {Nuckolls}}, \bibinfo {author} {\bibfnamefont {M.}~\bibnamefont {Oh}},
  \bibinfo {author} {\bibfnamefont {D.}~\bibnamefont {Wong}}, \bibinfo {author}
  {\bibfnamefont {B.}~\bibnamefont {Lian}}, \bibinfo {author} {\bibfnamefont
  {K.}~\bibnamefont {Watanabe}}, \bibinfo {author} {\bibfnamefont
  {T.}~\bibnamefont {Taniguchi}}, \bibinfo {author} {\bibfnamefont {B.~A.}\
  \bibnamefont {Bernevig}}, \ and\ \bibinfo {author} {\bibfnamefont
  {A.}~\bibnamefont {Yazdani}},\ }\href {\doibase 10.1038/s41586-020-3028-8}
  {\bibfield  {journal} {\bibinfo  {journal} {Nature}\ }\textbf {\bibinfo
  {volume} {588}},\ \bibinfo {pages} {610} (\bibinfo {year}
  {2020})}\BibitemShut {NoStop}%
\bibitem [{\citenamefont {Nuckolls}\ \emph {et~al.}(2023)\citenamefont
  {Nuckolls}, \citenamefont {Lee}, \citenamefont {Oh}, \citenamefont {Wong},
  \citenamefont {Soejima}, \citenamefont {Hong}, \citenamefont {Călugăru},
  \citenamefont {Herzog-Arbeitman}, \citenamefont {Bernevig}, \citenamefont
  {Watanabe}, \citenamefont {Taniguchi}, \citenamefont {Regnault},
  \citenamefont {Zaletel},\ and\ \citenamefont
  {Yazdani}}]{Nuckolls_2023_quantum}%
  \BibitemOpen
  \bibfield  {author} {\bibinfo {author} {\bibfnamefont {K.~P.}\ \bibnamefont
  {Nuckolls}}, \bibinfo {author} {\bibfnamefont {R.~L.}\ \bibnamefont {Lee}},
  \bibinfo {author} {\bibfnamefont {M.}~\bibnamefont {Oh}}, \bibinfo {author}
  {\bibfnamefont {D.}~\bibnamefont {Wong}}, \bibinfo {author} {\bibfnamefont
  {T.}~\bibnamefont {Soejima}}, \bibinfo {author} {\bibfnamefont {J.~P.}\
  \bibnamefont {Hong}}, \bibinfo {author} {\bibfnamefont {D.}~\bibnamefont
  {Călugăru}}, \bibinfo {author} {\bibfnamefont {J.}~\bibnamefont
  {Herzog-Arbeitman}}, \bibinfo {author} {\bibfnamefont {B.~A.}\ \bibnamefont
  {Bernevig}}, \bibinfo {author} {\bibfnamefont {K.}~\bibnamefont {Watanabe}},
  \bibinfo {author} {\bibfnamefont {T.}~\bibnamefont {Taniguchi}}, \bibinfo
  {author} {\bibfnamefont {N.}~\bibnamefont {Regnault}}, \bibinfo {author}
  {\bibfnamefont {M.~P.}\ \bibnamefont {Zaletel}}, \ and\ \bibinfo {author}
  {\bibfnamefont {A.}~\bibnamefont {Yazdani}},\ }\href {\doibase
  10.1038/s41586-023-06226-x} {\bibfield  {journal} {\bibinfo  {journal}
  {Nature}\ }\textbf {\bibinfo {volume} {620}},\ \bibinfo {pages} {525}
  (\bibinfo {year} {2023})}\BibitemShut {NoStop}%
\bibitem [{\citenamefont {Kim}\ \emph {et~al.}(2023)\citenamefont {Kim},
  \citenamefont {Choi}, \citenamefont {Lantagne-Hurtubise}, \citenamefont
  {Lewandowski}, \citenamefont {Thomson}, \citenamefont {Kong}, \citenamefont
  {Zhou}, \citenamefont {Baum}, \citenamefont {Zhang}, \citenamefont {Holleis},
  \citenamefont {Watanabe}, \citenamefont {Taniguchi}, \citenamefont {Young},
  \citenamefont {Alicea},\ and\ \citenamefont {Nadj-Perge}}]{kim_imaging_2023}%
  \BibitemOpen
  \bibfield  {author} {\bibinfo {author} {\bibfnamefont {H.}~\bibnamefont
  {Kim}}, \bibinfo {author} {\bibfnamefont {Y.}~\bibnamefont {Choi}}, \bibinfo
  {author} {\bibfnamefont {E.}~\bibnamefont {Lantagne-Hurtubise}}, \bibinfo
  {author} {\bibfnamefont {C.}~\bibnamefont {Lewandowski}}, \bibinfo {author}
  {\bibfnamefont {A.}~\bibnamefont {Thomson}}, \bibinfo {author} {\bibfnamefont
  {L.}~\bibnamefont {Kong}}, \bibinfo {author} {\bibfnamefont {H.}~\bibnamefont
  {Zhou}}, \bibinfo {author} {\bibfnamefont {E.}~\bibnamefont {Baum}}, \bibinfo
  {author} {\bibfnamefont {Y.}~\bibnamefont {Zhang}}, \bibinfo {author}
  {\bibfnamefont {L.}~\bibnamefont {Holleis}}, \bibinfo {author} {\bibfnamefont
  {K.}~\bibnamefont {Watanabe}}, \bibinfo {author} {\bibfnamefont
  {T.}~\bibnamefont {Taniguchi}}, \bibinfo {author} {\bibfnamefont {A.~F.}\
  \bibnamefont {Young}}, \bibinfo {author} {\bibfnamefont {J.}~\bibnamefont
  {Alicea}}, \ and\ \bibinfo {author} {\bibfnamefont {S.}~\bibnamefont
  {Nadj-Perge}},\ }\href {\doibase 10.1038/s41586-023-06663-8} {\bibfield
  {journal} {\bibinfo  {journal} {Nature}\ }\textbf {\bibinfo {volume} {623}},\
  \bibinfo {pages} {942} (\bibinfo {year} {2023})}\BibitemShut {NoStop}%
\bibitem [{\citenamefont {Yankowitz}\ \emph {et~al.}(2019)\citenamefont
  {Yankowitz}, \citenamefont {Chen}, \citenamefont {Polshyn}, \citenamefont
  {Zhang}, \citenamefont {Watanabe}, \citenamefont {Taniguchi}, \citenamefont
  {Graf}, \citenamefont {Young},\ and\ \citenamefont
  {Dean}}]{Yankowitz_2019_tuning}%
  \BibitemOpen
  \bibfield  {author} {\bibinfo {author} {\bibfnamefont {M.}~\bibnamefont
  {Yankowitz}}, \bibinfo {author} {\bibfnamefont {S.}~\bibnamefont {Chen}},
  \bibinfo {author} {\bibfnamefont {H.}~\bibnamefont {Polshyn}}, \bibinfo
  {author} {\bibfnamefont {Y.}~\bibnamefont {Zhang}}, \bibinfo {author}
  {\bibfnamefont {K.}~\bibnamefont {Watanabe}}, \bibinfo {author}
  {\bibfnamefont {T.}~\bibnamefont {Taniguchi}}, \bibinfo {author}
  {\bibfnamefont {D.}~\bibnamefont {Graf}}, \bibinfo {author} {\bibfnamefont
  {A.~F.}\ \bibnamefont {Young}}, \ and\ \bibinfo {author} {\bibfnamefont
  {C.~R.}\ \bibnamefont {Dean}},\ }\href {\doibase 10.1126/science.aav1910}
  {\bibfield  {journal} {\bibinfo  {journal} {Science}\ }\textbf {\bibinfo
  {volume} {363}},\ \bibinfo {pages} {1059} (\bibinfo {year}
  {2019})}\BibitemShut {NoStop}%
\bibitem [{\citenamefont {Arora}\ \emph {et~al.}(2020)\citenamefont {Arora},
  \citenamefont {Polski}, \citenamefont {Zhang}, \citenamefont {Thomson},
  \citenamefont {Choi}, \citenamefont {Kim}, \citenamefont {Lin}, \citenamefont
  {Wilson}, \citenamefont {Xu}, \citenamefont {Chu}, \citenamefont {Watanabe},
  \citenamefont {Taniguchi}, \citenamefont {Alicea},\ and\ \citenamefont
  {Nadj-Perge}}]{Arora_2020_superconductivity}%
  \BibitemOpen
  \bibfield  {author} {\bibinfo {author} {\bibfnamefont {H.~S.}\ \bibnamefont
  {Arora}}, \bibinfo {author} {\bibfnamefont {R.}~\bibnamefont {Polski}},
  \bibinfo {author} {\bibfnamefont {Y.}~\bibnamefont {Zhang}}, \bibinfo
  {author} {\bibfnamefont {A.}~\bibnamefont {Thomson}}, \bibinfo {author}
  {\bibfnamefont {Y.}~\bibnamefont {Choi}}, \bibinfo {author} {\bibfnamefont
  {H.}~\bibnamefont {Kim}}, \bibinfo {author} {\bibfnamefont {Z.}~\bibnamefont
  {Lin}}, \bibinfo {author} {\bibfnamefont {I.~Z.}\ \bibnamefont {Wilson}},
  \bibinfo {author} {\bibfnamefont {X.}~\bibnamefont {Xu}}, \bibinfo {author}
  {\bibfnamefont {J.-H.}\ \bibnamefont {Chu}}, \bibinfo {author} {\bibfnamefont
  {K.}~\bibnamefont {Watanabe}}, \bibinfo {author} {\bibfnamefont
  {T.}~\bibnamefont {Taniguchi}}, \bibinfo {author} {\bibfnamefont
  {J.}~\bibnamefont {Alicea}}, \ and\ \bibinfo {author} {\bibfnamefont
  {S.}~\bibnamefont {Nadj-Perge}},\ }\href {\doibase 10.1038/s41586-020-2473-8}
  {\bibfield  {journal} {\bibinfo  {journal} {Nature}\ }\textbf {\bibinfo
  {volume} {583}},\ \bibinfo {pages} {379} (\bibinfo {year}
  {2020})}\BibitemShut {NoStop}%
\bibitem [{\citenamefont {Saito}\ \emph {et~al.}(2020)\citenamefont {Saito},
  \citenamefont {Ge}, \citenamefont {Watanabe}, \citenamefont {Taniguchi},\
  and\ \citenamefont {Young}}]{Saito_independent_2020}%
  \BibitemOpen
  \bibfield  {author} {\bibinfo {author} {\bibfnamefont {Y.}~\bibnamefont
  {Saito}}, \bibinfo {author} {\bibfnamefont {J.}~\bibnamefont {Ge}}, \bibinfo
  {author} {\bibfnamefont {K.}~\bibnamefont {Watanabe}}, \bibinfo {author}
  {\bibfnamefont {T.}~\bibnamefont {Taniguchi}}, \ and\ \bibinfo {author}
  {\bibfnamefont {A.~F.}\ \bibnamefont {Young}},\ }\href {\doibase
  10.1038/s41567-020-0928-3} {\bibfield  {journal} {\bibinfo  {journal} {Nat.
  Phys.}\ }\textbf {\bibinfo {volume} {16}},\ \bibinfo {pages} {926} (\bibinfo
  {year} {2020})}\BibitemShut {NoStop}%
\bibitem [{\citenamefont {Stepanov}\ \emph {et~al.}(2020)\citenamefont
  {Stepanov}, \citenamefont {Das}, \citenamefont {Lu}, \citenamefont
  {Fahimniya}, \citenamefont {Watanabe}, \citenamefont {Taniguchi},
  \citenamefont {Koppens}, \citenamefont {Lischner}, \citenamefont {Levitov},\
  and\ \citenamefont {Efetov}}]{Stepanov_untying_2020}%
  \BibitemOpen
  \bibfield  {author} {\bibinfo {author} {\bibfnamefont {P.}~\bibnamefont
  {Stepanov}}, \bibinfo {author} {\bibfnamefont {I.}~\bibnamefont {Das}},
  \bibinfo {author} {\bibfnamefont {X.}~\bibnamefont {Lu}}, \bibinfo {author}
  {\bibfnamefont {A.}~\bibnamefont {Fahimniya}}, \bibinfo {author}
  {\bibfnamefont {K.}~\bibnamefont {Watanabe}}, \bibinfo {author}
  {\bibfnamefont {T.}~\bibnamefont {Taniguchi}}, \bibinfo {author}
  {\bibfnamefont {F.~H.~L.}\ \bibnamefont {Koppens}}, \bibinfo {author}
  {\bibfnamefont {J.}~\bibnamefont {Lischner}}, \bibinfo {author}
  {\bibfnamefont {L.}~\bibnamefont {Levitov}}, \ and\ \bibinfo {author}
  {\bibfnamefont {D.~K.}\ \bibnamefont {Efetov}},\ }\href {\doibase
  10.1038/s41586-020-2459-6} {\bibfield  {journal} {\bibinfo  {journal}
  {Nature}\ }\textbf {\bibinfo {volume} {583}},\ \bibinfo {pages} {375}
  (\bibinfo {year} {2020})}\BibitemShut {NoStop}%
\bibitem [{\citenamefont {Liu}\ \emph {et~al.}(2021)\citenamefont {Liu},
  \citenamefont {Wang}, \citenamefont {Watanabe}, \citenamefont {Taniguchi},
  \citenamefont {Vafek},\ and\ \citenamefont {Li}}]{Liu_tuning_2021}%
  \BibitemOpen
  \bibfield  {author} {\bibinfo {author} {\bibfnamefont {X.}~\bibnamefont
  {Liu}}, \bibinfo {author} {\bibfnamefont {Z.}~\bibnamefont {Wang}}, \bibinfo
  {author} {\bibfnamefont {K.}~\bibnamefont {Watanabe}}, \bibinfo {author}
  {\bibfnamefont {T.}~\bibnamefont {Taniguchi}}, \bibinfo {author}
  {\bibfnamefont {O.}~\bibnamefont {Vafek}}, \ and\ \bibinfo {author}
  {\bibfnamefont {J.~I.~A.}\ \bibnamefont {Li}},\ }\href {\doibase
  10.1126/science.abb8754} {\bibfield  {journal} {\bibinfo  {journal}
  {Science}\ }\textbf {\bibinfo {volume} {371}},\ \bibinfo {pages} {1261}
  (\bibinfo {year} {2021})}\BibitemShut {NoStop}%
\bibitem [{\citenamefont {Cao}\ \emph {et~al.}(2021{\natexlab{a}})\citenamefont
  {Cao}, \citenamefont {Rodan-Legrain}, \citenamefont {Park}, \citenamefont
  {Yuan}, \citenamefont {Watanabe}, \citenamefont {Taniguchi}, \citenamefont
  {Fernandes}, \citenamefont {Fu},\ and\ \citenamefont
  {Jarillo-Herrero}}]{Cao_2021_nematicity}%
  \BibitemOpen
  \bibfield  {author} {\bibinfo {author} {\bibfnamefont {Y.}~\bibnamefont
  {Cao}}, \bibinfo {author} {\bibfnamefont {D.}~\bibnamefont {Rodan-Legrain}},
  \bibinfo {author} {\bibfnamefont {J.~M.}\ \bibnamefont {Park}}, \bibinfo
  {author} {\bibfnamefont {N.~F.~Q.}\ \bibnamefont {Yuan}}, \bibinfo {author}
  {\bibfnamefont {K.}~\bibnamefont {Watanabe}}, \bibinfo {author}
  {\bibfnamefont {T.}~\bibnamefont {Taniguchi}}, \bibinfo {author}
  {\bibfnamefont {R.~M.}\ \bibnamefont {Fernandes}}, \bibinfo {author}
  {\bibfnamefont {L.}~\bibnamefont {Fu}}, \ and\ \bibinfo {author}
  {\bibfnamefont {P.}~\bibnamefont {Jarillo-Herrero}},\ }\href {\doibase
  10.1126/science.abc2836} {\bibfield  {journal} {\bibinfo  {journal}
  {Science}\ }\textbf {\bibinfo {volume} {372}},\ \bibinfo {pages} {264}
  (\bibinfo {year} {2021}{\natexlab{a}})}\BibitemShut {NoStop}%
\bibitem [{\citenamefont {Cao}\ \emph {et~al.}(2021{\natexlab{b}})\citenamefont
  {Cao}, \citenamefont {Park}, \citenamefont {Watanabe}, \citenamefont
  {Taniguchi},\ and\ \citenamefont {Jarillo-Herrero}}]{cao_pauli-limit_2021}%
  \BibitemOpen
  \bibfield  {author} {\bibinfo {author} {\bibfnamefont {Y.}~\bibnamefont
  {Cao}}, \bibinfo {author} {\bibfnamefont {J.~M.}\ \bibnamefont {Park}},
  \bibinfo {author} {\bibfnamefont {K.}~\bibnamefont {Watanabe}}, \bibinfo
  {author} {\bibfnamefont {T.}~\bibnamefont {Taniguchi}}, \ and\ \bibinfo
  {author} {\bibfnamefont {P.}~\bibnamefont {Jarillo-Herrero}},\ }\href
  {\doibase 10.1038/s41586-021-03685-y} {\bibfield  {journal} {\bibinfo
  {journal} {Nature}\ }\textbf {\bibinfo {volume} {595}},\ \bibinfo {pages}
  {526} (\bibinfo {year} {2021}{\natexlab{b}})}\BibitemShut {NoStop}%
\bibitem [{\citenamefont {Park}\ \emph {et~al.}(2021)\citenamefont {Park},
  \citenamefont {Cao}, \citenamefont {Watanabe}, \citenamefont {Taniguchi},\
  and\ \citenamefont {Jarillo-Herrero}}]{park_tunable_2021}%
  \BibitemOpen
  \bibfield  {author} {\bibinfo {author} {\bibfnamefont {J.~M.}\ \bibnamefont
  {Park}}, \bibinfo {author} {\bibfnamefont {Y.}~\bibnamefont {Cao}}, \bibinfo
  {author} {\bibfnamefont {K.}~\bibnamefont {Watanabe}}, \bibinfo {author}
  {\bibfnamefont {T.}~\bibnamefont {Taniguchi}}, \ and\ \bibinfo {author}
  {\bibfnamefont {P.}~\bibnamefont {Jarillo-Herrero}},\ }\href {\doibase
  10.1038/s41586-021-03192-0} {\bibfield  {journal} {\bibinfo  {journal}
  {Nature}\ }\textbf {\bibinfo {volume} {590}},\ \bibinfo {pages} {249}
  (\bibinfo {year} {2021})}\BibitemShut {NoStop}%
\bibitem [{\citenamefont {Hao}\ \emph {et~al.}(2021)\citenamefont {Hao},
  \citenamefont {Zimmerman}, \citenamefont {Ledwith}, \citenamefont {Khalaf},
  \citenamefont {Najafabadi}, \citenamefont {Watanabe}, \citenamefont
  {Taniguchi}, \citenamefont {Vishwanath},\ and\ \citenamefont
  {Kim}}]{hao_electric_2021}%
  \BibitemOpen
  \bibfield  {author} {\bibinfo {author} {\bibfnamefont {Z.}~\bibnamefont
  {Hao}}, \bibinfo {author} {\bibfnamefont {A.~M.}\ \bibnamefont {Zimmerman}},
  \bibinfo {author} {\bibfnamefont {P.}~\bibnamefont {Ledwith}}, \bibinfo
  {author} {\bibfnamefont {E.}~\bibnamefont {Khalaf}}, \bibinfo {author}
  {\bibfnamefont {D.~H.}\ \bibnamefont {Najafabadi}}, \bibinfo {author}
  {\bibfnamefont {K.}~\bibnamefont {Watanabe}}, \bibinfo {author}
  {\bibfnamefont {T.}~\bibnamefont {Taniguchi}}, \bibinfo {author}
  {\bibfnamefont {A.}~\bibnamefont {Vishwanath}}, \ and\ \bibinfo {author}
  {\bibfnamefont {P.}~\bibnamefont {Kim}},\ }\href {\doibase
  10.1126/science.abg0399} {\bibfield  {journal} {\bibinfo  {journal}
  {Science}\ }\textbf {\bibinfo {volume} {371}},\ \bibinfo {pages} {1133}
  (\bibinfo {year} {2021})}\BibitemShut {NoStop}%
\bibitem [{\citenamefont {Kim}\ \emph {et~al.}(2022)\citenamefont {Kim},
  \citenamefont {Choi}, \citenamefont {Lewandowski}, \citenamefont {Thomson},
  \citenamefont {Zhang}, \citenamefont {Polski}, \citenamefont {Watanabe},
  \citenamefont {Taniguchi}, \citenamefont {Alicea},\ and\ \citenamefont
  {Nadj-Perge}}]{kim_evidence_2022}%
  \BibitemOpen
  \bibfield  {author} {\bibinfo {author} {\bibfnamefont {H.}~\bibnamefont
  {Kim}}, \bibinfo {author} {\bibfnamefont {Y.}~\bibnamefont {Choi}}, \bibinfo
  {author} {\bibfnamefont {C.}~\bibnamefont {Lewandowski}}, \bibinfo {author}
  {\bibfnamefont {A.}~\bibnamefont {Thomson}}, \bibinfo {author} {\bibfnamefont
  {Y.}~\bibnamefont {Zhang}}, \bibinfo {author} {\bibfnamefont
  {R.}~\bibnamefont {Polski}}, \bibinfo {author} {\bibfnamefont
  {K.}~\bibnamefont {Watanabe}}, \bibinfo {author} {\bibfnamefont
  {T.}~\bibnamefont {Taniguchi}}, \bibinfo {author} {\bibfnamefont
  {J.}~\bibnamefont {Alicea}}, \ and\ \bibinfo {author} {\bibfnamefont
  {S.}~\bibnamefont {Nadj-Perge}},\ }\href {\doibase
  10.1038/s41586-022-04715-z} {\bibfield  {journal} {\bibinfo  {journal}
  {Nature}\ }\textbf {\bibinfo {volume} {606}},\ \bibinfo {pages} {494}
  (\bibinfo {year} {2022})}\BibitemShut {NoStop}%
\bibitem [{\citenamefont {Liu}\ \emph {et~al.}(2022{\natexlab{a}})\citenamefont
  {Liu}, \citenamefont {Zhang}, \citenamefont {Watanabe}, \citenamefont
  {Taniguchi},\ and\ \citenamefont {Li}}]{liu_isospin_2022}%
  \BibitemOpen
  \bibfield  {author} {\bibinfo {author} {\bibfnamefont {X.}~\bibnamefont
  {Liu}}, \bibinfo {author} {\bibfnamefont {N.~J.}\ \bibnamefont {Zhang}},
  \bibinfo {author} {\bibfnamefont {K.}~\bibnamefont {Watanabe}}, \bibinfo
  {author} {\bibfnamefont {T.}~\bibnamefont {Taniguchi}}, \ and\ \bibinfo
  {author} {\bibfnamefont {J.~I.~A.}\ \bibnamefont {Li}},\ }\href {\doibase
  10.1038/s41567-022-01515-0} {\bibfield  {journal} {\bibinfo  {journal} {Nat.
  Phys.}\ }\textbf {\bibinfo {volume} {18}},\ \bibinfo {pages} {522} (\bibinfo
  {year} {2022}{\natexlab{a}})}\BibitemShut {NoStop}%
\bibitem [{\citenamefont {Gao}\ \emph {et~al.}(2024)\citenamefont {Gao},
  \citenamefont {Jimeno-Pozo}, \citenamefont {Pantaleon}, \citenamefont
  {Codecido}, \citenamefont {Sharifi}, \citenamefont {Zhang}, \citenamefont
  {Liu}, \citenamefont {Watanabe}, \citenamefont {Taniguchi}, \citenamefont
  {Bockrath}, \citenamefont {Guinea},\ and\ \citenamefont
  {Lau}}]{gao_2024_doubleedgedrole}%
  \BibitemOpen
  \bibfield  {author} {\bibinfo {author} {\bibfnamefont {X.}~\bibnamefont
  {Gao}}, \bibinfo {author} {\bibfnamefont {A.}~\bibnamefont {Jimeno-Pozo}},
  \bibinfo {author} {\bibfnamefont {P.~A.}\ \bibnamefont {Pantaleon}}, \bibinfo
  {author} {\bibfnamefont {E.}~\bibnamefont {Codecido}}, \bibinfo {author}
  {\bibfnamefont {D.~L.}\ \bibnamefont {Sharifi}}, \bibinfo {author}
  {\bibfnamefont {Z.}~\bibnamefont {Zhang}}, \bibinfo {author} {\bibfnamefont
  {Y.}~\bibnamefont {Liu}}, \bibinfo {author} {\bibfnamefont {K.}~\bibnamefont
  {Watanabe}}, \bibinfo {author} {\bibfnamefont {T.}~\bibnamefont {Taniguchi}},
  \bibinfo {author} {\bibfnamefont {M.~W.}\ \bibnamefont {Bockrath}}, \bibinfo
  {author} {\bibfnamefont {F.}~\bibnamefont {Guinea}}, \ and\ \bibinfo {author}
  {\bibfnamefont {C.~N.}\ \bibnamefont {Lau}},\ }\href
  {https://arxiv.org/abs/2412.01578} {\enquote {\bibinfo {title} {Double-edged
  role of interactions in superconducting twisted bilayer graphene},}\ }
  (\bibinfo {year} {2024}),\ \Eprint {http://arxiv.org/abs/2412.01578}
  {arXiv:2412.01578 [cond-mat.mes-hall]} \BibitemShut {NoStop}%
\bibitem [{\citenamefont {Tanaka}\ \emph {et~al.}(2025)\citenamefont {Tanaka},
  \citenamefont {Wang}, \citenamefont {Dinh}, \citenamefont {Rodan-Legrain},
  \citenamefont {Zaman}, \citenamefont {Hays}, \citenamefont {Almanakly},
  \citenamefont {Kannan}, \citenamefont {Kim}, \citenamefont {Niedzielski},
  \citenamefont {Serniak}, \citenamefont {Schwartz}, \citenamefont {Watanabe},
  \citenamefont {Taniguchi}, \citenamefont {Orlando}, \citenamefont
  {Gustavsson}, \citenamefont {Grover}, \citenamefont {Jarillo-Herrero},\ and\
  \citenamefont {Oliver}}]{tanaka_superfluid_2025}%
  \BibitemOpen
  \bibfield  {author} {\bibinfo {author} {\bibfnamefont {M.}~\bibnamefont
  {Tanaka}}, \bibinfo {author} {\bibfnamefont {J.~r. h.-j.}\ \bibnamefont
  {Wang}}, \bibinfo {author} {\bibfnamefont {T.~H.}\ \bibnamefont {Dinh}},
  \bibinfo {author} {\bibfnamefont {D.}~\bibnamefont {Rodan-Legrain}}, \bibinfo
  {author} {\bibfnamefont {S.}~\bibnamefont {Zaman}}, \bibinfo {author}
  {\bibfnamefont {M.}~\bibnamefont {Hays}}, \bibinfo {author} {\bibfnamefont
  {A.}~\bibnamefont {Almanakly}}, \bibinfo {author} {\bibfnamefont
  {B.}~\bibnamefont {Kannan}}, \bibinfo {author} {\bibfnamefont {D.~K.}\
  \bibnamefont {Kim}}, \bibinfo {author} {\bibfnamefont {B.~M.}\ \bibnamefont
  {Niedzielski}}, \bibinfo {author} {\bibfnamefont {K.}~\bibnamefont
  {Serniak}}, \bibinfo {author} {\bibfnamefont {M.~E.}\ \bibnamefont
  {Schwartz}}, \bibinfo {author} {\bibfnamefont {K.}~\bibnamefont {Watanabe}},
  \bibinfo {author} {\bibfnamefont {T.}~\bibnamefont {Taniguchi}}, \bibinfo
  {author} {\bibfnamefont {T.~P.}\ \bibnamefont {Orlando}}, \bibinfo {author}
  {\bibfnamefont {S.}~\bibnamefont {Gustavsson}}, \bibinfo {author}
  {\bibfnamefont {J.~A.}\ \bibnamefont {Grover}}, \bibinfo {author}
  {\bibfnamefont {P.}~\bibnamefont {Jarillo-Herrero}}, \ and\ \bibinfo {author}
  {\bibfnamefont {W.~D.}\ \bibnamefont {Oliver}},\ }\href {\doibase
  10.1038/s41586-024-08494-7} {\bibfield  {journal} {\bibinfo  {journal}
  {Nature}\ }\textbf {\bibinfo {volume} {638}},\ \bibinfo {pages} {99}
  (\bibinfo {year} {2025})}\BibitemShut {NoStop}%
\bibitem [{\citenamefont {Banerjee}\ \emph {et~al.}(2025)\citenamefont
  {Banerjee}, \citenamefont {Hao}, \citenamefont {Kreidel}, \citenamefont
  {Ledwith}, \citenamefont {Phinney}, \citenamefont {Park}, \citenamefont
  {Zimmerman}, \citenamefont {Wesson}, \citenamefont {Watanabe}, \citenamefont
  {Taniguchi}, \citenamefont {Westervelt}, \citenamefont {Yacoby},
  \citenamefont {Jarillo-Herrero}, \citenamefont {Volkov}, \citenamefont
  {Vishwanath}, \citenamefont {Fong},\ and\ \citenamefont
  {Kim}}]{banerjee_superfluid_2025}%
  \BibitemOpen
  \bibfield  {author} {\bibinfo {author} {\bibfnamefont {A.}~\bibnamefont
  {Banerjee}}, \bibinfo {author} {\bibfnamefont {Z.}~\bibnamefont {Hao}},
  \bibinfo {author} {\bibfnamefont {M.}~\bibnamefont {Kreidel}}, \bibinfo
  {author} {\bibfnamefont {P.}~\bibnamefont {Ledwith}}, \bibinfo {author}
  {\bibfnamefont {I.}~\bibnamefont {Phinney}}, \bibinfo {author} {\bibfnamefont
  {J.~M.}\ \bibnamefont {Park}}, \bibinfo {author} {\bibfnamefont
  {A.}~\bibnamefont {Zimmerman}}, \bibinfo {author} {\bibfnamefont {M.~E.}\
  \bibnamefont {Wesson}}, \bibinfo {author} {\bibfnamefont {K.}~\bibnamefont
  {Watanabe}}, \bibinfo {author} {\bibfnamefont {T.}~\bibnamefont {Taniguchi}},
  \bibinfo {author} {\bibfnamefont {R.~M.}\ \bibnamefont {Westervelt}},
  \bibinfo {author} {\bibfnamefont {A.}~\bibnamefont {Yacoby}}, \bibinfo
  {author} {\bibfnamefont {P.}~\bibnamefont {Jarillo-Herrero}}, \bibinfo
  {author} {\bibfnamefont {P.~A.}\ \bibnamefont {Volkov}}, \bibinfo {author}
  {\bibfnamefont {A.}~\bibnamefont {Vishwanath}}, \bibinfo {author}
  {\bibfnamefont {K.~C.}\ \bibnamefont {Fong}}, \ and\ \bibinfo {author}
  {\bibfnamefont {P.}~\bibnamefont {Kim}},\ }\href {\doibase
  10.1038/s41586-024-08444-3} {\bibfield  {journal} {\bibinfo  {journal}
  {Nature}\ }\textbf {\bibinfo {volume} {638}},\ \bibinfo {pages} {93}
  (\bibinfo {year} {2025})}\BibitemShut {NoStop}%
\bibitem [{\citenamefont {Oh}\ \emph {et~al.}(2021)\citenamefont {Oh},
  \citenamefont {Nuckolls}, \citenamefont {Wong}, \citenamefont {Lee},
  \citenamefont {Liu}, \citenamefont {Watanabe}, \citenamefont {Taniguchi},\
  and\ \citenamefont {Yazdani}}]{Oh_2021_evidence}%
  \BibitemOpen
  \bibfield  {author} {\bibinfo {author} {\bibfnamefont {M.}~\bibnamefont
  {Oh}}, \bibinfo {author} {\bibfnamefont {K.~P.}\ \bibnamefont {Nuckolls}},
  \bibinfo {author} {\bibfnamefont {D.}~\bibnamefont {Wong}}, \bibinfo {author}
  {\bibfnamefont {R.~L.}\ \bibnamefont {Lee}}, \bibinfo {author} {\bibfnamefont
  {X.}~\bibnamefont {Liu}}, \bibinfo {author} {\bibfnamefont {K.}~\bibnamefont
  {Watanabe}}, \bibinfo {author} {\bibfnamefont {T.}~\bibnamefont {Taniguchi}},
  \ and\ \bibinfo {author} {\bibfnamefont {A.}~\bibnamefont {Yazdani}},\ }\href
  {\doibase 10.1038/s41586-021-04121-x} {\bibfield  {journal} {\bibinfo
  {journal} {Nature}\ }\textbf {\bibinfo {volume} {600}},\ \bibinfo {pages}
  {240} (\bibinfo {year} {2021})}\BibitemShut {NoStop}%
\bibitem [{\citenamefont {Park}\ \emph {et~al.}(2025)\citenamefont {Park},
  \citenamefont {Sun}, \citenamefont {Watanabe}, \citenamefont {Taniguchi},\
  and\ \citenamefont
  {Jarillo-Herrero}}]{park_2025_simultaneoustransporttunnelingspectroscopy}%
  \BibitemOpen
  \bibfield  {author} {\bibinfo {author} {\bibfnamefont {J.~M.}\ \bibnamefont
  {Park}}, \bibinfo {author} {\bibfnamefont {S.}~\bibnamefont {Sun}}, \bibinfo
  {author} {\bibfnamefont {K.}~\bibnamefont {Watanabe}}, \bibinfo {author}
  {\bibfnamefont {T.}~\bibnamefont {Taniguchi}}, \ and\ \bibinfo {author}
  {\bibfnamefont {P.}~\bibnamefont {Jarillo-Herrero}},\ }\href
  {https://arxiv.org/abs/2503.16410} {\enquote {\bibinfo {title} {Simultaneous
  transport and tunneling spectroscopy of moir\'e graphene: Distinct
  observation of the superconducting gap and signatures of nodal
  superconductivity},}\ } (\bibinfo {year} {2025}),\ \Eprint
  {http://arxiv.org/abs/2503.16410} {arXiv:2503.16410 [cond-mat.supr-con]}
  \BibitemShut {NoStop}%
\bibitem [{\citenamefont {Kim}\ \emph {et~al.}(2025)\citenamefont {Kim},
  \citenamefont {Rai}, \citenamefont {Crippa}, \citenamefont {Călugăru},
  \citenamefont {Hu}, \citenamefont {Choi}, \citenamefont {Kong}, \citenamefont
  {Baum}, \citenamefont {Zhang}, \citenamefont {Holleis}, \citenamefont
  {Watanabe}, \citenamefont {Taniguchi}, \citenamefont {Young}, \citenamefont
  {Bernevig}, \citenamefont {Valent\'i}, \citenamefont {Sangiovanni},
  \citenamefont {Wehling},\ and\ \citenamefont
  {Nadj-Perge}}]{kim_2025_resolvingintervalleygapsmanybody}%
  \BibitemOpen
  \bibfield  {author} {\bibinfo {author} {\bibfnamefont {H.}~\bibnamefont
  {Kim}}, \bibinfo {author} {\bibfnamefont {G.}~\bibnamefont {Rai}}, \bibinfo
  {author} {\bibfnamefont {L.}~\bibnamefont {Crippa}}, \bibinfo {author}
  {\bibfnamefont {D.}~\bibnamefont {Călugăru}}, \bibinfo {author}
  {\bibfnamefont {H.}~\bibnamefont {Hu}}, \bibinfo {author} {\bibfnamefont
  {Y.}~\bibnamefont {Choi}}, \bibinfo {author} {\bibfnamefont {L.}~\bibnamefont
  {Kong}}, \bibinfo {author} {\bibfnamefont {E.}~\bibnamefont {Baum}}, \bibinfo
  {author} {\bibfnamefont {Y.}~\bibnamefont {Zhang}}, \bibinfo {author}
  {\bibfnamefont {L.}~\bibnamefont {Holleis}}, \bibinfo {author} {\bibfnamefont
  {K.}~\bibnamefont {Watanabe}}, \bibinfo {author} {\bibfnamefont
  {T.}~\bibnamefont {Taniguchi}}, \bibinfo {author} {\bibfnamefont {A.~F.}\
  \bibnamefont {Young}}, \bibinfo {author} {\bibfnamefont {B.~A.}\ \bibnamefont
  {Bernevig}}, \bibinfo {author} {\bibfnamefont {R.}~\bibnamefont {Valent\'i}},
  \bibinfo {author} {\bibfnamefont {G.}~\bibnamefont {Sangiovanni}}, \bibinfo
  {author} {\bibfnamefont {T.}~\bibnamefont {Wehling}}, \ and\ \bibinfo
  {author} {\bibfnamefont {S.}~\bibnamefont {Nadj-Perge}},\ }\href
  {https://arxiv.org/abs/2505.17200} {\enquote {\bibinfo {title} {Resolving
  intervalley gaps and many-body resonances in moir\'e superconductor},}\ }
  (\bibinfo {year} {2025}),\ \Eprint {http://arxiv.org/abs/2505.17200}
  {arXiv:2505.17200 [cond-mat.supr-con]} \BibitemShut {NoStop}%
\bibitem [{\citenamefont {Polshyn}\ \emph {et~al.}(2019)\citenamefont
  {Polshyn}, \citenamefont {Yankowitz}, \citenamefont {Chen}, \citenamefont
  {Zhang}, \citenamefont {Watanabe}, \citenamefont {Taniguchi}, \citenamefont
  {Dean},\ and\ \citenamefont {Young}}]{Polshyn_2019_Large}%
  \BibitemOpen
  \bibfield  {author} {\bibinfo {author} {\bibfnamefont {H.}~\bibnamefont
  {Polshyn}}, \bibinfo {author} {\bibfnamefont {M.}~\bibnamefont {Yankowitz}},
  \bibinfo {author} {\bibfnamefont {S.}~\bibnamefont {Chen}}, \bibinfo {author}
  {\bibfnamefont {Y.}~\bibnamefont {Zhang}}, \bibinfo {author} {\bibfnamefont
  {K.}~\bibnamefont {Watanabe}}, \bibinfo {author} {\bibfnamefont
  {T.}~\bibnamefont {Taniguchi}}, \bibinfo {author} {\bibfnamefont {C.~R.}\
  \bibnamefont {Dean}}, \ and\ \bibinfo {author} {\bibfnamefont {A.~F.}\
  \bibnamefont {Young}},\ }\href {\doibase 10.1038/s41567-019-0596-3}
  {\bibfield  {journal} {\bibinfo  {journal} {Nat. Phys.}\ }\textbf {\bibinfo
  {volume} {15}},\ \bibinfo {pages} {1011} (\bibinfo {year}
  {2019})}\BibitemShut {NoStop}%
\bibitem [{\citenamefont {Cao}\ \emph {et~al.}(2020)\citenamefont {Cao},
  \citenamefont {Chowdhury}, \citenamefont {Rodan-Legrain}, \citenamefont
  {Rubies-Bigorda}, \citenamefont {Watanabe}, \citenamefont {Taniguchi},
  \citenamefont {Senthil},\ and\ \citenamefont
  {Jarillo-Herrero}}]{Cao_2020_strange}%
  \BibitemOpen
  \bibfield  {author} {\bibinfo {author} {\bibfnamefont {Y.}~\bibnamefont
  {Cao}}, \bibinfo {author} {\bibfnamefont {D.}~\bibnamefont {Chowdhury}},
  \bibinfo {author} {\bibfnamefont {D.}~\bibnamefont {Rodan-Legrain}}, \bibinfo
  {author} {\bibfnamefont {O.}~\bibnamefont {Rubies-Bigorda}}, \bibinfo
  {author} {\bibfnamefont {K.}~\bibnamefont {Watanabe}}, \bibinfo {author}
  {\bibfnamefont {T.}~\bibnamefont {Taniguchi}}, \bibinfo {author}
  {\bibfnamefont {T.}~\bibnamefont {Senthil}}, \ and\ \bibinfo {author}
  {\bibfnamefont {P.}~\bibnamefont {Jarillo-Herrero}},\ }\href {\doibase
  10.1103/PhysRevLett.124.076801} {\bibfield  {journal} {\bibinfo  {journal}
  {Phys. Rev. Lett.}\ }\textbf {\bibinfo {volume} {124}},\ \bibinfo {pages}
  {076801} (\bibinfo {year} {2020})}\BibitemShut {NoStop}%
\bibitem [{\citenamefont {Jaoui}\ \emph {et~al.}(2022)\citenamefont {Jaoui},
  \citenamefont {Das}, \citenamefont {Di~Battista}, \citenamefont
  {Díez-Mérida}, \citenamefont {Lu}, \citenamefont {Watanabe}, \citenamefont
  {Taniguchi}, \citenamefont {Ishizuka}, \citenamefont {Levitov},\ and\
  \citenamefont {Efetov}}]{Jaoui_2022_quantum}%
  \BibitemOpen
  \bibfield  {author} {\bibinfo {author} {\bibfnamefont {A.}~\bibnamefont
  {Jaoui}}, \bibinfo {author} {\bibfnamefont {I.}~\bibnamefont {Das}}, \bibinfo
  {author} {\bibfnamefont {G.}~\bibnamefont {Di~Battista}}, \bibinfo {author}
  {\bibfnamefont {J.}~\bibnamefont {Díez-Mérida}}, \bibinfo {author}
  {\bibfnamefont {X.}~\bibnamefont {Lu}}, \bibinfo {author} {\bibfnamefont
  {K.}~\bibnamefont {Watanabe}}, \bibinfo {author} {\bibfnamefont
  {T.}~\bibnamefont {Taniguchi}}, \bibinfo {author} {\bibfnamefont
  {H.}~\bibnamefont {Ishizuka}}, \bibinfo {author} {\bibfnamefont
  {L.}~\bibnamefont {Levitov}}, \ and\ \bibinfo {author} {\bibfnamefont
  {D.~K.}\ \bibnamefont {Efetov}},\ }\href {\doibase
  10.1038/s41567-022-01556-5} {\bibfield  {journal} {\bibinfo  {journal} {Nat.
  Phys.}\ }\textbf {\bibinfo {volume} {18}},\ \bibinfo {pages} {633} (\bibinfo
  {year} {2022})}\BibitemShut {NoStop}%
\bibitem [{\citenamefont {Zou}\ \emph {et~al.}(2018)\citenamefont {Zou},
  \citenamefont {Po}, \citenamefont {Vishwanath},\ and\ \citenamefont
  {Senthil}}]{Zou_2018_band}%
  \BibitemOpen
  \bibfield  {author} {\bibinfo {author} {\bibfnamefont {L.}~\bibnamefont
  {Zou}}, \bibinfo {author} {\bibfnamefont {H.~C.}\ \bibnamefont {Po}},
  \bibinfo {author} {\bibfnamefont {A.}~\bibnamefont {Vishwanath}}, \ and\
  \bibinfo {author} {\bibfnamefont {T.}~\bibnamefont {Senthil}},\ }\href
  {\doibase 10.1103/PhysRevB.98.085435} {\bibfield  {journal} {\bibinfo
  {journal} {Phys. Rev. B}\ }\textbf {\bibinfo {volume} {98}},\ \bibinfo
  {pages} {085435} (\bibinfo {year} {2018})}\BibitemShut {NoStop}%
\bibitem [{\citenamefont {Song}\ \emph {et~al.}(2019)\citenamefont {Song},
  \citenamefont {Wang}, \citenamefont {Shi}, \citenamefont {Li}, \citenamefont
  {Fang},\ and\ \citenamefont {Bernevig}}]{song_all_2019}%
  \BibitemOpen
  \bibfield  {author} {\bibinfo {author} {\bibfnamefont {Z.}~\bibnamefont
  {Song}}, \bibinfo {author} {\bibfnamefont {Z.}~\bibnamefont {Wang}}, \bibinfo
  {author} {\bibfnamefont {W.}~\bibnamefont {Shi}}, \bibinfo {author}
  {\bibfnamefont {G.}~\bibnamefont {Li}}, \bibinfo {author} {\bibfnamefont
  {C.}~\bibnamefont {Fang}}, \ and\ \bibinfo {author} {\bibfnamefont {B.~A.}\
  \bibnamefont {Bernevig}},\ }\href {\doibase 10.1103/PhysRevLett.123.036401}
  {\bibfield  {journal} {\bibinfo  {journal} {Phys. Rev. Lett.}\ }\textbf
  {\bibinfo {volume} {123}},\ \bibinfo {pages} {036401} (\bibinfo {year}
  {2019})}\BibitemShut {NoStop}%
\bibitem [{\citenamefont {Po}\ \emph {et~al.}(2019)\citenamefont {Po},
  \citenamefont {Zou}, \citenamefont {Senthil},\ and\ \citenamefont
  {Vishwanath}}]{Po_2019_faithful}%
  \BibitemOpen
  \bibfield  {author} {\bibinfo {author} {\bibfnamefont {H.~C.}\ \bibnamefont
  {Po}}, \bibinfo {author} {\bibfnamefont {L.}~\bibnamefont {Zou}}, \bibinfo
  {author} {\bibfnamefont {T.}~\bibnamefont {Senthil}}, \ and\ \bibinfo
  {author} {\bibfnamefont {A.}~\bibnamefont {Vishwanath}},\ }\href {\doibase
  10.1103/PhysRevB.99.195455} {\bibfield  {journal} {\bibinfo  {journal} {Phys.
  Rev. B}\ }\textbf {\bibinfo {volume} {99}},\ \bibinfo {pages} {195455}
  (\bibinfo {year} {2019})}\BibitemShut {NoStop}%
\bibitem [{\citenamefont {Ahn}\ \emph {et~al.}(2019)\citenamefont {Ahn},
  \citenamefont {Park},\ and\ \citenamefont {Yang}}]{Ahn_2019_NNfail}%
  \BibitemOpen
  \bibfield  {author} {\bibinfo {author} {\bibfnamefont {J.}~\bibnamefont
  {Ahn}}, \bibinfo {author} {\bibfnamefont {S.}~\bibnamefont {Park}}, \ and\
  \bibinfo {author} {\bibfnamefont {B.-J.}\ \bibnamefont {Yang}},\ }\href
  {\doibase 10.1103/PhysRevX.9.021013} {\bibfield  {journal} {\bibinfo
  {journal} {Phys. Rev. X}\ }\textbf {\bibinfo {volume} {9}},\ \bibinfo {pages}
  {021013} (\bibinfo {year} {2019})}\BibitemShut {NoStop}%
\bibitem [{\citenamefont {Wang}\ \emph {et~al.}(2021)\citenamefont {Wang},
  \citenamefont {Zheng}, \citenamefont {Millis},\ and\ \citenamefont
  {Cano}}]{wang_chiral_2021}%
  \BibitemOpen
  \bibfield  {author} {\bibinfo {author} {\bibfnamefont {J.}~\bibnamefont
  {Wang}}, \bibinfo {author} {\bibfnamefont {Y.}~\bibnamefont {Zheng}},
  \bibinfo {author} {\bibfnamefont {A.~J.}\ \bibnamefont {Millis}}, \ and\
  \bibinfo {author} {\bibfnamefont {J.}~\bibnamefont {Cano}},\ }\href {\doibase
  10.1103/PhysRevResearch.3.023155} {\bibfield  {journal} {\bibinfo  {journal}
  {Phys. Rev. Res.}\ }\textbf {\bibinfo {volume} {3}},\ \bibinfo {pages}
  {023155} (\bibinfo {year} {2021})}\BibitemShut {NoStop}%
\bibitem [{\citenamefont {Song}\ and\ \citenamefont
  {Bernevig}(2022)}]{song_magic-angle_2022}%
  \BibitemOpen
  \bibfield  {author} {\bibinfo {author} {\bibfnamefont {Z.-D.}\ \bibnamefont
  {Song}}\ and\ \bibinfo {author} {\bibfnamefont {B.~A.}\ \bibnamefont
  {Bernevig}},\ }\href {\doibase 10.1103/PhysRevLett.129.047601} {\bibfield
  {journal} {\bibinfo  {journal} {Phys. Rev. Lett.}\ }\textbf {\bibinfo
  {volume} {129}},\ \bibinfo {pages} {047601} (\bibinfo {year}
  {2022})}\BibitemShut {NoStop}%
\bibitem [{\citenamefont {Shi}\ and\ \citenamefont
  {Dai}(2022)}]{shi_heavy-fermion_2022}%
  \BibitemOpen
  \bibfield  {author} {\bibinfo {author} {\bibfnamefont {H.}~\bibnamefont
  {Shi}}\ and\ \bibinfo {author} {\bibfnamefont {X.}~\bibnamefont {Dai}},\
  }\href {\doibase 10.1103/PhysRevB.106.245129} {\bibfield  {journal} {\bibinfo
   {journal} {Phys. Rev. B}\ }\textbf {\bibinfo {volume} {106}},\ \bibinfo
  {pages} {245129} (\bibinfo {year} {2022})}\BibitemShut {NoStop}%
\bibitem [{\citenamefont {Yu}\ \emph {et~al.}(2023)\citenamefont {Yu},
  \citenamefont {Xie}, \citenamefont {Bernevig},\ and\ \citenamefont
  {Das~Sarma}}]{Yu_2023_THF_TSTG}%
  \BibitemOpen
  \bibfield  {author} {\bibinfo {author} {\bibfnamefont {J.}~\bibnamefont
  {Yu}}, \bibinfo {author} {\bibfnamefont {M.}~\bibnamefont {Xie}}, \bibinfo
  {author} {\bibfnamefont {B.~A.}\ \bibnamefont {Bernevig}}, \ and\ \bibinfo
  {author} {\bibfnamefont {S.}~\bibnamefont {Das~Sarma}},\ }\href {\doibase
  10.1103/PhysRevB.108.035129} {\bibfield  {journal} {\bibinfo  {journal}
  {Phys. Rev. B}\ }\textbf {\bibinfo {volume} {108}},\ \bibinfo {pages}
  {035129} (\bibinfo {year} {2023})}\BibitemShut {NoStop}%
\bibitem [{\citenamefont {Singh}\ \emph {et~al.}(2024)\citenamefont {Singh},
  \citenamefont {Chew}, \citenamefont {Herzog-Arbeitman}, \citenamefont
  {Bernevig},\ and\ \citenamefont {Vafek}}]{singh_topological_2024}%
  \BibitemOpen
  \bibfield  {author} {\bibinfo {author} {\bibfnamefont {K.}~\bibnamefont
  {Singh}}, \bibinfo {author} {\bibfnamefont {A.}~\bibnamefont {Chew}},
  \bibinfo {author} {\bibfnamefont {J.}~\bibnamefont {Herzog-Arbeitman}},
  \bibinfo {author} {\bibfnamefont {B.~A.}\ \bibnamefont {Bernevig}}, \ and\
  \bibinfo {author} {\bibfnamefont {O.}~\bibnamefont {Vafek}},\ }\href
  {\doibase 10.1038/s41467-024-49531-3} {\bibfield  {journal} {\bibinfo
  {journal} {Nat. Commun.}\ }\textbf {\bibinfo {volume} {15}},\ \bibinfo
  {pages} {5257} (\bibinfo {year} {2024})}\BibitemShut {NoStop}%
\bibitem [{\citenamefont {Herzog-Arbeitman}\ \emph
  {et~al.}(2025{\natexlab{a}})\citenamefont {Herzog-Arbeitman}, \citenamefont
  {Yu}, \citenamefont {C\ifmmode \u{a}\else \u{a}\fi{}lug\ifmmode~\u{a}\else
  \u{a}\fi{}ru}, \citenamefont {Hu}, \citenamefont {Regnault}, \citenamefont
  {Vafek}, \citenamefont {Kang},\ and\ \citenamefont
  {Bernevig}}]{herzog_2025_efficient}%
  \BibitemOpen
  \bibfield  {author} {\bibinfo {author} {\bibfnamefont {J.}~\bibnamefont
  {Herzog-Arbeitman}}, \bibinfo {author} {\bibfnamefont {J.}~\bibnamefont
  {Yu}}, \bibinfo {author} {\bibfnamefont {D.}~\bibnamefont {C\ifmmode
  \u{a}\else \u{a}\fi{}lug\ifmmode~\u{a}\else \u{a}\fi{}ru}}, \bibinfo {author}
  {\bibfnamefont {H.}~\bibnamefont {Hu}}, \bibinfo {author} {\bibfnamefont
  {N.}~\bibnamefont {Regnault}}, \bibinfo {author} {\bibfnamefont
  {O.}~\bibnamefont {Vafek}}, \bibinfo {author} {\bibfnamefont
  {J.}~\bibnamefont {Kang}}, \ and\ \bibinfo {author} {\bibfnamefont {B.~A.}\
  \bibnamefont {Bernevig}},\ }\href {\doibase 10.1103/xv3m-vtlr} {\bibfield
  {journal} {\bibinfo  {journal} {Phys. Rev. B}\ }\textbf {\bibinfo {volume}
  {112}},\ \bibinfo {pages} {125128} (\bibinfo {year}
  {2025}{\natexlab{a}})}\BibitemShut {NoStop}%
\bibitem [{\citenamefont {Chou}\ and\ \citenamefont
  {Das~Sarma}(2023{\natexlab{a}})}]{Chou_2023_Kondo}%
  \BibitemOpen
  \bibfield  {author} {\bibinfo {author} {\bibfnamefont {Y.-Z.}\ \bibnamefont
  {Chou}}\ and\ \bibinfo {author} {\bibfnamefont {S.}~\bibnamefont
  {Das~Sarma}},\ }\href {\doibase 10.1103/PhysRevLett.131.026501} {\bibfield
  {journal} {\bibinfo  {journal} {Phys. Rev. Lett.}\ }\textbf {\bibinfo
  {volume} {131}},\ \bibinfo {pages} {026501} (\bibinfo {year}
  {2023}{\natexlab{a}})}\BibitemShut {NoStop}%
\bibitem [{\citenamefont {Zhou}\ \emph {et~al.}(2024)\citenamefont {Zhou},
  \citenamefont {Wang}, \citenamefont {Tong},\ and\ \citenamefont
  {Song}}]{zhou_kondo_2024}%
  \BibitemOpen
  \bibfield  {author} {\bibinfo {author} {\bibfnamefont {G.-D.}\ \bibnamefont
  {Zhou}}, \bibinfo {author} {\bibfnamefont {Y.-J.}\ \bibnamefont {Wang}},
  \bibinfo {author} {\bibfnamefont {N.}~\bibnamefont {Tong}}, \ and\ \bibinfo
  {author} {\bibfnamefont {Z.-D.}\ \bibnamefont {Song}},\ }\href {\doibase
  10.1103/PhysRevB.109.045419} {\bibfield  {journal} {\bibinfo  {journal}
  {Phys. Rev. B}\ }\textbf {\bibinfo {volume} {109}},\ \bibinfo {pages}
  {045419} (\bibinfo {year} {2024})}\BibitemShut {NoStop}%
\bibitem [{\citenamefont {Rai}\ \emph {et~al.}(2024)\citenamefont {Rai},
  \citenamefont {Crippa}, \citenamefont {C\ifmmode \u{a}\else
  \u{a}\fi{}lug\ifmmode~\u{a}\else \u{a}\fi{}ru}, \citenamefont {Hu},
  \citenamefont {Paoletti}, \citenamefont {de' Medici}, \citenamefont
  {Georges}, \citenamefont {Bernevig}, \citenamefont {Valent\'{\i}},
  \citenamefont {Sangiovanni},\ and\ \citenamefont {Wehling}}]{Rai_2023_DMFT}%
  \BibitemOpen
  \bibfield  {author} {\bibinfo {author} {\bibfnamefont {G.}~\bibnamefont
  {Rai}}, \bibinfo {author} {\bibfnamefont {L.}~\bibnamefont {Crippa}},
  \bibinfo {author} {\bibfnamefont {D.}~\bibnamefont {C\ifmmode \u{a}\else
  \u{a}\fi{}lug\ifmmode~\u{a}\else \u{a}\fi{}ru}}, \bibinfo {author}
  {\bibfnamefont {H.}~\bibnamefont {Hu}}, \bibinfo {author} {\bibfnamefont
  {F.}~\bibnamefont {Paoletti}}, \bibinfo {author} {\bibfnamefont
  {L.}~\bibnamefont {de' Medici}}, \bibinfo {author} {\bibfnamefont
  {A.}~\bibnamefont {Georges}}, \bibinfo {author} {\bibfnamefont {B.~A.}\
  \bibnamefont {Bernevig}}, \bibinfo {author} {\bibfnamefont {R.}~\bibnamefont
  {Valent\'{\i}}}, \bibinfo {author} {\bibfnamefont {G.}~\bibnamefont
  {Sangiovanni}}, \ and\ \bibinfo {author} {\bibfnamefont {T.}~\bibnamefont
  {Wehling}},\ }\href {\doibase 10.1103/PhysRevX.14.031045} {\bibfield
  {journal} {\bibinfo  {journal} {Phys. Rev. X}\ }\textbf {\bibinfo {volume}
  {14}},\ \bibinfo {pages} {031045} (\bibinfo {year} {2024})}\BibitemShut
  {NoStop}%
\bibitem [{\citenamefont {Hu}\ \emph {et~al.}(2023{\natexlab{a}})\citenamefont
  {Hu}, \citenamefont {Bernevig},\ and\ \citenamefont
  {Tsvelik}}]{Hu_2023_Kondo}%
  \BibitemOpen
  \bibfield  {author} {\bibinfo {author} {\bibfnamefont {H.}~\bibnamefont
  {Hu}}, \bibinfo {author} {\bibfnamefont {B.~A.}\ \bibnamefont {Bernevig}}, \
  and\ \bibinfo {author} {\bibfnamefont {A.~M.}\ \bibnamefont {Tsvelik}},\
  }\href {\doibase 10.1103/PhysRevLett.131.026502} {\bibfield  {journal}
  {\bibinfo  {journal} {Phys. Rev. Lett.}\ }\textbf {\bibinfo {volume} {131}},\
  \bibinfo {pages} {026502} (\bibinfo {year} {2023}{\natexlab{a}})}\BibitemShut
  {NoStop}%
\bibitem [{\citenamefont {Hu}\ \emph {et~al.}(2023{\natexlab{b}})\citenamefont
  {Hu}, \citenamefont {Rai}, \citenamefont {Crippa}, \citenamefont
  {Herzog-Arbeitman}, \citenamefont {C\ifmmode \u{a}\else
  \u{a}\fi{}lug\ifmmode~\u{a}\else \u{a}\fi{}ru}, \citenamefont {Wehling},
  \citenamefont {Sangiovanni}, \citenamefont {Valent\'{\i}}, \citenamefont
  {Tsvelik},\ and\ \citenamefont {Bernevig}}]{Hu_2023_Symmetric}%
  \BibitemOpen
  \bibfield  {author} {\bibinfo {author} {\bibfnamefont {H.}~\bibnamefont
  {Hu}}, \bibinfo {author} {\bibfnamefont {G.}~\bibnamefont {Rai}}, \bibinfo
  {author} {\bibfnamefont {L.}~\bibnamefont {Crippa}}, \bibinfo {author}
  {\bibfnamefont {J.}~\bibnamefont {Herzog-Arbeitman}}, \bibinfo {author}
  {\bibfnamefont {D.}~\bibnamefont {C\ifmmode \u{a}\else
  \u{a}\fi{}lug\ifmmode~\u{a}\else \u{a}\fi{}ru}}, \bibinfo {author}
  {\bibfnamefont {T.}~\bibnamefont {Wehling}}, \bibinfo {author} {\bibfnamefont
  {G.}~\bibnamefont {Sangiovanni}}, \bibinfo {author} {\bibfnamefont
  {R.}~\bibnamefont {Valent\'{\i}}}, \bibinfo {author} {\bibfnamefont {A.~M.}\
  \bibnamefont {Tsvelik}}, \ and\ \bibinfo {author} {\bibfnamefont {B.~A.}\
  \bibnamefont {Bernevig}},\ }\href {\doibase 10.1103/PhysRevLett.131.166501}
  {\bibfield  {journal} {\bibinfo  {journal} {Phys. Rev. Lett.}\ }\textbf
  {\bibinfo {volume} {131}},\ \bibinfo {pages} {166501} (\bibinfo {year}
  {2023}{\natexlab{b}})}\BibitemShut {NoStop}%
\bibitem [{\citenamefont {Chou}\ and\ \citenamefont
  {Das~Sarma}(2023{\natexlab{b}})}]{Chou_2023_scaling}%
  \BibitemOpen
  \bibfield  {author} {\bibinfo {author} {\bibfnamefont {Y.-Z.}\ \bibnamefont
  {Chou}}\ and\ \bibinfo {author} {\bibfnamefont {S.}~\bibnamefont
  {Das~Sarma}},\ }\href {\doibase 10.1103/PhysRevB.108.125106} {\bibfield
  {journal} {\bibinfo  {journal} {Phys. Rev. B}\ }\textbf {\bibinfo {volume}
  {108}},\ \bibinfo {pages} {125106} (\bibinfo {year}
  {2023}{\natexlab{b}})}\BibitemShut {NoStop}%
\bibitem [{\citenamefont {Datta}\ \emph {et~al.}(2023)\citenamefont {Datta},
  \citenamefont {Calderón}, \citenamefont {Camjayi},\ and\ \citenamefont
  {Bascones}}]{Datta_2023_heavy}%
  \BibitemOpen
  \bibfield  {author} {\bibinfo {author} {\bibfnamefont {A.}~\bibnamefont
  {Datta}}, \bibinfo {author} {\bibfnamefont {M.~J.}\ \bibnamefont
  {Calderón}}, \bibinfo {author} {\bibfnamefont {A.}~\bibnamefont {Camjayi}},
  \ and\ \bibinfo {author} {\bibfnamefont {E.}~\bibnamefont {Bascones}},\
  }\href {\doibase 10.1038/s41467-023-40754-4} {\bibfield  {journal} {\bibinfo
  {journal} {Nat. Commun.}\ }\textbf {\bibinfo {volume} {14}},\ \bibinfo
  {pages} {5036} (\bibinfo {year} {2023})}\BibitemShut {NoStop}%
\bibitem [{\citenamefont {Lau}\ and\ \citenamefont
  {Coleman}(2025)}]{Lau_2023_topological}%
  \BibitemOpen
  \bibfield  {author} {\bibinfo {author} {\bibfnamefont {L.~L.}\ \bibnamefont
  {Lau}}\ and\ \bibinfo {author} {\bibfnamefont {P.}~\bibnamefont {Coleman}},\
  }\href {\doibase 10.1103/PhysRevX.15.021028} {\bibfield  {journal} {\bibinfo
  {journal} {Phys. Rev. X}\ }\textbf {\bibinfo {volume} {15}},\ \bibinfo
  {pages} {021028} (\bibinfo {year} {2025})}\BibitemShut {NoStop}%
\bibitem [{\citenamefont {Călugăru}\ \emph {et~al.}(2024)\citenamefont
  {Călugăru}, \citenamefont {Hu}, \citenamefont {Merino}, \citenamefont
  {Regnault}, \citenamefont {Efetov},\ and\ \citenamefont
  {Bernevig}}]{calugaru_thermoelectric_2024}%
  \BibitemOpen
  \bibfield  {author} {\bibinfo {author} {\bibfnamefont {D.}~\bibnamefont
  {Călugăru}}, \bibinfo {author} {\bibfnamefont {H.}~\bibnamefont {Hu}},
  \bibinfo {author} {\bibfnamefont {R.~L.}\ \bibnamefont {Merino}}, \bibinfo
  {author} {\bibfnamefont {N.}~\bibnamefont {Regnault}}, \bibinfo {author}
  {\bibfnamefont {D.~K.}\ \bibnamefont {Efetov}}, \ and\ \bibinfo {author}
  {\bibfnamefont {B.~A.}\ \bibnamefont {Bernevig}},\ }\href
  {http://arxiv.org/abs/2402.14057} {\enquote {\bibinfo {title} {The
  {Thermoelectric} {Effect} and {Its} {Natural} {Heavy} {Fermion} {Explanation}
  in {Twisted} {Bilayer} and {Trilayer} {Graphene}},}\ } (\bibinfo {year}
  {2024}),\ \bibinfo {note} {arXiv:2402.14057 [cond-mat]}\BibitemShut {NoStop}%
\bibitem [{\citenamefont {Herzog-Arbeitman}\ \emph
  {et~al.}(2025{\natexlab{b}})\citenamefont {Herzog-Arbeitman}, \citenamefont
  {Călugăru}, \citenamefont {Hu}, \citenamefont {Yu}, \citenamefont
  {Regnault}, \citenamefont {Kang}, \citenamefont {Bernevig},\ and\
  \citenamefont {Vafek}}]{herzog_2025_kekule}%
  \BibitemOpen
  \bibfield  {author} {\bibinfo {author} {\bibfnamefont {J.}~\bibnamefont
  {Herzog-Arbeitman}}, \bibinfo {author} {\bibfnamefont {D.}~\bibnamefont
  {Călugăru}}, \bibinfo {author} {\bibfnamefont {H.}~\bibnamefont {Hu}},
  \bibinfo {author} {\bibfnamefont {J.}~\bibnamefont {Yu}}, \bibinfo {author}
  {\bibfnamefont {N.}~\bibnamefont {Regnault}}, \bibinfo {author}
  {\bibfnamefont {J.}~\bibnamefont {Kang}}, \bibinfo {author} {\bibfnamefont
  {B.~A.}\ \bibnamefont {Bernevig}}, \ and\ \bibinfo {author} {\bibfnamefont
  {O.}~\bibnamefont {Vafek}},\ }\href {\doibase 10.1103/rr5g-3js8} {\bibfield
  {journal} {\bibinfo  {journal} {Phys. Rev. B}\ }\textbf {\bibinfo {volume}
  {112}},\ \bibinfo {pages} {125129} (\bibinfo {year}
  {2025}{\natexlab{b}})}\BibitemShut {NoStop}%
\bibitem [{\citenamefont {Crippa}\ \emph {et~al.}(2025)\citenamefont {Crippa},
  \citenamefont {Rai}, \citenamefont {Călugăru}, \citenamefont {Hu},
  \citenamefont {Herzog-Arbeitman}, \citenamefont {Bernevig}, \citenamefont
  {Valentí}, \citenamefont {Sangiovanni},\ and\ \citenamefont
  {Wehling}}]{crippa_2025_dynamicalcorrelation}%
  \BibitemOpen
  \bibfield  {author} {\bibinfo {author} {\bibfnamefont {L.}~\bibnamefont
  {Crippa}}, \bibinfo {author} {\bibfnamefont {G.}~\bibnamefont {Rai}},
  \bibinfo {author} {\bibfnamefont {D.}~\bibnamefont {Călugăru}}, \bibinfo
  {author} {\bibfnamefont {H.}~\bibnamefont {Hu}}, \bibinfo {author}
  {\bibfnamefont {J.}~\bibnamefont {Herzog-Arbeitman}}, \bibinfo {author}
  {\bibfnamefont {B.~A.}\ \bibnamefont {Bernevig}}, \bibinfo {author}
  {\bibfnamefont {R.}~\bibnamefont {Valentí}}, \bibinfo {author}
  {\bibfnamefont {G.}~\bibnamefont {Sangiovanni}}, \ and\ \bibinfo {author}
  {\bibfnamefont {T.}~\bibnamefont {Wehling}},\ }\href
  {https://arxiv.org/abs/2509.19436} {\enquote {\bibinfo {title} {Dynamical
  correlation effects in twisted bilayer graphene under strain and lattice
  relaxation},}\ } (\bibinfo {year} {2025}),\ \Eprint
  {http://arxiv.org/abs/2509.19436} {arXiv:2509.19436 [cond-mat.str-el]}
  \BibitemShut {NoStop}%
\bibitem [{\citenamefont {Călugăru}\ \emph {et~al.}(2025)\citenamefont
  {Călugăru}, \citenamefont {Hu}, \citenamefont {Crippa}, \citenamefont
  {Rai}, \citenamefont {Regnault}, \citenamefont {Wehling}, \citenamefont
  {Valentí}, \citenamefont {Sangiovanni},\ and\ \citenamefont
  {Bernevig}}]{calugaru_2025_obtainingspectral}%
  \BibitemOpen
  \bibfield  {author} {\bibinfo {author} {\bibfnamefont {D.}~\bibnamefont
  {Călugăru}}, \bibinfo {author} {\bibfnamefont {H.}~\bibnamefont {Hu}},
  \bibinfo {author} {\bibfnamefont {L.}~\bibnamefont {Crippa}}, \bibinfo
  {author} {\bibfnamefont {G.}~\bibnamefont {Rai}}, \bibinfo {author}
  {\bibfnamefont {N.}~\bibnamefont {Regnault}}, \bibinfo {author}
  {\bibfnamefont {T.~O.}\ \bibnamefont {Wehling}}, \bibinfo {author}
  {\bibfnamefont {R.}~\bibnamefont {Valentí}}, \bibinfo {author}
  {\bibfnamefont {G.}~\bibnamefont {Sangiovanni}}, \ and\ \bibinfo {author}
  {\bibfnamefont {B.~A.}\ \bibnamefont {Bernevig}},\ }\href
  {https://arxiv.org/abs/2509.18256} {\enquote {\bibinfo {title} {Obtaining the
  spectral function of moir\'e graphene heavy-fermions using iterative
  perturbation theory},}\ } (\bibinfo {year} {2025}),\ \Eprint
  {http://arxiv.org/abs/2509.18256} {arXiv:2509.18256 [cond-mat.str-el]}
  \BibitemShut {NoStop}%
\bibitem [{\citenamefont {Rozen}\ \emph {et~al.}(2021)\citenamefont {Rozen},
  \citenamefont {Park}, \citenamefont {Zondiner}, \citenamefont {Cao},
  \citenamefont {Rodan-Legrain}, \citenamefont {Taniguchi}, \citenamefont
  {Watanabe}, \citenamefont {Oreg}, \citenamefont {Stern}, \citenamefont
  {Berg}, \citenamefont {Jarillo-Herrero},\ and\ \citenamefont
  {Ilani}}]{Rozen_2021_entropic}%
  \BibitemOpen
  \bibfield  {author} {\bibinfo {author} {\bibfnamefont {A.}~\bibnamefont
  {Rozen}}, \bibinfo {author} {\bibfnamefont {J.~M.}\ \bibnamefont {Park}},
  \bibinfo {author} {\bibfnamefont {U.}~\bibnamefont {Zondiner}}, \bibinfo
  {author} {\bibfnamefont {Y.}~\bibnamefont {Cao}}, \bibinfo {author}
  {\bibfnamefont {D.}~\bibnamefont {Rodan-Legrain}}, \bibinfo {author}
  {\bibfnamefont {T.}~\bibnamefont {Taniguchi}}, \bibinfo {author}
  {\bibfnamefont {K.}~\bibnamefont {Watanabe}}, \bibinfo {author}
  {\bibfnamefont {Y.}~\bibnamefont {Oreg}}, \bibinfo {author} {\bibfnamefont
  {A.}~\bibnamefont {Stern}}, \bibinfo {author} {\bibfnamefont
  {E.}~\bibnamefont {Berg}}, \bibinfo {author} {\bibfnamefont {P.}~\bibnamefont
  {Jarillo-Herrero}}, \ and\ \bibinfo {author} {\bibfnamefont {S.}~\bibnamefont
  {Ilani}},\ }\href {\doibase 10.1038/s41586-021-03319-3} {\bibfield  {journal}
  {\bibinfo  {journal} {Nature}\ }\textbf {\bibinfo {volume} {592}},\ \bibinfo
  {pages} {214} (\bibinfo {year} {2021})}\BibitemShut {NoStop}%
\bibitem [{\citenamefont {Saito}\ \emph {et~al.}(2021)\citenamefont {Saito},
  \citenamefont {Yang}, \citenamefont {Ge}, \citenamefont {Liu}, \citenamefont
  {Taniguchi}, \citenamefont {Watanabe}, \citenamefont {Li}, \citenamefont
  {Berg},\ and\ \citenamefont {Young}}]{Saito_2021_isospin}%
  \BibitemOpen
  \bibfield  {author} {\bibinfo {author} {\bibfnamefont {Y.}~\bibnamefont
  {Saito}}, \bibinfo {author} {\bibfnamefont {F.}~\bibnamefont {Yang}},
  \bibinfo {author} {\bibfnamefont {J.}~\bibnamefont {Ge}}, \bibinfo {author}
  {\bibfnamefont {X.}~\bibnamefont {Liu}}, \bibinfo {author} {\bibfnamefont
  {T.}~\bibnamefont {Taniguchi}}, \bibinfo {author} {\bibfnamefont
  {K.}~\bibnamefont {Watanabe}}, \bibinfo {author} {\bibfnamefont {J.~I.~A.}\
  \bibnamefont {Li}}, \bibinfo {author} {\bibfnamefont {E.}~\bibnamefont
  {Berg}}, \ and\ \bibinfo {author} {\bibfnamefont {A.~F.}\ \bibnamefont
  {Young}},\ }\href {\doibase 10.1038/s41586-021-03409-2} {\bibfield  {journal}
  {\bibinfo  {journal} {Nature}\ }\textbf {\bibinfo {volume} {592}},\ \bibinfo
  {pages} {220} (\bibinfo {year} {2021})}\BibitemShut {NoStop}%
\bibitem [{\citenamefont {Zhang}\ \emph {et~al.}(2025)\citenamefont {Zhang},
  \citenamefont {Wu}, \citenamefont {Călugăru}, \citenamefont {Hu},
  \citenamefont {Taniguchi}, \citenamefont {Wanatabe}, \citenamefont
  {Bernevig},\ and\ \citenamefont {Andrei}}]{zhang_2025_heavyfermions}%
  \BibitemOpen
  \bibfield  {author} {\bibinfo {author} {\bibfnamefont {Z.}~\bibnamefont
  {Zhang}}, \bibinfo {author} {\bibfnamefont {S.}~\bibnamefont {Wu}}, \bibinfo
  {author} {\bibfnamefont {D.}~\bibnamefont {Călugăru}}, \bibinfo {author}
  {\bibfnamefont {H.}~\bibnamefont {Hu}}, \bibinfo {author} {\bibfnamefont
  {T.}~\bibnamefont {Taniguchi}}, \bibinfo {author} {\bibfnamefont
  {K.}~\bibnamefont {Wanatabe}}, \bibinfo {author} {\bibfnamefont {A.~B.}\
  \bibnamefont {Bernevig}}, \ and\ \bibinfo {author} {\bibfnamefont {E.~Y.}\
  \bibnamefont {Andrei}},\ }\href {https://arxiv.org/abs/2503.17875} {\enquote
  {\bibinfo {title} {Heavy fermions, mass renormalization and local moments in
  magic-angle twisted bilayer graphene via planar tunneling spectroscopy},}\ }
  (\bibinfo {year} {2025}),\ \Eprint {http://arxiv.org/abs/2503.17875}
  {arXiv:2503.17875 [cond-mat.mes-hall]} \BibitemShut {NoStop}%
\bibitem [{\citenamefont {Zondiner}\ \emph {et~al.}(2020)\citenamefont
  {Zondiner}, \citenamefont {Rozen}, \citenamefont {Rodan-Legrain},
  \citenamefont {Cao}, \citenamefont {Queiroz}, \citenamefont {Taniguchi},
  \citenamefont {Watanabe}, \citenamefont {Oreg}, \citenamefont {von Oppen},
  \citenamefont {Stern}, \citenamefont {Berg}, \citenamefont
  {Jarillo-Herrero},\ and\ \citenamefont {Ilani}}]{Zondiner_2020_cascade}%
  \BibitemOpen
  \bibfield  {author} {\bibinfo {author} {\bibfnamefont {U.}~\bibnamefont
  {Zondiner}}, \bibinfo {author} {\bibfnamefont {A.}~\bibnamefont {Rozen}},
  \bibinfo {author} {\bibfnamefont {D.}~\bibnamefont {Rodan-Legrain}}, \bibinfo
  {author} {\bibfnamefont {Y.}~\bibnamefont {Cao}}, \bibinfo {author}
  {\bibfnamefont {R.}~\bibnamefont {Queiroz}}, \bibinfo {author} {\bibfnamefont
  {T.}~\bibnamefont {Taniguchi}}, \bibinfo {author} {\bibfnamefont
  {K.}~\bibnamefont {Watanabe}}, \bibinfo {author} {\bibfnamefont
  {Y.}~\bibnamefont {Oreg}}, \bibinfo {author} {\bibfnamefont {F.}~\bibnamefont
  {von Oppen}}, \bibinfo {author} {\bibfnamefont {A.}~\bibnamefont {Stern}},
  \bibinfo {author} {\bibfnamefont {E.}~\bibnamefont {Berg}}, \bibinfo {author}
  {\bibfnamefont {P.}~\bibnamefont {Jarillo-Herrero}}, \ and\ \bibinfo {author}
  {\bibfnamefont {S.}~\bibnamefont {Ilani}},\ }\href {\doibase
  10.1038/s41586-020-2373-y} {\bibfield  {journal} {\bibinfo  {journal}
  {Nature}\ }\textbf {\bibinfo {volume} {582}},\ \bibinfo {pages} {203}
  (\bibinfo {year} {2020})}\BibitemShut {NoStop}%
\bibitem [{\citenamefont {Wong}\ \emph {et~al.}(2020)\citenamefont {Wong},
  \citenamefont {Nuckolls}, \citenamefont {Oh}, \citenamefont {Lian},
  \citenamefont {Xie}, \citenamefont {Jeon}, \citenamefont {Watanabe},
  \citenamefont {Taniguchi}, \citenamefont {Bernevig},\ and\ \citenamefont
  {Yazdani}}]{Wong_2020_cascade}%
  \BibitemOpen
  \bibfield  {author} {\bibinfo {author} {\bibfnamefont {D.}~\bibnamefont
  {Wong}}, \bibinfo {author} {\bibfnamefont {K.~P.}\ \bibnamefont {Nuckolls}},
  \bibinfo {author} {\bibfnamefont {M.}~\bibnamefont {Oh}}, \bibinfo {author}
  {\bibfnamefont {B.}~\bibnamefont {Lian}}, \bibinfo {author} {\bibfnamefont
  {Y.}~\bibnamefont {Xie}}, \bibinfo {author} {\bibfnamefont {S.}~\bibnamefont
  {Jeon}}, \bibinfo {author} {\bibfnamefont {K.}~\bibnamefont {Watanabe}},
  \bibinfo {author} {\bibfnamefont {T.}~\bibnamefont {Taniguchi}}, \bibinfo
  {author} {\bibfnamefont {B.~A.}\ \bibnamefont {Bernevig}}, \ and\ \bibinfo
  {author} {\bibfnamefont {A.}~\bibnamefont {Yazdani}},\ }\href {\doibase
  10.1038/s41586-020-2339-0} {\bibfield  {journal} {\bibinfo  {journal}
  {Nature}\ }\textbf {\bibinfo {volume} {582}},\ \bibinfo {pages} {198}
  (\bibinfo {year} {2020})}\BibitemShut {NoStop}%
\bibitem [{\citenamefont {Hu}\ \emph {et~al.}(2024)\citenamefont {Hu},
  \citenamefont {Liang}, \citenamefont {Li}, \citenamefont {Shi}, \citenamefont
  {Dai},\ and\ \citenamefont {Xu}}]{hu_2024_linkcascade}%
  \BibitemOpen
  \bibfield  {author} {\bibinfo {author} {\bibfnamefont {Q.}~\bibnamefont
  {Hu}}, \bibinfo {author} {\bibfnamefont {S.}~\bibnamefont {Liang}}, \bibinfo
  {author} {\bibfnamefont {X.}~\bibnamefont {Li}}, \bibinfo {author}
  {\bibfnamefont {H.}~\bibnamefont {Shi}}, \bibinfo {author} {\bibfnamefont
  {X.}~\bibnamefont {Dai}}, \ and\ \bibinfo {author} {\bibfnamefont
  {Y.}~\bibnamefont {Xu}},\ }\href {https://arxiv.org/abs/2406.08734} {\enquote
  {\bibinfo {title} {Link between cascade transitions and correlated chern
  insulators in magic-angle twisted bilayer graphene},}\ } (\bibinfo {year}
  {2024}),\ \Eprint {http://arxiv.org/abs/2406.08734} {arXiv:2406.08734
  [cond-mat.mes-hall]} \BibitemShut {NoStop}%
\bibitem [{\citenamefont {Merino}\ \emph {et~al.}(2025)\citenamefont {Merino},
  \citenamefont {Călugăru}, \citenamefont {Hu}, \citenamefont
  {Díez-Mérida}, \citenamefont {Díez-Carlón}, \citenamefont {Taniguchi},
  \citenamefont {Watanabe}, \citenamefont {Seifert}, \citenamefont {Bernevig},\
  and\ \citenamefont {Efetov}}]{merino_interplay_2025}%
  \BibitemOpen
  \bibfield  {author} {\bibinfo {author} {\bibfnamefont {R.~L.}\ \bibnamefont
  {Merino}}, \bibinfo {author} {\bibfnamefont {D.}~\bibnamefont {Călugăru}},
  \bibinfo {author} {\bibfnamefont {H.}~\bibnamefont {Hu}}, \bibinfo {author}
  {\bibfnamefont {J.}~\bibnamefont {Díez-Mérida}}, \bibinfo {author}
  {\bibfnamefont {A.}~\bibnamefont {Díez-Carlón}}, \bibinfo {author}
  {\bibfnamefont {T.}~\bibnamefont {Taniguchi}}, \bibinfo {author}
  {\bibfnamefont {K.}~\bibnamefont {Watanabe}}, \bibinfo {author}
  {\bibfnamefont {P.}~\bibnamefont {Seifert}}, \bibinfo {author} {\bibfnamefont
  {B.~A.}\ \bibnamefont {Bernevig}}, \ and\ \bibinfo {author} {\bibfnamefont
  {D.~K.}\ \bibnamefont {Efetov}},\ }\href {\doibase
  10.1038/s41567-025-02912-x} {\bibfield  {journal} {\bibinfo  {journal} {Nat.
  Phys.}\ }\textbf {\bibinfo {volume} {21}},\ \bibinfo {pages} {1078} (\bibinfo
  {year} {2025})}\BibitemShut {NoStop}%
\bibitem [{\citenamefont {Xiao}\ \emph {et~al.}(2025)\citenamefont {Xiao},
  \citenamefont {Inbar}, \citenamefont {Birkbeck}, \citenamefont {Gershon},
  \citenamefont {Zamir}, \citenamefont {Taniguchi}, \citenamefont {Watanabe},
  \citenamefont {Berg},\ and\ \citenamefont {Ilani}}]{xiao_2025_interacting}%
  \BibitemOpen
  \bibfield  {author} {\bibinfo {author} {\bibfnamefont {J.}~\bibnamefont
  {Xiao}}, \bibinfo {author} {\bibfnamefont {A.}~\bibnamefont {Inbar}},
  \bibinfo {author} {\bibfnamefont {J.}~\bibnamefont {Birkbeck}}, \bibinfo
  {author} {\bibfnamefont {N.}~\bibnamefont {Gershon}}, \bibinfo {author}
  {\bibfnamefont {Y.}~\bibnamefont {Zamir}}, \bibinfo {author} {\bibfnamefont
  {T.}~\bibnamefont {Taniguchi}}, \bibinfo {author} {\bibfnamefont
  {K.}~\bibnamefont {Watanabe}}, \bibinfo {author} {\bibfnamefont
  {E.}~\bibnamefont {Berg}}, \ and\ \bibinfo {author} {\bibfnamefont
  {S.}~\bibnamefont {Ilani}},\ }\href {https://arxiv.org/abs/2506.20738}
  {\enquote {\bibinfo {title} {The interacting energy bands of magic angle
  twisted bilayer graphene revealed by the quantum twisting microscope},}\ }
  (\bibinfo {year} {2025}),\ \Eprint {http://arxiv.org/abs/2506.20738}
  {arXiv:2506.20738 [cond-mat.mes-hall]} \BibitemShut {NoStop}%
\bibitem [{\citenamefont {Wu}\ \emph {et~al.}(2018{\natexlab{b}})\citenamefont
  {Wu}, \citenamefont {MacDonald},\ and\ \citenamefont
  {Martin}}]{Wu_2018_SCop}%
  \BibitemOpen
  \bibfield  {author} {\bibinfo {author} {\bibfnamefont {F.}~\bibnamefont
  {Wu}}, \bibinfo {author} {\bibfnamefont {A.~H.}\ \bibnamefont {MacDonald}}, \
  and\ \bibinfo {author} {\bibfnamefont {I.}~\bibnamefont {Martin}},\ }\href
  {\doibase 10.1103/PhysRevLett.121.257001} {\bibfield  {journal} {\bibinfo
  {journal} {Phys. Rev. Lett.}\ }\textbf {\bibinfo {volume} {121}},\ \bibinfo
  {pages} {257001} (\bibinfo {year} {2018}{\natexlab{b}})}\BibitemShut
  {NoStop}%
\bibitem [{\citenamefont {Lian}\ \emph {et~al.}(2019)\citenamefont {Lian},
  \citenamefont {Wang},\ and\ \citenamefont {Bernevig}}]{Lian_2019_SCac}%
  \BibitemOpen
  \bibfield  {author} {\bibinfo {author} {\bibfnamefont {B.}~\bibnamefont
  {Lian}}, \bibinfo {author} {\bibfnamefont {Z.}~\bibnamefont {Wang}}, \ and\
  \bibinfo {author} {\bibfnamefont {B.~A.}\ \bibnamefont {Bernevig}},\ }\href
  {\doibase 10.1103/PhysRevLett.122.257002} {\bibfield  {journal} {\bibinfo
  {journal} {Phys. Rev. Lett.}\ }\textbf {\bibinfo {volume} {122}},\ \bibinfo
  {pages} {257002} (\bibinfo {year} {2019})}\BibitemShut {NoStop}%
\bibitem [{\citenamefont {Koshino}\ and\ \citenamefont
  {Nam}(2020)}]{Koshino_epc_2020}%
  \BibitemOpen
  \bibfield  {author} {\bibinfo {author} {\bibfnamefont {M.}~\bibnamefont
  {Koshino}}\ and\ \bibinfo {author} {\bibfnamefont {N.~N.~T.}\ \bibnamefont
  {Nam}},\ }\href {\doibase 10.1103/PhysRevB.101.195425} {\bibfield  {journal}
  {\bibinfo  {journal} {Phys. Rev. B}\ }\textbf {\bibinfo {volume} {101}},\
  \bibinfo {pages} {195425} (\bibinfo {year} {2020})}\BibitemShut {NoStop}%
\bibitem [{\citenamefont {Cea}\ and\ \citenamefont
  {Guinea}(2021)}]{Cea_2021_coulomb}%
  \BibitemOpen
  \bibfield  {author} {\bibinfo {author} {\bibfnamefont {T.}~\bibnamefont
  {Cea}}\ and\ \bibinfo {author} {\bibfnamefont {F.}~\bibnamefont {Guinea}},\
  }\href {\doibase 10.1073/pnas.2107874118} {\bibfield  {journal} {\bibinfo
  {journal} {Proceedings of the National Academy of Sciences}\ }\textbf
  {\bibinfo {volume} {118}},\ \bibinfo {pages} {e2107874118} (\bibinfo {year}
  {2021})}\BibitemShut {NoStop}%
\bibitem [{\citenamefont {Liu}\ \emph {et~al.}(2022{\natexlab{b}})\citenamefont
  {Liu}, \citenamefont {Peng}, \citenamefont {Sun},\ and\ \citenamefont
  {Liu}}]{liu_moire_2022}%
  \BibitemOpen
  \bibfield  {author} {\bibinfo {author} {\bibfnamefont {X.}~\bibnamefont
  {Liu}}, \bibinfo {author} {\bibfnamefont {R.}~\bibnamefont {Peng}}, \bibinfo
  {author} {\bibfnamefont {Z.}~\bibnamefont {Sun}}, \ and\ \bibinfo {author}
  {\bibfnamefont {J.}~\bibnamefont {Liu}},\ }\href {\doibase
  10.1021/acs.nanolett.2c02010} {\bibfield  {journal} {\bibinfo  {journal}
  {Nano Letters}\ }\textbf {\bibinfo {volume} {22}},\ \bibinfo {pages} {7791}
  (\bibinfo {year} {2022}{\natexlab{b}})}\BibitemShut {NoStop}%
\bibitem [{\citenamefont {Liu}\ \emph {et~al.}(2024)\citenamefont {Liu},
  \citenamefont {Chen}, \citenamefont {Yazdani},\ and\ \citenamefont
  {Bernevig}}]{Liu_2023_electronkphonon}%
  \BibitemOpen
  \bibfield  {author} {\bibinfo {author} {\bibfnamefont {C.-X.}\ \bibnamefont
  {Liu}}, \bibinfo {author} {\bibfnamefont {Y.}~\bibnamefont {Chen}}, \bibinfo
  {author} {\bibfnamefont {A.}~\bibnamefont {Yazdani}}, \ and\ \bibinfo
  {author} {\bibfnamefont {B.~A.}\ \bibnamefont {Bernevig}},\ }\href {\doibase
  10.1103/PhysRevB.110.045133} {\bibfield  {journal} {\bibinfo  {journal}
  {Phys. Rev. B}\ }\textbf {\bibinfo {volume} {110}},\ \bibinfo {pages}
  {045133} (\bibinfo {year} {2024})}\BibitemShut {NoStop}%
\bibitem [{\citenamefont {Zhu}\ and\ \citenamefont
  {Devereaux}(2025)}]{zhu_2025_microscopictheory}%
  \BibitemOpen
  \bibfield  {author} {\bibinfo {author} {\bibfnamefont {Z.}~\bibnamefont
  {Zhu}}\ and\ \bibinfo {author} {\bibfnamefont {T.~P.}\ \bibnamefont
  {Devereaux}},\ }\href {https://arxiv.org/abs/2407.03293} {\enquote {\bibinfo
  {title} {Microscopic theory for electron-phonon coupling in twisted bilayer
  graphene},}\ } (\bibinfo {year} {2025}),\ \Eprint
  {http://arxiv.org/abs/2407.03293} {arXiv:2407.03293 [cond-mat.mes-hall]}
  \BibitemShut {NoStop}%
\bibitem [{\citenamefont {Lau}\ \emph {et~al.}(2025)\citenamefont {Lau},
  \citenamefont {Gleis}, \citenamefont {Kaplan}, \citenamefont {Chandra},\ and\
  \citenamefont {Coleman}}]{lau_oscillate_2025}%
  \BibitemOpen
  \bibfield  {author} {\bibinfo {author} {\bibfnamefont {L.~L.~H.}\
  \bibnamefont {Lau}}, \bibinfo {author} {\bibfnamefont {A.}~\bibnamefont
  {Gleis}}, \bibinfo {author} {\bibfnamefont {D.}~\bibnamefont {Kaplan}},
  \bibinfo {author} {\bibfnamefont {P.}~\bibnamefont {Chandra}}, \ and\
  \bibinfo {author} {\bibfnamefont {P.}~\bibnamefont {Coleman}},\ }\href
  {\doibase 10.1103/xxyt-4bql} {\bibfield  {journal} {\bibinfo  {journal}
  {Phys. Rev. B}\ }\textbf {\bibinfo {volume} {111}},\ \bibinfo {pages}
  {245149} (\bibinfo {year} {2025})}\BibitemShut {NoStop}%
\bibitem [{\citenamefont {Dodaro}\ \emph {et~al.}(2018)\citenamefont {Dodaro},
  \citenamefont {Kivelson}, \citenamefont {Schattner}, \citenamefont {Sun},\
  and\ \citenamefont {Wang}}]{Dodaro_2018_phases}%
  \BibitemOpen
  \bibfield  {author} {\bibinfo {author} {\bibfnamefont {J.~F.}\ \bibnamefont
  {Dodaro}}, \bibinfo {author} {\bibfnamefont {S.~A.}\ \bibnamefont
  {Kivelson}}, \bibinfo {author} {\bibfnamefont {Y.}~\bibnamefont {Schattner}},
  \bibinfo {author} {\bibfnamefont {X.~Q.}\ \bibnamefont {Sun}}, \ and\
  \bibinfo {author} {\bibfnamefont {C.}~\bibnamefont {Wang}},\ }\href {\doibase
  10.1103/PhysRevB.98.075154} {\bibfield  {journal} {\bibinfo  {journal} {Phys.
  Rev. B}\ }\textbf {\bibinfo {volume} {98}},\ \bibinfo {pages} {075154}
  (\bibinfo {year} {2018})}\BibitemShut {NoStop}%
\bibitem [{\citenamefont {Angeli}\ \emph {et~al.}(2019)\citenamefont {Angeli},
  \citenamefont {Tosatti},\ and\ \citenamefont
  {Fabrizio}}]{Angeli_2019_valleyJT}%
  \BibitemOpen
  \bibfield  {author} {\bibinfo {author} {\bibfnamefont {M.}~\bibnamefont
  {Angeli}}, \bibinfo {author} {\bibfnamefont {E.}~\bibnamefont {Tosatti}}, \
  and\ \bibinfo {author} {\bibfnamefont {M.}~\bibnamefont {Fabrizio}},\ }\href
  {\doibase 10.1103/PhysRevX.9.041010} {\bibfield  {journal} {\bibinfo
  {journal} {Phys. Rev. X}\ }\textbf {\bibinfo {volume} {9}},\ \bibinfo {pages}
  {041010} (\bibinfo {year} {2019})}\BibitemShut {NoStop}%
\bibitem [{\citenamefont {Blason}\ and\ \citenamefont
  {Fabrizio}(2022)}]{Andrea_2022_local}%
  \BibitemOpen
  \bibfield  {author} {\bibinfo {author} {\bibfnamefont {A.}~\bibnamefont
  {Blason}}\ and\ \bibinfo {author} {\bibfnamefont {M.}~\bibnamefont
  {Fabrizio}},\ }\href {\doibase 10.1103/PhysRevB.106.235112} {\bibfield
  {journal} {\bibinfo  {journal} {Phys. Rev. B}\ }\textbf {\bibinfo {volume}
  {106}},\ \bibinfo {pages} {235112} (\bibinfo {year} {2022})}\BibitemShut
  {NoStop}%
\bibitem [{\citenamefont {Wang}\ \emph {et~al.}(2025)\citenamefont {Wang},
  \citenamefont {Zhou}, \citenamefont {Lian},\ and\ \citenamefont
  {Song}}]{wang_2025_epc}%
  \BibitemOpen
  \bibfield  {author} {\bibinfo {author} {\bibfnamefont {Y.-J.}\ \bibnamefont
  {Wang}}, \bibinfo {author} {\bibfnamefont {G.-D.}\ \bibnamefont {Zhou}},
  \bibinfo {author} {\bibfnamefont {B.}~\bibnamefont {Lian}}, \ and\ \bibinfo
  {author} {\bibfnamefont {Z.-D.}\ \bibnamefont {Song}},\ }\href {\doibase
  10.1103/PhysRevB.111.035110} {\bibfield  {journal} {\bibinfo  {journal}
  {Phys. Rev. B}\ }\textbf {\bibinfo {volume} {111}},\ \bibinfo {pages}
  {035110} (\bibinfo {year} {2025})}\BibitemShut {NoStop}%
\bibitem [{\citenamefont {Shi}\ \emph {et~al.}(2025)\citenamefont {Shi},
  \citenamefont {Miao},\ and\ \citenamefont {Dai}}]{shi_2025_optical}%
  \BibitemOpen
  \bibfield  {author} {\bibinfo {author} {\bibfnamefont {H.}~\bibnamefont
  {Shi}}, \bibinfo {author} {\bibfnamefont {W.}~\bibnamefont {Miao}}, \ and\
  \bibinfo {author} {\bibfnamefont {X.}~\bibnamefont {Dai}},\ }\href {\doibase
  10.1103/PhysRevB.111.155126} {\bibfield  {journal} {\bibinfo  {journal}
  {Phys. Rev. B}\ }\textbf {\bibinfo {volume} {111}},\ \bibinfo {pages}
  {155126} (\bibinfo {year} {2025})}\BibitemShut {NoStop}%
\bibitem [{\citenamefont {Wang}\ \emph {et~al.}(2024)\citenamefont {Wang},
  \citenamefont {Zhou}, \citenamefont {Peng}, \citenamefont {Lian},\ and\
  \citenamefont {Song}}]{wang_molecular_2024}%
  \BibitemOpen
  \bibfield  {author} {\bibinfo {author} {\bibfnamefont {Y.-J.}\ \bibnamefont
  {Wang}}, \bibinfo {author} {\bibfnamefont {G.-D.}\ \bibnamefont {Zhou}},
  \bibinfo {author} {\bibfnamefont {S.-Y.}\ \bibnamefont {Peng}}, \bibinfo
  {author} {\bibfnamefont {B.}~\bibnamefont {Lian}}, \ and\ \bibinfo {author}
  {\bibfnamefont {Z.-D.}\ \bibnamefont {Song}},\ }\href {\doibase
  10.1103/PhysRevLett.133.146001} {\bibfield  {journal} {\bibinfo  {journal}
  {Phys. Rev. Lett.}\ }\textbf {\bibinfo {volume} {133}},\ \bibinfo {pages}
  {146001} (\bibinfo {year} {2024})}\BibitemShut {NoStop}%
\bibitem [{\citenamefont {Youn}\ \emph {et~al.}(2024)\citenamefont {Youn},
  \citenamefont {Goh}, \citenamefont {Zhou}, \citenamefont {Song},\ and\
  \citenamefont {Lee}}]{youn_hundness_2024}%
  \BibitemOpen
  \bibfield  {author} {\bibinfo {author} {\bibfnamefont {S.}~\bibnamefont
  {Youn}}, \bibinfo {author} {\bibfnamefont {B.}~\bibnamefont {Goh}}, \bibinfo
  {author} {\bibfnamefont {G.-D.}\ \bibnamefont {Zhou}}, \bibinfo {author}
  {\bibfnamefont {Z.-D.}\ \bibnamefont {Song}}, \ and\ \bibinfo {author}
  {\bibfnamefont {S.-S.~B.}\ \bibnamefont {Lee}},\ }\href {\doibase
  10.48550/arXiv.2412.03108} {\enquote {\bibinfo {title} {Hundness in twisted
  bilayer graphene: correlated gaps and pairing},}\ } (\bibinfo {year}
  {2024}),\ \bibinfo {note} {arXiv:2412.03108 [cond-mat]}\BibitemShut {NoStop}%
\bibitem [{\citenamefont {Parker}\ \emph {et~al.}(2021)\citenamefont {Parker},
  \citenamefont {Soejima}, \citenamefont {Hauschild}, \citenamefont {Zaletel},\
  and\ \citenamefont {Bultinck}}]{Parker_2021_strain}%
  \BibitemOpen
  \bibfield  {author} {\bibinfo {author} {\bibfnamefont {D.~E.}\ \bibnamefont
  {Parker}}, \bibinfo {author} {\bibfnamefont {T.}~\bibnamefont {Soejima}},
  \bibinfo {author} {\bibfnamefont {J.}~\bibnamefont {Hauschild}}, \bibinfo
  {author} {\bibfnamefont {M.~P.}\ \bibnamefont {Zaletel}}, \ and\ \bibinfo
  {author} {\bibfnamefont {N.}~\bibnamefont {Bultinck}},\ }\href {\doibase
  10.1103/PhysRevLett.127.027601} {\bibfield  {journal} {\bibinfo  {journal}
  {Phys. Rev. Lett.}\ }\textbf {\bibinfo {volume} {127}},\ \bibinfo {pages}
  {027601} (\bibinfo {year} {2021})}\BibitemShut {NoStop}%
\bibitem [{\citenamefont {Wagner}\ \emph {et~al.}(2022)\citenamefont {Wagner},
  \citenamefont {Kwan}, \citenamefont {Bultinck}, \citenamefont {Simon},\ and\
  \citenamefont {Parameswaran}}]{Wagner_2022_global}%
  \BibitemOpen
  \bibfield  {author} {\bibinfo {author} {\bibfnamefont {G.}~\bibnamefont
  {Wagner}}, \bibinfo {author} {\bibfnamefont {Y.~H.}\ \bibnamefont {Kwan}},
  \bibinfo {author} {\bibfnamefont {N.}~\bibnamefont {Bultinck}}, \bibinfo
  {author} {\bibfnamefont {S.~H.}\ \bibnamefont {Simon}}, \ and\ \bibinfo
  {author} {\bibfnamefont {S.~A.}\ \bibnamefont {Parameswaran}},\ }\href
  {\doibase 10.1103/PhysRevLett.128.156401} {\bibfield  {journal} {\bibinfo
  {journal} {Phys. Rev. Lett.}\ }\textbf {\bibinfo {volume} {128}},\ \bibinfo
  {pages} {156401} (\bibinfo {year} {2022})}\BibitemShut {NoStop}%
\bibitem [{\citenamefont {Wang}\ \emph {et~al.}(2026)\citenamefont {Wang},
  \citenamefont {Zhou}, \citenamefont {Jung}, \citenamefont {Youn},
  \citenamefont {Lee},\ and\ \citenamefont {Song}}]{wang_bosonization_2025}%
  \BibitemOpen
  \bibfield  {author} {\bibinfo {author} {\bibfnamefont {Y.-J.}\ \bibnamefont
  {Wang}}, \bibinfo {author} {\bibfnamefont {G.-D.}\ \bibnamefont {Zhou}},
  \bibinfo {author} {\bibfnamefont {H.}~\bibnamefont {Jung}}, \bibinfo {author}
  {\bibfnamefont {S.}~\bibnamefont {Youn}}, \bibinfo {author} {\bibfnamefont
  {S.-S.~B.}\ \bibnamefont {Lee}}, \ and\ \bibinfo {author} {\bibfnamefont
  {Z.-D.}\ \bibnamefont {Song}},\ }\href {https://arxiv.org/abs/2601.16525}
  {\enquote {\bibinfo {title} {{Bosonization} {Solution} to {Spin-Valley}
  {Kondo} {Problem}: {Finite-Size} {Spectrum} and {Renormalization} {Group}
  {Analysis}},}\ } (\bibinfo {year} {2026}),\ \Eprint
  {http://arxiv.org/abs/2601.16525} {arXiv:2601.16525 [cond-mat.str-el]}
  \BibitemShut {NoStop}%
\bibitem [{sup()}]{supplement}%
  \BibitemOpen
  \href@noop {} {}\bibinfo {note} {See Supplemental Material for detailed
  information on quantum impurity model, spectral function and correlation
  self-energy {\it ans\"atze}, effective interactions, and NRG
  verification.}\BibitemShut {Stop}%
\bibitem [{\citenamefont {Andrei}\ \emph {et~al.}(1983)\citenamefont {Andrei},
  \citenamefont {Furuya},\ and\ \citenamefont
  {Lowenstein}}]{Andrei_1983_Solution}%
  \BibitemOpen
  \bibfield  {author} {\bibinfo {author} {\bibfnamefont {N.}~\bibnamefont
  {Andrei}}, \bibinfo {author} {\bibfnamefont {K.}~\bibnamefont {Furuya}}, \
  and\ \bibinfo {author} {\bibfnamefont {J.~H.}\ \bibnamefont {Lowenstein}},\
  }\href {\doibase 10.1103/RevModPhys.55.331} {\bibfield  {journal} {\bibinfo
  {journal} {Rev. Mod. Phys.}\ }\textbf {\bibinfo {volume} {55}},\ \bibinfo
  {pages} {331} (\bibinfo {year} {1983})}\BibitemShut {NoStop}%
\bibitem [{\citenamefont {Parcollet}\ \emph {et~al.}(1998)\citenamefont
  {Parcollet}, \citenamefont {Georges}, \citenamefont {Kotliar},\ and\
  \citenamefont {Sengupta}}]{Parcollet_1998_overscreened}%
  \BibitemOpen
  \bibfield  {author} {\bibinfo {author} {\bibfnamefont {O.}~\bibnamefont
  {Parcollet}}, \bibinfo {author} {\bibfnamefont {A.}~\bibnamefont {Georges}},
  \bibinfo {author} {\bibfnamefont {G.}~\bibnamefont {Kotliar}}, \ and\
  \bibinfo {author} {\bibfnamefont {A.}~\bibnamefont {Sengupta}},\ }\href
  {\doibase 10.1103/PhysRevB.58.3794} {\bibfield  {journal} {\bibinfo
  {journal} {Phys. Rev. B}\ }\textbf {\bibinfo {volume} {58}},\ \bibinfo
  {pages} {3794} (\bibinfo {year} {1998})}\BibitemShut {NoStop}%
\bibitem [{\citenamefont {Nozières}\ and\ \citenamefont
  {Blandin}(1980)}]{nozieres_kondo_1980}%
  \BibitemOpen
  \bibfield  {author} {\bibinfo {author} {\bibfnamefont {P.}~\bibnamefont
  {Nozières}}\ and\ \bibinfo {author} {\bibfnamefont {A.}~\bibnamefont
  {Blandin}},\ }\href {\doibase 10.1051/jphys:01980004103019300} {\bibfield
  {journal} {\bibinfo  {journal} {Journal de Physique}\ }\textbf {\bibinfo
  {volume} {41}},\ \bibinfo {pages} {193} (\bibinfo {year} {1980})}\BibitemShut
  {NoStop}%
\bibitem [{\citenamefont {Hewson}(1993{\natexlab{a}})}]{hewson_fermi_1993}%
  \BibitemOpen
  \bibfield  {author} {\bibinfo {author} {\bibfnamefont {A.~C.}\ \bibnamefont
  {Hewson}},\ }\href {\doibase 10.1088/0953-8984/5/34/014} {\bibfield
  {journal} {\bibinfo  {journal} {Journal of Physics: Condensed Matter}\
  }\textbf {\bibinfo {volume} {5}},\ \bibinfo {pages} {6277} (\bibinfo {year}
  {1993}{\natexlab{a}})}\BibitemShut {NoStop}%
\bibitem [{\citenamefont
  {Hewson}(1993{\natexlab{b}})}]{hewson_renormalized_1993}%
  \BibitemOpen
  \bibfield  {author} {\bibinfo {author} {\bibfnamefont {A.~C.}\ \bibnamefont
  {Hewson}},\ }\href {\doibase 10.1103/PhysRevLett.70.4007} {\bibfield
  {journal} {\bibinfo  {journal} {Phys. Rev. Lett.}\ }\textbf {\bibinfo
  {volume} {70}},\ \bibinfo {pages} {4007} (\bibinfo {year}
  {1993}{\natexlab{b}})}\BibitemShut {NoStop}%
\bibitem [{\citenamefont {Hewson}\ \emph {et~al.}(2004)\citenamefont {Hewson},
  \citenamefont {Oguri},\ and\ \citenamefont
  {Meyer}}]{hewson_renormalized_2004}%
  \BibitemOpen
  \bibfield  {author} {\bibinfo {author} {\bibfnamefont {A.~C.}\ \bibnamefont
  {Hewson}}, \bibinfo {author} {\bibfnamefont {A.}~\bibnamefont {Oguri}}, \
  and\ \bibinfo {author} {\bibfnamefont {D.}~\bibnamefont {Meyer}},\ }\href
  {\doibase 10.1140/epjb/e2004-00256-0} {\bibfield  {journal} {\bibinfo
  {journal} {The European Physical Journal B}\ }\textbf {\bibinfo {volume}
  {40}},\ \bibinfo {pages} {177} (\bibinfo {year} {2004})}\BibitemShut
  {NoStop}%
\bibitem [{\citenamefont {Nishikawa}\ \emph {et~al.}(2010)\citenamefont
  {Nishikawa}, \citenamefont {Crow},\ and\ \citenamefont
  {Hewson}}]{nishikawa_renormalized_2010}%
  \BibitemOpen
  \bibfield  {author} {\bibinfo {author} {\bibfnamefont {Y.}~\bibnamefont
  {Nishikawa}}, \bibinfo {author} {\bibfnamefont {D.~J.~G.}\ \bibnamefont
  {Crow}}, \ and\ \bibinfo {author} {\bibfnamefont {A.~C.}\ \bibnamefont
  {Hewson}},\ }\href {\doibase 10.1103/PhysRevB.82.115123} {\bibfield
  {journal} {\bibinfo  {journal} {Phys. Rev. B}\ }\textbf {\bibinfo {volume}
  {82}},\ \bibinfo {pages} {115123} (\bibinfo {year} {2010})}\BibitemShut
  {NoStop}%
\bibitem [{\citenamefont
  {Yamada}(1975{\natexlab{a}})}]{yamada_perturbation_1975}%
  \BibitemOpen
  \bibfield  {author} {\bibinfo {author} {\bibfnamefont {K.}~\bibnamefont
  {Yamada}},\ }\href {\doibase 10.1143/PTP.53.970} {\bibfield  {journal}
  {\bibinfo  {journal} {Progress of Theoretical Physics}\ }\textbf {\bibinfo
  {volume} {53}},\ \bibinfo {pages} {970} (\bibinfo {year}
  {1975}{\natexlab{a}})}\BibitemShut {NoStop}%
\bibitem [{\citenamefont
  {Yamada}(1975{\natexlab{b}})}]{yamada_perturbation_1975-1}%
  \BibitemOpen
  \bibfield  {author} {\bibinfo {author} {\bibfnamefont {K.}~\bibnamefont
  {Yamada}},\ }\href {\doibase 10.1143/PTP.54.316} {\bibfield  {journal}
  {\bibinfo  {journal} {Progress of Theoretical Physics}\ }\textbf {\bibinfo
  {volume} {54}},\ \bibinfo {pages} {316} (\bibinfo {year}
  {1975}{\natexlab{b}})}\BibitemShut {NoStop}%
\bibitem [{\citenamefont {Yoshimori}(1976)}]{yoshimori_perturbation_1976}%
  \BibitemOpen
  \bibfield  {author} {\bibinfo {author} {\bibfnamefont {A.}~\bibnamefont
  {Yoshimori}},\ }\href {\doibase 10.1143/PTP.55.67} {\bibfield  {journal}
  {\bibinfo  {journal} {Progress of Theoretical Physics}\ }\textbf {\bibinfo
  {volume} {55}},\ \bibinfo {pages} {67} (\bibinfo {year} {1976})}\BibitemShut
  {NoStop}%
\bibitem [{\citenamefont {G.~Toulouse}(1969)}]{toulouse_1969}%
  \BibitemOpen
  \bibfield  {author} {\bibinfo {author} {\bibfnamefont {C.~R.}\ \bibnamefont
  {G.~Toulouse}},\ }\href@noop {} {\bibfield  {journal} {\bibinfo  {journal}
  {Acad. Sci.}\ }\textbf {\bibinfo {volume} {268}},\ \bibinfo {pages} {1200}
  (\bibinfo {year} {1969})}\BibitemShut {NoStop}%
\bibitem [{\citenamefont {Emery}\ and\ \citenamefont
  {Kivelson}(1992)}]{Mapping_Emery_1992}%
  \BibitemOpen
  \bibfield  {author} {\bibinfo {author} {\bibfnamefont {V.~J.}\ \bibnamefont
  {Emery}}\ and\ \bibinfo {author} {\bibfnamefont {S.}~\bibnamefont
  {Kivelson}},\ }\href {\doibase 10.1103/PhysRevB.46.10812} {\bibfield
  {journal} {\bibinfo  {journal} {Phys. Rev. B}\ }\textbf {\bibinfo {volume}
  {46}},\ \bibinfo {pages} {10812} (\bibinfo {year} {1992})}\BibitemShut
  {NoStop}%
\bibitem [{\citenamefont {Kotliar}\ and\ \citenamefont
  {Si}(1996)}]{Kotliar_toulouse_1996}%
  \BibitemOpen
  \bibfield  {author} {\bibinfo {author} {\bibfnamefont {G.}~\bibnamefont
  {Kotliar}}\ and\ \bibinfo {author} {\bibfnamefont {Q.}~\bibnamefont {Si}},\
  }\href {\doibase 10.1103/PhysRevB.53.12373} {\bibfield  {journal} {\bibinfo
  {journal} {Phys. Rev. B}\ }\textbf {\bibinfo {volume} {53}},\ \bibinfo
  {pages} {12373} (\bibinfo {year} {1996})}\BibitemShut {NoStop}%
\bibitem [{\citenamefont {von Delft}\ and\ \citenamefont
  {Schoeller}(1998)}]{von_delft_bosonization_1998}%
  \BibitemOpen
  \bibfield  {author} {\bibinfo {author} {\bibfnamefont {J.}~\bibnamefont {von
  Delft}}\ and\ \bibinfo {author} {\bibfnamefont {H.}~\bibnamefont
  {Schoeller}},\ }\href {\doibase 10.1002/andp.19985100401} {\bibfield
  {journal} {\bibinfo  {journal} {Annalen der Physik}\ }\textbf {\bibinfo
  {volume} {510}},\ \bibinfo {pages} {225} (\bibinfo {year}
  {1998})}\BibitemShut {NoStop}%
\bibitem [{\citenamefont {von Delft}\ \emph {et~al.}(1998)\citenamefont {von
  Delft}, \citenamefont {Zar\'and},\ and\ \citenamefont
  {Fabrizio}}]{vonDelft_1998_finitesize}%
  \BibitemOpen
  \bibfield  {author} {\bibinfo {author} {\bibfnamefont {J.}~\bibnamefont {von
  Delft}}, \bibinfo {author} {\bibfnamefont {G.}~\bibnamefont {Zar\'and}}, \
  and\ \bibinfo {author} {\bibfnamefont {M.}~\bibnamefont {Fabrizio}},\ }\href
  {\doibase 10.1103/PhysRevLett.81.196} {\bibfield  {journal} {\bibinfo
  {journal} {Phys. Rev. Lett.}\ }\textbf {\bibinfo {volume} {81}},\ \bibinfo
  {pages} {196} (\bibinfo {year} {1998})}\BibitemShut {NoStop}%
\bibitem [{\citenamefont {Zar\'and}\ and\ \citenamefont {von
  Delft}(2000)}]{zarand_analytical_2000}%
  \BibitemOpen
  \bibfield  {author} {\bibinfo {author} {\bibfnamefont {G.}~\bibnamefont
  {Zar\'and}}\ and\ \bibinfo {author} {\bibfnamefont {J.}~\bibnamefont {von
  Delft}},\ }\href {\doibase 10.1103/PhysRevB.61.6918} {\bibfield  {journal}
  {\bibinfo  {journal} {Phys. Rev. B}\ }\textbf {\bibinfo {volume} {61}},\
  \bibinfo {pages} {6918} (\bibinfo {year} {2000})}\BibitemShut {NoStop}%
\bibitem [{\citenamefont {Giamarchi}(2003)}]{giamarchi2003quantum}%
  \BibitemOpen
  \bibfield  {author} {\bibinfo {author} {\bibfnamefont {T.}~\bibnamefont
  {Giamarchi}},\ }\enquote {\bibinfo {title} {11.2 impurities in fermi
  liquids},}\ in\ \href@noop {} {\emph {\bibinfo {booktitle} {Quantum physics
  in one dimension}}},\ Vol.\ \bibinfo {volume} {121}\ (\bibinfo  {publisher}
  {Clarendon press},\ \bibinfo {year} {2003})\ Chap.~\bibinfo {chapter}
  {11}\BibitemShut {NoStop}%
\bibitem [{\citenamefont {Schiller}\ and\ \citenamefont
  {De~Leo}(2008)}]{Schiller_phasediagram_2008}%
  \BibitemOpen
  \bibfield  {author} {\bibinfo {author} {\bibfnamefont {A.}~\bibnamefont
  {Schiller}}\ and\ \bibinfo {author} {\bibfnamefont {L.}~\bibnamefont
  {De~Leo}},\ }\href {\doibase 10.1103/PhysRevB.77.075114} {\bibfield
  {journal} {\bibinfo  {journal} {Phys. Rev. B}\ }\textbf {\bibinfo {volume}
  {77}},\ \bibinfo {pages} {075114} (\bibinfo {year} {2008})}\BibitemShut
  {NoStop}%
\bibitem [{\citenamefont {Krishnan}\ and\ \citenamefont
  {Metlitski}(2024)}]{Krishnan_2024_kondo}%
  \BibitemOpen
  \bibfield  {author} {\bibinfo {author} {\bibfnamefont {A.}~\bibnamefont
  {Krishnan}}\ and\ \bibinfo {author} {\bibfnamefont {M.~A.}\ \bibnamefont
  {Metlitski}},\ }\href {https://arxiv.org/abs/2408.12650} {\enquote {\bibinfo
  {title} {The {Kondo} impurity in the large spin limit},}\ } (\bibinfo {year}
  {2024}),\ \Eprint {http://arxiv.org/abs/2408.12650} {arXiv:2408.12650
  [cond-mat.str-el]} \BibitemShut {NoStop}%
\bibitem [{\citenamefont {Galpin}\ \emph {et~al.}(2005)\citenamefont {Galpin},
  \citenamefont {Logan},\ and\ \citenamefont
  {Krishnamurthy}}]{Galpin_2005_quantumphasetransition}%
  \BibitemOpen
  \bibfield  {author} {\bibinfo {author} {\bibfnamefont {M.~R.}\ \bibnamefont
  {Galpin}}, \bibinfo {author} {\bibfnamefont {D.~E.}\ \bibnamefont {Logan}}, \
  and\ \bibinfo {author} {\bibfnamefont {H.~R.}\ \bibnamefont
  {Krishnamurthy}},\ }\href {\doibase 10.1103/PhysRevLett.94.186406} {\bibfield
   {journal} {\bibinfo  {journal} {Phys. Rev. Lett.}\ }\textbf {\bibinfo
  {volume} {94}},\ \bibinfo {pages} {186406} (\bibinfo {year}
  {2005})}\BibitemShut {NoStop}%
\bibitem [{\citenamefont {Galpin}\ \emph {et~al.}(2006)\citenamefont {Galpin},
  \citenamefont {Logan},\ and\ \citenamefont
  {Krishnamurthy}}]{Galpin_2006_renormalization}%
  \BibitemOpen
  \bibfield  {author} {\bibinfo {author} {\bibfnamefont {M.~R.}\ \bibnamefont
  {Galpin}}, \bibinfo {author} {\bibfnamefont {D.~E.}\ \bibnamefont {Logan}}, \
  and\ \bibinfo {author} {\bibfnamefont {H.~R.}\ \bibnamefont
  {Krishnamurthy}},\ }\href {\doibase 10.1088/0953-8984/18/29/001} {\bibfield
  {journal} {\bibinfo  {journal} {Journal of Physics: Condensed Matter}\
  }\textbf {\bibinfo {volume} {18}},\ \bibinfo {pages} {6545} (\bibinfo {year}
  {2006})}\BibitemShut {NoStop}%
\bibitem [{\citenamefont {Nishikawa}\ \emph {et~al.}(2012)\citenamefont
  {Nishikawa}, \citenamefont {Crow},\ and\ \citenamefont
  {Hewson}}]{nishikawa_convergence_2012}%
  \BibitemOpen
  \bibfield  {author} {\bibinfo {author} {\bibfnamefont {Y.}~\bibnamefont
  {Nishikawa}}, \bibinfo {author} {\bibfnamefont {D.~J.~G.}\ \bibnamefont
  {Crow}}, \ and\ \bibinfo {author} {\bibfnamefont {A.~C.}\ \bibnamefont
  {Hewson}},\ }\href {\doibase 10.1103/PhysRevLett.108.056402} {\bibfield
  {journal} {\bibinfo  {journal} {Phys. Rev. Lett.}\ }\textbf {\bibinfo
  {volume} {108}},\ \bibinfo {pages} {056402} (\bibinfo {year}
  {2012})}\BibitemShut {NoStop}%
\bibitem [{\citenamefont {Costi}(2000)}]{Costi_2000_Kondo}%
  \BibitemOpen
  \bibfield  {author} {\bibinfo {author} {\bibfnamefont {T.~A.}\ \bibnamefont
  {Costi}},\ }\href {\doibase 10.1103/PhysRevLett.85.1504} {\bibfield
  {journal} {\bibinfo  {journal} {Phys. Rev. Lett.}\ }\textbf {\bibinfo
  {volume} {85}},\ \bibinfo {pages} {1504} (\bibinfo {year}
  {2000})}\BibitemShut {NoStop}%
\bibitem [{\citenamefont {Bulla}\ \emph {et~al.}(2008)\citenamefont {Bulla},
  \citenamefont {Costi},\ and\ \citenamefont {Pruschke}}]{Bulla_2008_nrg}%
  \BibitemOpen
  \bibfield  {author} {\bibinfo {author} {\bibfnamefont {R.}~\bibnamefont
  {Bulla}}, \bibinfo {author} {\bibfnamefont {T.~A.}\ \bibnamefont {Costi}}, \
  and\ \bibinfo {author} {\bibfnamefont {T.}~\bibnamefont {Pruschke}},\ }\href
  {\doibase 10.1103/RevModPhys.80.395} {\bibfield  {journal} {\bibinfo
  {journal} {Rev. Mod. Phys.}\ }\textbf {\bibinfo {volume} {80}},\ \bibinfo
  {pages} {395} (\bibinfo {year} {2008})}\BibitemShut {NoStop}%
\bibitem [{\citenamefont {Moca}\ \emph {et~al.}(2019)\citenamefont {Moca},
  \citenamefont {Chirla}, \citenamefont {D\'ora},\ and\ \citenamefont
  {Zar\'and}}]{Moca_2019_quantumcriticality}%
  \BibitemOpen
  \bibfield  {author} {\bibinfo {author} {\bibfnamefont {C.~u. u. u. u.~P.}\
  \bibnamefont {Moca}}, \bibinfo {author} {\bibfnamefont {R.}~\bibnamefont
  {Chirla}}, \bibinfo {author} {\bibfnamefont {B.}~\bibnamefont {D\'ora}}, \
  and\ \bibinfo {author} {\bibfnamefont {G.}~\bibnamefont {Zar\'and}},\ }\href
  {\doibase 10.1103/PhysRevLett.123.136803} {\bibfield  {journal} {\bibinfo
  {journal} {Phys. Rev. Lett.}\ }\textbf {\bibinfo {volume} {123}},\ \bibinfo
  {pages} {136803} (\bibinfo {year} {2019})}\BibitemShut {NoStop}%
\bibitem [{\citenamefont {Fabrizio}\ \emph {et~al.}(2003)\citenamefont
  {Fabrizio}, \citenamefont {Ho}, \citenamefont {Leo},\ and\ \citenamefont
  {Santoro}}]{fabrizio_nontrivial_2003}%
  \BibitemOpen
  \bibfield  {author} {\bibinfo {author} {\bibfnamefont {M.}~\bibnamefont
  {Fabrizio}}, \bibinfo {author} {\bibfnamefont {A.~F.}\ \bibnamefont {Ho}},
  \bibinfo {author} {\bibfnamefont {L.~D.}\ \bibnamefont {Leo}}, \ and\
  \bibinfo {author} {\bibfnamefont {G.~E.}\ \bibnamefont {Santoro}},\ }\href
  {\doibase 10.1103/PhysRevLett.91.246402} {\bibfield  {journal} {\bibinfo
  {journal} {Phys. Rev. Lett.}\ }\textbf {\bibinfo {volume} {91}},\ \bibinfo
  {pages} {246402} (\bibinfo {year} {2003})}\BibitemShut {NoStop}%
\bibitem [{\citenamefont {Leo}\ and\ \citenamefont
  {Fabrizio}(2004)}]{leo_spectral_2004}%
  \BibitemOpen
  \bibfield  {author} {\bibinfo {author} {\bibfnamefont {L.~D.}\ \bibnamefont
  {Leo}}\ and\ \bibinfo {author} {\bibfnamefont {M.}~\bibnamefont {Fabrizio}},\
  }\href {\doibase 10.1103/PhysRevB.69.245114} {\bibfield  {journal} {\bibinfo
  {journal} {Phys. Rev. B}\ }\textbf {\bibinfo {volume} {69}},\ \bibinfo
  {pages} {245114} (\bibinfo {year} {2004})}\BibitemShut {NoStop}%
\bibitem [{\citenamefont {Jayaprakash}\ \emph {et~al.}(1981)\citenamefont
  {Jayaprakash}, \citenamefont {Krishna-murthy},\ and\ \citenamefont
  {Wilkins}}]{Jayaprakash_1981_2IK}%
  \BibitemOpen
  \bibfield  {author} {\bibinfo {author} {\bibfnamefont {C.}~\bibnamefont
  {Jayaprakash}}, \bibinfo {author} {\bibfnamefont {H.~R.}\ \bibnamefont
  {Krishna-murthy}}, \ and\ \bibinfo {author} {\bibfnamefont {J.~W.}\
  \bibnamefont {Wilkins}},\ }\href {\doibase 10.1103/PhysRevLett.47.737}
  {\bibfield  {journal} {\bibinfo  {journal} {Phys. Rev. Lett.}\ }\textbf
  {\bibinfo {volume} {47}},\ \bibinfo {pages} {737} (\bibinfo {year}
  {1981})}\BibitemShut {NoStop}%
\bibitem [{\citenamefont {Jones}\ and\ \citenamefont
  {Varma}(1987)}]{Jones_1987_study}%
  \BibitemOpen
  \bibfield  {author} {\bibinfo {author} {\bibfnamefont {B.~A.}\ \bibnamefont
  {Jones}}\ and\ \bibinfo {author} {\bibfnamefont {C.~M.}\ \bibnamefont
  {Varma}},\ }\href {\doibase 10.1103/PhysRevLett.58.843} {\bibfield  {journal}
  {\bibinfo  {journal} {Phys. Rev. Lett.}\ }\textbf {\bibinfo {volume} {58}},\
  \bibinfo {pages} {843} (\bibinfo {year} {1987})}\BibitemShut {NoStop}%
\bibitem [{\citenamefont {Jones}\ \emph {et~al.}(1988)\citenamefont {Jones},
  \citenamefont {Varma},\ and\ \citenamefont {Wilkins}}]{Jones_1988_lowT}%
  \BibitemOpen
  \bibfield  {author} {\bibinfo {author} {\bibfnamefont {B.~A.}\ \bibnamefont
  {Jones}}, \bibinfo {author} {\bibfnamefont {C.~M.}\ \bibnamefont {Varma}}, \
  and\ \bibinfo {author} {\bibfnamefont {J.~W.}\ \bibnamefont {Wilkins}},\
  }\href {\doibase 10.1103/PhysRevLett.61.125} {\bibfield  {journal} {\bibinfo
  {journal} {Phys. Rev. Lett.}\ }\textbf {\bibinfo {volume} {61}},\ \bibinfo
  {pages} {125} (\bibinfo {year} {1988})}\BibitemShut {NoStop}%
\bibitem [{\citenamefont {Jones}\ and\ \citenamefont
  {Varma}(1989)}]{Jones_1989_critical}%
  \BibitemOpen
  \bibfield  {author} {\bibinfo {author} {\bibfnamefont {B.~A.}\ \bibnamefont
  {Jones}}\ and\ \bibinfo {author} {\bibfnamefont {C.~M.}\ \bibnamefont
  {Varma}},\ }\href {\doibase 10.1103/PhysRevB.40.324} {\bibfield  {journal}
  {\bibinfo  {journal} {Phys. Rev. B}\ }\textbf {\bibinfo {volume} {40}},\
  \bibinfo {pages} {324} (\bibinfo {year} {1989})}\BibitemShut {NoStop}%
\bibitem [{\citenamefont {Affleck}\ and\ \citenamefont
  {Ludwig}(1992)}]{Affleck_1992_exactcriticaltheory}%
  \BibitemOpen
  \bibfield  {author} {\bibinfo {author} {\bibfnamefont {I.}~\bibnamefont
  {Affleck}}\ and\ \bibinfo {author} {\bibfnamefont {A.~W.~W.}\ \bibnamefont
  {Ludwig}},\ }\href {\doibase 10.1103/PhysRevLett.68.1046} {\bibfield
  {journal} {\bibinfo  {journal} {Phys. Rev. Lett.}\ }\textbf {\bibinfo
  {volume} {68}},\ \bibinfo {pages} {1046} (\bibinfo {year}
  {1992})}\BibitemShut {NoStop}%
\bibitem [{\citenamefont {Gan}(1995)}]{Gan_1995_mapping}%
  \BibitemOpen
  \bibfield  {author} {\bibinfo {author} {\bibfnamefont {J.}~\bibnamefont
  {Gan}},\ }\href {\doibase 10.1103/PhysRevLett.74.2583} {\bibfield  {journal}
  {\bibinfo  {journal} {Phys. Rev. Lett.}\ }\textbf {\bibinfo {volume} {74}},\
  \bibinfo {pages} {2583} (\bibinfo {year} {1995})}\BibitemShut {NoStop}%
\bibitem [{\citenamefont {Mitchell}\ and\ \citenamefont
  {Sela}(2012)}]{mitchell_universal_2012}%
  \BibitemOpen
  \bibfield  {author} {\bibinfo {author} {\bibfnamefont {A.~K.}\ \bibnamefont
  {Mitchell}}\ and\ \bibinfo {author} {\bibfnamefont {E.}~\bibnamefont
  {Sela}},\ }\href {\doibase 10.1103/PhysRevB.85.235127} {\bibfield  {journal}
  {\bibinfo  {journal} {Phys. Rev. B}\ }\textbf {\bibinfo {volume} {85}},\
  \bibinfo {pages} {235127} (\bibinfo {year} {2012})}\BibitemShut {NoStop}%
\bibitem [{\citenamefont {Mitchell}\ \emph {et~al.}(2012)\citenamefont
  {Mitchell}, \citenamefont {Sela},\ and\ \citenamefont
  {Logan}}]{mitchell_2channel_2012}%
  \BibitemOpen
  \bibfield  {author} {\bibinfo {author} {\bibfnamefont {A.~K.}\ \bibnamefont
  {Mitchell}}, \bibinfo {author} {\bibfnamefont {E.}~\bibnamefont {Sela}}, \
  and\ \bibinfo {author} {\bibfnamefont {D.~E.}\ \bibnamefont {Logan}},\ }\href
  {\doibase 10.1103/PhysRevLett.108.086405} {\bibfield  {journal} {\bibinfo
  {journal} {Phys. Rev. Lett.}\ }\textbf {\bibinfo {volume} {108}},\ \bibinfo
  {pages} {086405} (\bibinfo {year} {2012})}\BibitemShut {NoStop}%
\bibitem [{\citenamefont {Youn}\ \emph {et~al.}()\citenamefont {Youn},
  \citenamefont {Goh}, \citenamefont {Zhou}, \citenamefont {Wang},
  \citenamefont {Song},\ and\ \citenamefont {Lee}}]{youn_hundness_2025}%
  \BibitemOpen
  \bibfield  {author} {\bibinfo {author} {\bibfnamefont {S.}~\bibnamefont
  {Youn}}, \bibinfo {author} {\bibfnamefont {B.}~\bibnamefont {Goh}}, \bibinfo
  {author} {\bibfnamefont {G.-D.}\ \bibnamefont {Zhou}}, \bibinfo {author}
  {\bibfnamefont {Y.-J.}\ \bibnamefont {Wang}}, \bibinfo {author}
  {\bibfnamefont {Z.-D.}\ \bibnamefont {Song}}, \ and\ \bibinfo {author}
  {\bibfnamefont {S.-S.~B.}\ \bibnamefont {Lee}},\ }\href@noop {} {\enquote
  {\bibinfo {title} {Hundness in twisted bilayer graphene: phase classification
  and quantum criticalities},}\ }\bibinfo {note} {In preparation}\BibitemShut
  {NoStop}%
\bibitem [{\citenamefont {Zhang}\ and\ \citenamefont
  {Mao}(2020)}]{zhang_2020_spinliquids}%
  \BibitemOpen
  \bibfield  {author} {\bibinfo {author} {\bibfnamefont {Y.-H.}\ \bibnamefont
  {Zhang}}\ and\ \bibinfo {author} {\bibfnamefont {D.}~\bibnamefont {Mao}},\
  }\href {\doibase 10.1103/PhysRevB.101.035122} {\bibfield  {journal} {\bibinfo
   {journal} {Phys. Rev. B}\ }\textbf {\bibinfo {volume} {101}},\ \bibinfo
  {pages} {035122} (\bibinfo {year} {2020})}\BibitemShut {NoStop}%
\bibitem [{\citenamefont {Zhao}\ \emph
  {et~al.}(2025{\natexlab{a}})\citenamefont {Zhao}, \citenamefont {Zhou},\ and\
  \citenamefont {Zhang}}]{zhao_mixed_2025}%
  \BibitemOpen
  \bibfield  {author} {\bibinfo {author} {\bibfnamefont {J.-Y.}\ \bibnamefont
  {Zhao}}, \bibinfo {author} {\bibfnamefont {B.}~\bibnamefont {Zhou}}, \ and\
  \bibinfo {author} {\bibfnamefont {Y.-H.}\ \bibnamefont {Zhang}},\ }\href
  {\doibase 10.48550/arXiv.2507.00139} {\enquote {\bibinfo {title} {Mixed
  valence {Mott} insulator and composite excitation in twisted bilayer
  graphene},}\ } (\bibinfo {year} {2025}{\natexlab{a}}),\ \bibinfo {note}
  {arXiv:2507.00139 [cond-mat]}\BibitemShut {NoStop}%
\bibitem [{\citenamefont {Zhao}\ and\ \citenamefont
  {Zhang}(2025)}]{zhao_2025_rvb}%
  \BibitemOpen
  \bibfield  {author} {\bibinfo {author} {\bibfnamefont {J.-Y.}\ \bibnamefont
  {Zhao}}\ and\ \bibinfo {author} {\bibfnamefont {Y.-H.}\ \bibnamefont
  {Zhang}},\ }\href {https://arxiv.org/abs/2510.26801} {\enquote {\bibinfo
  {title} {Resonating-valence-bond superconductor from small fermi surface in
  twisted bilayer graphene},}\ } (\bibinfo {year} {2025}),\ \Eprint
  {http://arxiv.org/abs/2510.26801} {arXiv:2510.26801 [cond-mat.str-el]}
  \BibitemShut {NoStop}%
\bibitem [{\citenamefont {Hubbard}(1964)}]{Hubbard_2}%
  \BibitemOpen
  \bibfield  {author} {\bibinfo {author} {\bibfnamefont {J.}~\bibnamefont
  {Hubbard}},\ }\href {\doibase 10.1098/rspa.1964.0019} {\bibfield  {journal}
  {\bibinfo  {journal} {Proc. R. Soc. Lond. A}\ }\textbf {\bibinfo {volume}
  {277}},\ \bibinfo {pages} {237} (\bibinfo {year} {1964})}\BibitemShut
  {NoStop}%
\bibitem [{\citenamefont {Lichtenstein}\ and\ \citenamefont
  {Katsnelson}(1998)}]{Lichtenstein_abinitio_1998}%
  \BibitemOpen
  \bibfield  {author} {\bibinfo {author} {\bibfnamefont {A.~I.}\ \bibnamefont
  {Lichtenstein}}\ and\ \bibinfo {author} {\bibfnamefont {M.~I.}\ \bibnamefont
  {Katsnelson}},\ }\href {\doibase 10.1103/PhysRevB.57.6884} {\bibfield
  {journal} {\bibinfo  {journal} {Phys. Rev. B}\ }\textbf {\bibinfo {volume}
  {57}},\ \bibinfo {pages} {6884} (\bibinfo {year} {1998})}\BibitemShut
  {NoStop}%
\bibitem [{\citenamefont {Hu}\ \emph {et~al.}(2025)\citenamefont {Hu},
  \citenamefont {Song},\ and\ \citenamefont {Bernevig}}]{hu_projected_2025}%
  \BibitemOpen
  \bibfield  {author} {\bibinfo {author} {\bibfnamefont {H.}~\bibnamefont
  {Hu}}, \bibinfo {author} {\bibfnamefont {Z.-D.}\ \bibnamefont {Song}}, \ and\
  \bibinfo {author} {\bibfnamefont {B.~A.}\ \bibnamefont {Bernevig}},\ }\href
  {\doibase 10.48550/arXiv.2502.14039} {\enquote {\bibinfo {title} {Projected
  and {Solvable} {Topological} {Heavy} {Fermion} {Model} of {Twisted} {Bilayer}
  {Graphene}},}\ } (\bibinfo {year} {2025}),\ \bibinfo {note} {arXiv:2502.14039
  [cond-mat]}\BibitemShut {NoStop}%
\bibitem [{\citenamefont {Ledwith}\ \emph {et~al.}(2025)\citenamefont
  {Ledwith}, \citenamefont {Dong}, \citenamefont {Vishwanath},\ and\
  \citenamefont {Khalaf}}]{ledwith_nonlocal_2024}%
  \BibitemOpen
  \bibfield  {author} {\bibinfo {author} {\bibfnamefont {P.~J.}\ \bibnamefont
  {Ledwith}}, \bibinfo {author} {\bibfnamefont {J.}~\bibnamefont {Dong}},
  \bibinfo {author} {\bibfnamefont {A.}~\bibnamefont {Vishwanath}}, \ and\
  \bibinfo {author} {\bibfnamefont {E.}~\bibnamefont {Khalaf}},\ }\href
  {\doibase 10.1103/PhysRevX.15.021087} {\bibfield  {journal} {\bibinfo
  {journal} {Phys. Rev. X}\ }\textbf {\bibinfo {volume} {15}},\ \bibinfo
  {pages} {021087} (\bibinfo {year} {2025})}\BibitemShut {NoStop}%
\bibitem [{\citenamefont {Zhao}\ \emph
  {et~al.}(2025{\natexlab{b}})\citenamefont {Zhao}, \citenamefont {Zhou},\ and\
  \citenamefont {Zhang}}]{zhao_topological_mott_2025}%
  \BibitemOpen
  \bibfield  {author} {\bibinfo {author} {\bibfnamefont {J.-Y.}\ \bibnamefont
  {Zhao}}, \bibinfo {author} {\bibfnamefont {B.}~\bibnamefont {Zhou}}, \ and\
  \bibinfo {author} {\bibfnamefont {Y.-H.}\ \bibnamefont {Zhang}},\ }\href
  {\doibase 10.1103/9n8v-7rx2} {\bibfield  {journal} {\bibinfo  {journal}
  {Phys. Rev. B}\ }\textbf {\bibinfo {volume} {112}},\ \bibinfo {pages}
  {085107} (\bibinfo {year} {2025}{\natexlab{b}})}\BibitemShut {NoStop}%
\bibitem [{\citenamefont {Krishna-murthy}\ \emph
  {et~al.}(1980{\natexlab{a}})\citenamefont {Krishna-murthy}, \citenamefont
  {Wilkins},\ and\ \citenamefont
  {Wilson}}]{krishna-murthy_renormalization-group_1980-1}%
  \BibitemOpen
  \bibfield  {author} {\bibinfo {author} {\bibfnamefont {H.~R.}\ \bibnamefont
  {Krishna-murthy}}, \bibinfo {author} {\bibfnamefont {J.~W.}\ \bibnamefont
  {Wilkins}}, \ and\ \bibinfo {author} {\bibfnamefont {K.~G.}\ \bibnamefont
  {Wilson}},\ }\href {\doibase 10.1103/PhysRevB.21.1003} {\bibfield  {journal}
  {\bibinfo  {journal} {Physical Review B}\ }\textbf {\bibinfo {volume} {21}},\
  \bibinfo {pages} {1003} (\bibinfo {year} {1980}{\natexlab{a}})}\BibitemShut
  {NoStop}%
\bibitem [{\citenamefont {Lee}\ and\ \citenamefont
  {Weichselbaum}(2016)}]{lee_adaptive_2016}%
  \BibitemOpen
  \bibfield  {author} {\bibinfo {author} {\bibfnamefont {S.-S.~B.}\
  \bibnamefont {Lee}}\ and\ \bibinfo {author} {\bibfnamefont {A.}~\bibnamefont
  {Weichselbaum}},\ }\href {\doibase 10.1103/PhysRevB.94.235127} {\bibfield
  {journal} {\bibinfo  {journal} {Phys. Rev. B}\ }\textbf {\bibinfo {volume}
  {94}},\ \bibinfo {pages} {235127} (\bibinfo {year} {2016})}\BibitemShut
  {NoStop}%
\bibitem [{\citenamefont {Lee}\ \emph {et~al.}(2017)\citenamefont {Lee},
  \citenamefont {von Delft},\ and\ \citenamefont
  {Weichselbaum}}]{lee_doublon-holon_2017}%
  \BibitemOpen
  \bibfield  {author} {\bibinfo {author} {\bibfnamefont {S.-S.~B.}\
  \bibnamefont {Lee}}, \bibinfo {author} {\bibfnamefont {J.}~\bibnamefont {von
  Delft}}, \ and\ \bibinfo {author} {\bibfnamefont {A.}~\bibnamefont
  {Weichselbaum}},\ }\href {\doibase 10.1103/PhysRevLett.119.236402} {\bibfield
   {journal} {\bibinfo  {journal} {Phys. Rev. Lett.}\ }\textbf {\bibinfo
  {volume} {119}},\ \bibinfo {pages} {236402} (\bibinfo {year}
  {2017})}\BibitemShut {NoStop}%
\bibitem [{\citenamefont
  {Weichselbaum}(2024{\natexlab{a}})}]{10.21468/SciPostPhysCodeb.40}%
  \BibitemOpen
  \bibfield  {author} {\bibinfo {author} {\bibfnamefont {A.}~\bibnamefont
  {Weichselbaum}},\ }\href {\doibase 10.21468/SciPostPhysCodeb.40} {\bibfield
  {journal} {\bibinfo  {journal} {SciPost Phys. Codebases}\ ,\ \bibinfo {pages}
  {40}} (\bibinfo {year} {2024}{\natexlab{a}})}\BibitemShut {NoStop}%
\bibitem [{\citenamefont
  {Weichselbaum}(2024{\natexlab{b}})}]{10.21468/SciPostPhysCodeb.40-r4.0}%
  \BibitemOpen
  \bibfield  {author} {\bibinfo {author} {\bibfnamefont {A.}~\bibnamefont
  {Weichselbaum}},\ }\href {\doibase 10.21468/SciPostPhysCodeb.40-r4.0}
  {\bibfield  {journal} {\bibinfo  {journal} {SciPost Phys. Codebases}\ ,\
  \bibinfo {pages} {40}} (\bibinfo {year} {2024}{\natexlab{b}})}\BibitemShut
  {NoStop}%
\bibitem [{\citenamefont {Yoshida}\ \emph {et~al.}(1990)\citenamefont
  {Yoshida}, \citenamefont {Whitaker},\ and\ \citenamefont
  {Oliveira}}]{yoshida_renormalization-group_1990}%
  \BibitemOpen
  \bibfield  {author} {\bibinfo {author} {\bibfnamefont {M.}~\bibnamefont
  {Yoshida}}, \bibinfo {author} {\bibfnamefont {M.~A.}\ \bibnamefont
  {Whitaker}}, \ and\ \bibinfo {author} {\bibfnamefont {L.~N.}\ \bibnamefont
  {Oliveira}},\ }\href {\doibase 10.1103/PhysRevB.41.9403} {\bibfield
  {journal} {\bibinfo  {journal} {Phys. Rev. B}\ }\textbf {\bibinfo {volume}
  {41}},\ \bibinfo {pages} {9403} (\bibinfo {year} {1990})}\BibitemShut
  {NoStop}%
\bibitem [{\citenamefont {Campo}\ and\ \citenamefont
  {Oliveira}(2005)}]{campo_alternative_2005}%
  \BibitemOpen
  \bibfield  {author} {\bibinfo {author} {\bibfnamefont {V.~L.}\ \bibnamefont
  {Campo}}\ and\ \bibinfo {author} {\bibfnamefont {L.~N.}\ \bibnamefont
  {Oliveira}},\ }\href {\doibase 10.1103/PhysRevB.72.104432} {\bibfield
  {journal} {\bibinfo  {journal} {Phys. Rev. B}\ }\textbf {\bibinfo {volume}
  {72}},\ \bibinfo {pages} {104432} (\bibinfo {year} {2005})}\BibitemShut
  {NoStop}%
\bibitem [{\citenamefont {Žitko}\ and\ \citenamefont
  {Pruschke}(2009)}]{zitko_energy_2009}%
  \BibitemOpen
  \bibfield  {author} {\bibinfo {author} {\bibfnamefont {R.}~\bibnamefont
  {Žitko}}\ and\ \bibinfo {author} {\bibfnamefont {T.}~\bibnamefont
  {Pruschke}},\ }\href {\doibase 10.1103/PhysRevB.79.085106} {\bibfield
  {journal} {\bibinfo  {journal} {Phys. Rev. B}\ }\textbf {\bibinfo {volume}
  {79}},\ \bibinfo {pages} {085106} (\bibinfo {year} {2009})}\BibitemShut
  {NoStop}%
\bibitem [{\citenamefont {Georges}\ \emph {et~al.}(1996)\citenamefont
  {Georges}, \citenamefont {Kotliar}, \citenamefont {Krauth},\ and\
  \citenamefont {Rozenberg}}]{Georges1996}%
  \BibitemOpen
  \bibfield  {author} {\bibinfo {author} {\bibfnamefont {A.}~\bibnamefont
  {Georges}}, \bibinfo {author} {\bibfnamefont {G.}~\bibnamefont {Kotliar}},
  \bibinfo {author} {\bibfnamefont {W.}~\bibnamefont {Krauth}}, \ and\ \bibinfo
  {author} {\bibfnamefont {M.~J.}\ \bibnamefont {Rozenberg}},\ }\href {\doibase
  10.1103/RevModPhys.68.13} {\bibfield  {journal} {\bibinfo  {journal} {Rev.
  Mod. Phys.}\ }\textbf {\bibinfo {volume} {68}},\ \bibinfo {pages} {13}
  (\bibinfo {year} {1996})}\BibitemShut {NoStop}%
\bibitem [{\citenamefont {Costi}\ \emph {et~al.}(1994)\citenamefont {Costi},
  \citenamefont {Hewson},\ and\ \citenamefont {Zlatic}}]{costi_transport_1994}%
  \BibitemOpen
  \bibfield  {author} {\bibinfo {author} {\bibfnamefont {T.~A.}\ \bibnamefont
  {Costi}}, \bibinfo {author} {\bibfnamefont {A.~C.}\ \bibnamefont {Hewson}}, \
  and\ \bibinfo {author} {\bibfnamefont {V.}~\bibnamefont {Zlatic}},\ }\href
  {\doibase 10.1088/0953-8984/6/13/013} {\bibfield  {journal} {\bibinfo
  {journal} {Journal of Physics: Condensed Matter}\ }\textbf {\bibinfo {volume}
  {6}},\ \bibinfo {pages} {2519} (\bibinfo {year} {1994})}\BibitemShut
  {NoStop}%
\bibitem [{\citenamefont {Krishna-murthy}\ \emph
  {et~al.}(1980{\natexlab{b}})\citenamefont {Krishna-murthy}, \citenamefont
  {Wilkins},\ and\ \citenamefont
  {Wilson}}]{krishna-murthy_renormalization-group_1980}%
  \BibitemOpen
  \bibfield  {author} {\bibinfo {author} {\bibfnamefont {H.}~\bibnamefont
  {Krishna-murthy}}, \bibinfo {author} {\bibfnamefont {J.}~\bibnamefont
  {Wilkins}}, \ and\ \bibinfo {author} {\bibfnamefont {K.}~\bibnamefont
  {Wilson}},\ }\href {\doibase 10.1103/PhysRevB.21.1044} {\bibfield  {journal}
  {\bibinfo  {journal} {Phys. Rev. B}\ }\textbf {\bibinfo {volume} {21}},\
  \bibinfo {pages} {1044} (\bibinfo {year} {1980}{\natexlab{b}})}\BibitemShut
  {NoStop}%
\end{thebibliography}
\end{document}